\documentclass[12pt,letterpaper,oneside, openany]{report}
\usepackage{amsfonts} 
\usepackage{amssymb}  
\usepackage{graphicx} 
\usepackage{amsmath}  
\usepackage[english]{babel}
\usepackage[latin1]{inputenc}
\usepackage{latexsym}
\usepackage[amssymb]{SIunits} 
\usepackage{color}
\usepackage[]{dsfont}
\usepackage[]{subfigure}
\usepackage{fancyhdr}
\usepackage[T1]{fontenc}
\usepackage{geometry}

\begin{document}
\pagestyle{empty}
\begin{titlepage}
\begin{center}
{\large \bf INSTITUTE OF CYBERNETICS MATHEMATICS AND PHYSICS}
\end{center}

\begin{center}
\vspace{0.3cm}

{\large \bf DEPARTMENT OF THEORETICAL PHYSICS}

\vspace{3.5cm}

\begin{figure}[h!t]
\begin{center}
\includegraphics[width=0.25\textwidth]{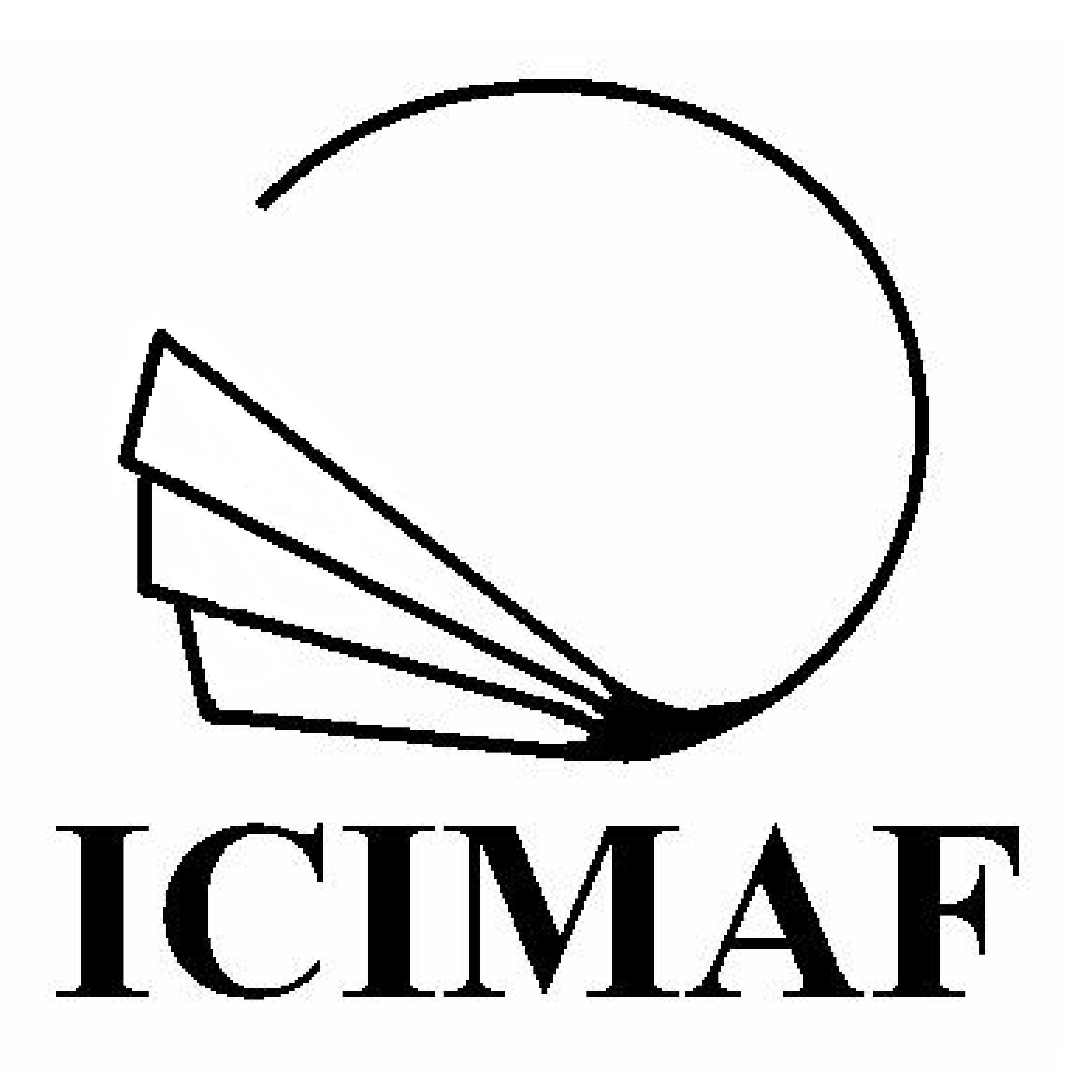}
\end{center}
\end{figure}

\vspace{0.5in}

\begin{center}
{\Large \bf STRANGELETS UNDER STRONG MAGNETIC FIELDS}
\end{center}
\vspace{0.5in} {\large \it presented thesis for the degree in}\\
{\large \bf Master in Physical Sciences}\\
\end{center}

\vspace{0.6in}

{\large \bf Author:}\hspace{0.3cm}{\large BSc. Ernesto López Fune.} \\

{\large \bf Supervisor:} \hspace{0.3cm}{\large{Dra. Aurora Pérez
Martínez.}}

\begin{center}
\Large{Havana. Cuba.}\\
\Large{2011}
\end{center}

\end{titlepage}

\newpage
\voffset=1.8in \thispagestyle{empty} \hoffset=3.6in
\noindent {\large{ \emph{...dedicada a papi, Jessy y Jenny}}}\\

\newpage
\hoffset=0in \voffset=1in \vspace{3in} \thispagestyle{empty} {\large
\bf    \emph{Acknowledgments:}}

\vspace{0.5in}

\emph{This thesis is the result of many efforts during these two intense years of graduate from the School of Physics. I want to first thank my supervisor Aurorita, for the dedication, patience; by inserting new, fresh and interesting ideas to my research topics... for being an excellent researcher and friend. I want to thank very much my family and dearest friends. To my work colleagues from ICIMAF: Hugo, Cabo, Elizabeth, Zochil, Augusto, Alain Ulacia and Alexander, for welcoming me in their group and stay during these two warm years. To my classmates, both, those who are and those who are abroad; to our research team: Richard, Daryel. To all thank you very much.}

\newpage
\hoffset=0in \voffset=1in \vspace{3in} \thispagestyle{empty}
\vspace{0.5in}

\section*{Resumen}

En la presente tesis se estudiar\'an tres propiedades fundamentales de los conglomerados de materia formada por quarks $u, d$ y $s$ llamados strangelets: la energía por barión, el radio y la carga eléctrica; todo en presencia de campos magn\'eticos intensos y temperatura finita. Dos casos nos ocupar\'a la atenci\'on: strangelets en la fase desapareada, donde no existe restricci\'on del n\'umero de sabores de quarks, y un caso particular de la fase superconductora de color, donde si existe dicha restricci\'on y una energ\'ia de gap adicional. Estudiaremos la estabilidad de los strangelets, medida por la energ\'ia por bari\'on, para comparar con la del $^{56}\text{Fe}:$ el is\'otopo más estable que existe en la Naturaleza. Empleamos el formalismo de Gota L\'iquida del Modelo de Bag del MIT para describir la interacci\'on entre los quarks. Llegamos a la conclusión de que los efectos del campo tienden a disminuir la energía por barión de los strangelets y la temperatura produce el efecto contrario. Se muestra además que los strangelets en la fase superconductora de color son más estables que aquellos en la fase desapareada, para una energ\'ia de gap del orden de $100\;\text{MeV}$. El radio de los strangelets muestra un comportamiento an\'alogo al de los n\'ucleos, con respecto al n\'umero bari\'onico y presenta poca variaci\'on tanto con el campo como con la temperatura. Se obtiene que la presencia de campos magnéticos modifican los valores de la carga el\'ectrica con respecto al caso no magnetizado, siendo estas mayores (menores) para los strangelets en la fase desapareada (superconductora).

\section*{Abstract}
In this thesis is studied three fundamental properties of clusters of matter made of quarks $u,d$ and $s$ called strangelets: the energy per baryon, the radius and the electric charge, all in the presence of intense magnetic fields and finite temperature. Two cases will take our attention: unpaired phase strangelets, where there is no restriction  to the number of flavors of quarks, and a particular case of the color superconducting phase, where exists a restriction to the quark numbers and an additional energy gap. We study the stability of strangelets, measured by the energy per baryon, to compare later with that of the $^{56}\text{Fe}:$ the most stable isotope known in nature. We use the Liquid Drop formalism of the Bag Model MIT to describe the interaction between quarks. We conclude that the field effects tend to decrease the energy per baryon of strangelets and temperature produces the opposite effect. It is also shown that strangelets in the color superconducting phase are more stable than those in the unpaired phase for an energy gap of about $100\, \text{MeV}$. The radius of strangelets shows an analogous behavior with the baryon number, as that of the nuclei, and shows small variations with the magnetic field and temperature. It is obtained that the presence of magnetic fields modify the values of the electric charge regarding the non-magnetized case, being these higher (lower) for strangelets in the unpaired phase (superconducting).

\newpage
\voffset=0in 
\pagestyle{fancy}
\pagenumbering{roman} \setcounter{page}{1}
\tableofcontents \pagebreak \pagenumbering{arabic}
\setcounter{page}{1}


\section{General Introduction}\label{cap0}

One of the main achievements of Theoretical Physics of the last century, is the formulation of the Standard Model of Particle Physics; among its contributions, includes the idea that the ``visible'' matter of our Universe is composed of fermions as building blocks. These interact through the so-called \emph{gauge bosons}: the \emph{photon} (electromagnetic interaction), the \emph{bosons $W^{\pm}$ and $Z^{0}$} (weak interaction), and 8 \emph{gluons g} (strong interaction) \footnote{The Standard Model is not a complete theory of the fundamental interactions because it does not include gravity (the fourth interaction), whose boson gauge is the \emph{graviton} $G$. There is also a large number of numerical parameters, including mass and coupling constants to be introduced in the theory, instead of being derived from first principles.}. Within the group of fermions are the quarks and leptons (and their antiparticles), which, according to the Standard Model, are grouped into three generations, as shown in the following table:

\vspace{0.2cm}
\begin{table}[h!t]
\centering
\begin{tabular}{|c|ccc|cc|}
\hline
& &\multicolumn{2}{c}{\;\;\;\;\;\;\;\;Leptons} \vline& \multicolumn{2}{c}{Quarks}\vline\\
 \cline{2-6}

Generation & & Name&Symbol&Name&Symbol\\
\hline
$1^{ra}$ &  &Electron neutrino& $\nu_{e^{-}}$ &Up &$u$ \\
         &  &Electron  & $e^{-}$&Down & $d$ \\

\hline
$2^{da}$&  &Muon neutrino& $\nu_{\mu}$ &Charm &$c$ \\
         &  &Muon  & $\mu$&Strange & $s$ \\

\hline
$3^{ra}$ &  &Tau neutrino& $\nu_{\tau}$ &Top &$t$ \\
         &  &Tau lepton  & $\tau$&Bottom & $b$ \\
\hline
\end{tabular}
\caption{\label{tableME} Leptons and quarks: building block of the Standard Model.} \vspace{0.2cm}
\end{table}

\begin{figure}[h!t]
\centering
\includegraphics[width=0.45\textwidth]{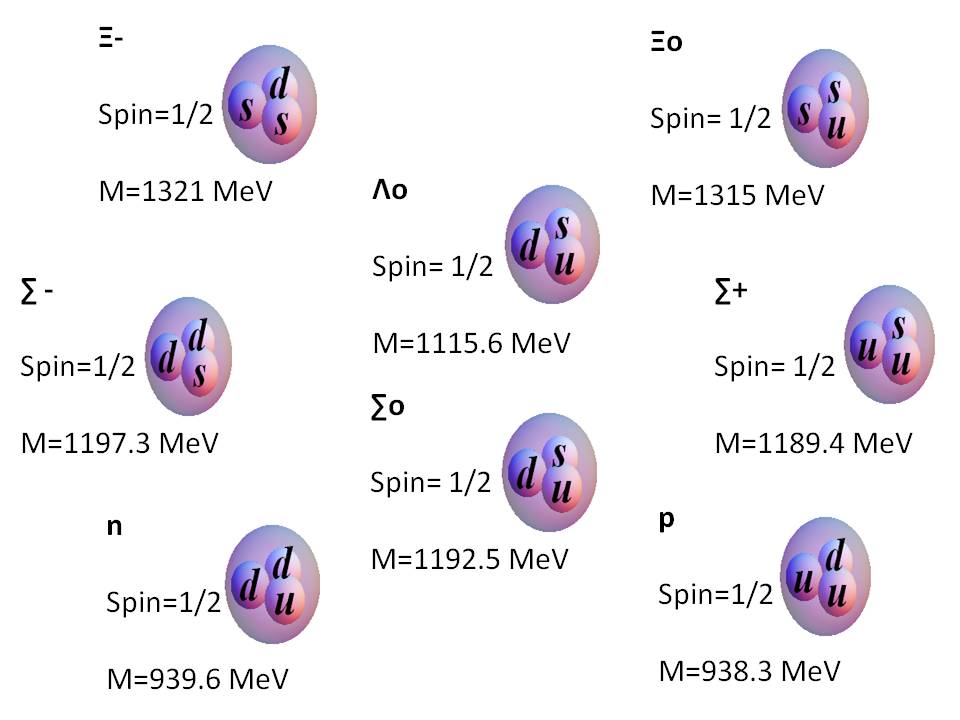}
\includegraphics[width=0.45\textwidth]{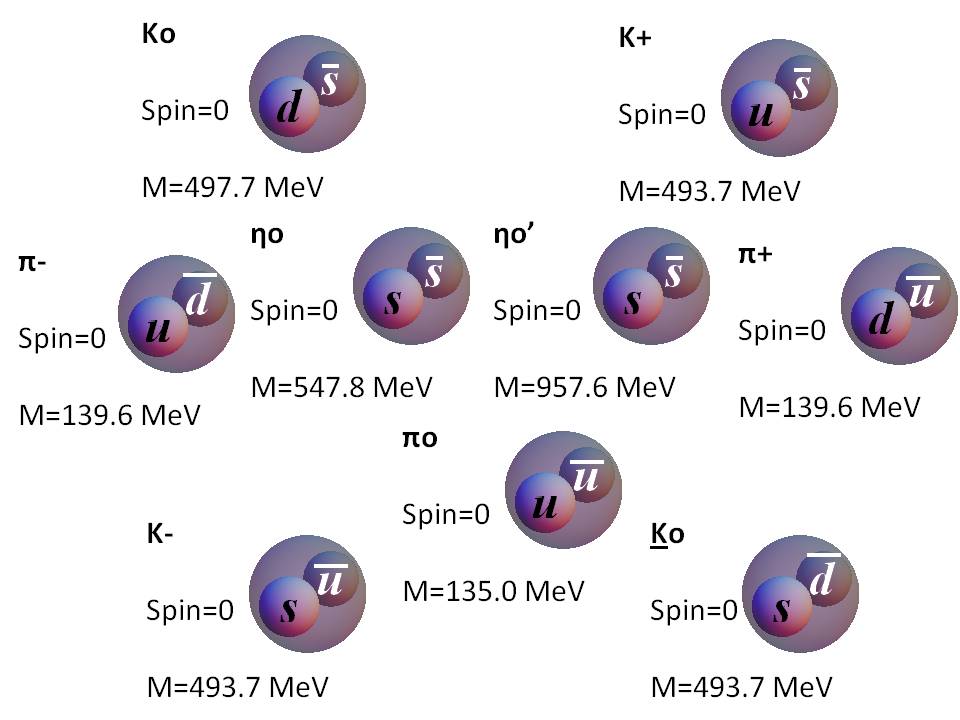}\\
\caption{\textbf{\small{Eightfold way construction of the baryon octet and meson nonet via quark combinations:}} \footnotesize{In the figure is illustrated, in the left panel, the spin $1/2$ baryon octet, in which are included the neutron, proton and the hyperons $\Xi^{\pm},\Lambda^{o},\Sigma^{\pm},\Sigma^{o}$, which are heavier and unstable particles. The right panel shows the quark composition of the spin $0$ meson nonet; among them, are included the pion triplet $(\pi^{0},\pi^{-},\pi^{+}),$ very well know in Nuclear Physics.}}\label{barionmeson}
\end{figure}

Quarks and antiquarks, through the three fundamental interactions of the Standard Model, can form states of two and three particles called \emph{hadrons}. The latter are divided into two subfamilies: the \emph{baryons} (three quarks combinations) and \emph{mesons} (combinations of a quark and an antiquark). Among the best known hadrons are: the \emph{protons} and \emph{neutron}, which as we know, are in atomic nuclei; see Fig.(\ref{barionmeson}).

\newpage


The interaction arising between quarks to form hadrons, is know as Strong Interaction which is mediated through the exchange of gluons, and mathematically it is described by the Quantum Chromodynamics Field Theory (QCD). QCD is a quantum field theory, whose gauge group \footnote{The $QCD$ equations are invariant under transformations of the gauge group SU(3).} is non-abelian \cite{Yndurain}. This makes QCD a highly non-linear theory that has special features very different from those of Quantum Electrodynamics (QED). In particular, QCD has two very important properties: \emph{asymptotic freedom} and \emph{color confinement}. Asymptotic freedom is the property by which the quarks interaction disappears in the limit of infinite momentum. In other words, at very high energies, quarks behave as quasi-free particles; therefore the interaction could be studied using perturbative methods. The theoretical description of asymptotic freedom was developed by scientists David Gross, Frank Wilczek \cite{GW} and David Politzer \cite{Politzer}, who received for such a work the Nobel Prize for Physics in 2004. In the years $'60s$, the asymptotic freedom was first observed experimentally at Stanford Linear Accelerator (SLAC) at energies in the range $1\text{GeV}-1\text {TeV} $ \cite{SB}.

By contrast, for intermediate or low energies ($\lesssim 1 \text{GeV}$), the QCD equations of motion, in their natural form, are non-linear due to the non-abelian character of the gauge group. This is the regime where the second feature of QCD plays a fundamental role: color confinement; the interaction energy between quarks increases with the separation between them. These two particular features of QCD, make possible to conceive, at least theoretically, the fact that at very high densities and/or temperatures, quarks and gluons can be unconfined or quasi-free, because the ``intensity'' of the strong interaction decreases with the increasing energy.

\bigskip

Color confinement is responsible for not being able to observe free quarks in Nature, but in the form hadrons; however, numerical calculations of non-perturbative QCD, predict the existence of a critical temperature $T_{c}\simeq150\sim180 \text{MeV},$ above which, they lose their individuality, producing an unconfined quark and gluon plasma (QGP). However, there's a lack of observational evidence that such a plasma exists in Nature, but currently there are developing experiments to produce such a state. Among the most important experiments are found, the the Large Hadron Collider (Large Hadron Collider: LHC) at CERN and Ion Collider Relativistic Heavy (Relativistic Heavy Ion Collider: RHIC) \cite{RHICI} in Brookhaven, New Jersey; in which very high temperatures will be achieved, producing a kind of ``modern version'' of the early universe.

While those experiments are waiting to be realized, on the other hand, astrophysicists continue to search for information coming from astrophysical and cosmological origins, that could prove the existence of the free QGP in Nature. In astrophysical environments, for example: inside Neutron Stars (NS), matter is so compressed that could reach densities higher than normal nuclear density $n_{b}=3.93\times10^{14}$ ~ g cm$^{-3}$. These objects therefore constitute stellar laboratories ``par excellence'' \cite{Baym:2006rq}.

\bigskip

Due to the high densities in the NS nuclei, a phase transition of the neutrons fluid (and other hadrons) could occur, to a gaseous phase of unconfined quarks and gluons, as suggested by QCD; in this way, they could explain some of the anomalous \emph{X-rays} bursts observations, high rotational speeds and low surface temperatures, found in pulsars and NSs that do not match the predictions made by the canonical models. It would take place then in the neighboring regions to the stellar core, a conversion of the hadronic matter, in particular, neutrons and protons into Strange Quark Matter (SQM): matter made of quarks $ u $, $ d $ and $ s $ in thermodynamic equilibrium with gluons. This phase transition was first conjectured in 1971 by A. R. Bodmer ~\cite{Bodmer:1971we}; and if it's true, one could expect the discovery of even more exotic compact objects than the NSs: stars formed by quarks or Quark Stars (QSs). This conjecture also ensures that the SQM at high density, low temperature and zero pressure, is the most stable state of matter in Nature; even more stable than the iron isotope $^{56}$Fe.

Theoretical studies of QCD, support the idea that at very high densities and low temperatures, the SQM could transit to a new phase of even lower energy: the \emph{color superconductivity} phase. This state has similar properties to ordinary superconductivity as it is formed by analogous ``Cooper pairs'' due to the color interactions between quarks, with the appearance of the corresponding energy gap. It stands in color superconductivity, the Color-Flavor Phase-Locked (CFL) ~\cite{Bailin:1983bm, Alford:2001zr}, where the quarks color and flavor charges are correlated to form a symmetrical pattern.

\bigskip

If Bodmer's conjecture is true, one might expect that atomic nuclei, consisting of protons and neutrons, constitute meta-stable and relatively long lifetime states of certain fragments of the SQM called \emph{strangelets} \cite {Farhi:1984qu}. Those ``chunks of strange matter'', made by $ u, d $ and $ s $ quarks, in equilibrium through mechanisms of the weak interactions, were first studied in \cite{Farhi:1984qu}, as bounded states of the SQM with baryon numbers $A\leq 10^{7}$. The discovery of these states may be the decisive proof that there's SQM and/or free QGP in Nature. Possible sources of production of strangelets could be: the collision of Compact Objects, Supernovae Explosion, Heavy Ion Relativistic Colliders as the LHC and RHIC ~\cite{RHICII}, or even from Cosmic Rays emitted during the primary formation of the SQs ~\cite{Klingenberg:2001qs, Finch:2006pq}.

\bigskip

There are no doubts about of the importance of the magnetic field and temperature in the stability of the SQM; both in the astrophysical context \cite{Felipe:2007vb, Felipe:2008cm, Martinez:2010sf, Ferrer:2010wz, Chakrabarty:1996te}, as well as in the accelerators of relativistic particles environment. Indirect measurements show the existence of very strong magnetic fields, of the order of $10^{13}-10^{15}$~G, which may be even higher $\simeq10^{19}\;\text{G}$ in Compact Objects nuclei such as pulsars, magnetars, NSs, and \emph{X}~ray emitting sources \cite{Duncan:1992hi, Kouveliotou:1998ze}. In this context, the effects of temperature are dispensable compared to magnetic field and high densities; however, temperature plays an important role in environments low densities as heavy ion colliders, as well as strong magnetic fields.

Strangelets, as being objects of great interest to both, the astrophysics and experiments in particle accelerators community, to their study in the presence of strong magnetic fields and taking the effects of temperature, we will focus in this thesis. To describe the interaction between quarks, a phenomenological model of QCD will be used: the MIT Bag Model \footnote{Massachusetts Institute of Technology.}, in particular, the Liquid Drop Model formalism, developed in Refs \cite{Farhi:1984qu, Madsen:1998uh} and used successfully in Refs ~\cite{Madsen:1998uh, Chao:1995bk, Paulucci:2008jd, Wen:2005uf} to study the strangelets at zero magnetic field. With the inclusion of the magnetic field and temperature, the results can be applied both in the astrophysical context as well as in heavy ion colliders. I will take into account also the effects of magnetic field and temperature on the surface properties of the quark and gluon gas, with the aim of studying the size of strangelets, their electric charge and stability,  thereof measured by their energy per baryon. Predate to this study, there are the works of \cite{Farhi:1984qu, Madsen:1998uh, Chao:1995bk, Paulucci:2008jd, Wen:2005uf, Gilson:1993zs, Madsen:2001fu}, where the basic properties of strangelets were studied at finite temperature and zero magnetic field. Both studies took into account the hydrostatic equilibrium of strangelets, to obtain stable or metastable configurations depending on a set of characteristic parameters and in particular the phases of the SQM of what they are made of. In this thesis, I will analyze separately strangelets formed by the SQM in the unpaired and paired phases CFL ~\cite{Paulucci:2008jd, Madsen:2001fu}, both in the presence of strong magnetic fields.

The thesis is organized as follows. The first chapter is introductory and it is devoted to discuss some of the important properties of the QCD and the most common phenomenological models used in the description of the strong interactions; which will then be applied to my study. The MIT Bag model will be described and explained the Liquid Drop Model formalism; for which turns out to be easier to incorporate the effects of the magnetic field and temperature, as well as obtaining equilibrium configurations of strangelets by minimizing the free energy with respect to the volume. I provide a brief discussion on the stability of the SQM and the possible phases in which it could be found.

The second chapter is reserved to discuss the main properties of strangelets and the most common models used to study them; in particular, I will focus on the Liquid Drop Model formalism of the MIT Bag Model, which will guide our study. I will discuss those thermodynamic expressions that will be useful to study the energy per baryon, the radius and the Debye screening of the electric charge within strangelets; and the contribution of the Coulomb energy. Finally, I will cite some of the experiments that are being carried out to detect strangelets in Nature.

The third chapter is devoted to discuss the role of the magnetic field on strangelets: the modifications that it produces on the surface and curvature terms respectively; while the original part of the thesis is reserved for the fourth chapter, in which I will write the mechanical equilibrium conditions and write the equations that one must solve to obtain stable configurations of strangelets. First I will do it for strangelets made of  Magnetized Strange Quark Matter (MSQM or unpaired phase) and then for strangelets of SQM in the Magnetized CFL (MCFL) phase. The results and discussion of the main physical properties for both types of strangelets will be shown. Posteriorly, I will present the conclusions and recommendations of this thesis, as well as directions for future work.
\chapter{The Strange Quark Matter}\label{cap1}

In this first chapter the key concepts for our study will be introduced. A \emph{grosso modo} description will be given of the main properties of QCD and some of the consequences of this theory; then I consider one of the alternative phenomenological models of QCD: the MIT Bag model. Following this, I will discuss about the hypothesis of the existence of the SQM, its stability and the color superconductivity states in which it could be found.

\section{Quantum Chromodynamics and Phenomenological Models}\label{sec1.1}

The existence of quarks, idea proposed by Murray Gell-Mann and Kazuhiko Nishijima in the early $'60s $ of the last century, allows a better understanding of the hadronic spectrum, describing them as combinations of quarks: particles of spin $1/2$. These, like the leptons of the Standard Model (see Table 1), have certain ``charges'' that make them susceptible to the fundamental interactions:

\begin{itemize}
\item Each quark can be found in three states of ``color charge'': ``red'', ``green'' or ``blue''; then they undergo the strong interaction. None of the leptons possess color charge, in this sense they are said to be ``white'' and thus are deprived of such interaction.

\item The $u,$ $c$ and $t$ quarks carry an electric charge of \footnote{In units of the positron electric charge: $e=0.302818$ in CGS system.}  $+\frac{2}{3}$, and the $d,$ $s$ and $b,$ quarks of $-\frac{1}{3};$ therefore, they are susceptible to the electromagnetic interaction. The electron, muon and tau lepton carry $-1$ of electric charge, which also interact via electromagnetic field as well. None of the neutrinos carry electric charge.

\item The ``flavor'', as well as the isotopic spin of quarks and leptons, allow them to interact weakly.
\end{itemize}

\bigskip

The dynamics of the interaction between quarks, through the exchange of gluons is studied by the QCD. It is a non-abelian quantum field theory, whose symmetry group is SU(3) with Lagrangian density given by:
\begin{equation}\label{LagrangianQCD}
\mathcal{L}_\mathrm{CDC}=\bar{\psi}\left(i\gamma^\mu D_\mu-\hat{m}\right)\psi-\frac{1}{4}G^a_{\mu \nu}G^{\mu \nu}_a,
\end{equation}

\noindent where $G^{a}_{\mu}(x)$ and $\psi(x)$ represent the gauge fields or gluons and the quark flavor fields $(u,d,s,c,t,b)$, respectively; $\hat{m}=\text{diag}(m_{u},m_{d},m_{s},\ldots)$ is the mass matrix, $\gamma^{\mu}$ are the Dirac matrices, while the covariant derivative is given by
\begin{equation}\label{DCovQCD}
D_{\mu}=\partial_{\mu}-igT_{a}G^{a}_{\mu},
\end{equation}
\noindent which contains the details of the strong interaction, being $g$ the corresponding coupling constant and $a$, indexing the color charge. The Gell-Mann matrices $T_{a}$, or SU(3) generators, satisfy the corresponding commutation relations
\begin{align}
\left[T_{a},T_{b}\right]&=f_{abc}T^{c},\label{ConmGellman}\\
f^{123}&=1, f^{147}=f^{165}=f^{246}=f^{257}=f^{345}=f^{376}=\dfrac{1}{2}, f^{458}=f^{678}=\dfrac{\sqrt{3}}{2}\label{StructConsts},
\end{align}
\noindent being $f_{abc}$ the totally antisymmetric structure constants in their three indices\footnote{In Eq.(\ref{StructConsts}) are shown the non-zero structure constants.} The gluon field tensor $G^a_{\mu\nu}(x)$ is expressed via $G^{a}_{\mu}(x)$ through the usual relation:
\begin{equation}\label{gluonstrength}
G^a_{\mu\nu}=\partial_\mu G^a_{\nu}-\partial_\nu
G^a_\mu-gf^{a}_{bc}G^b_\mu G^c_\nu.
\end{equation}

The precense of a third term in Eq.(\ref{gluonstrength}), leads to non-linear equations of motion for the fields $G^a_{\mu}(x)$ and $\psi(x)$, which rarely have analytic solutions; neither using perturbation theory since at the energy scales nowadays accessible to experiments, the strong coupling constant evaluates nearly $g^{2}\sim\alpha_{c}\simeq1,$ in contrast to QED in which $e^{2}\sim\alpha_{\gamma}\simeq1/137;$ for this reason, one has to use alternative phenomenological perturbative methods.

On the other hand, there's a subtlety in QCD due to the non-abelian property of its gauge group: the asymptotic freedom. Since not all the structure constants are zero, in Eq.(\ref{LagrangianQCD}) there appears a term proportional to $g^{2},$ which in the one-loop approximation \cite{MBuballa2005} one can be compute the coupling constant and gives
\begin{equation}
\dfrac{g^{2}}{4\pi}\simeq\dfrac{4\pi}{(11-\frac{4}{3}N_{f})\ln(\mu^{2}/\Lambda_{QCD}^{2})},
\end{equation}
\noindent where $N_{f}$ is the quark flavor number, $\mu$ is a scale chemical potential and $\Lambda_{QCD}\sim200-300\;\text{MeV}$ is the renormalization cutoff. From this, one can notice that at high densities $\sim\mu^{3}$, $g^{2}$ is almost negligible, reason why quarks can be treated as quasi-free particles.

Another less obvious property of the QCD is the color confinement, although it is accepted as intrinsic in the theory, this idea has only intuitive and phenomenological character, as there's no formal evidence, at least by now, obtained from the corresponding equations. This intuitive feature, is a phenomenon caused by the strong coupling between quarks, where non-perturbative effects predominate at low energy scales.

\bigskip
\bigskip

An alternative method to give solution to the QCD equations, is based in a space-time discrete as a lattice. In particular, for QCD, the lattice model associated is called in the scientific literature as Lattice QCD. It simulates a discrete space-time rather than continuous, which introduces a threshold value in the linear momentum of the order of $1/\rho$, where $\rho$ is the grid spacing. The latter connections are made through straight lines forming a net in space-time, whose vertices are occupied by quarks, and the gluons exchange occurs through these connections. As the lattice spacing decreases, a more realistic description of the problem is obtained. This new description allows to alternatively study properties of QCD, and is by far, the best approach existing to solve in a non-perturbative way the equations of this theory \cite{Lepage2003}.

One of the main achievements of these techniques, is predicting a ``deconfinement'' transition of the quark phase inside hadrons, to form the QGP at temperatures close to $T=170\;\text{MeV}$ \cite{Heller}. The main limitation is the high computational cost that involves to solve a dynamical system of nonlinear partial differential equations of second order. Such computational cost increases as it grows the number of quarks in each specific problem; for this reason, Lattice QCD is used only in problems of QCD where prevail low densities and high temperatures. Therefore, the main applications of these techniques are found in studies and experiments of very high energy phenomena, such as, collisions between ions and heavy nuclei in particle accelerators.

\bigskip

The ``material impossibility'' to describe the properties of QCD in the regime of high densities, has allowed to formulate two different categories of alternative phenomenological models, and applied them to the wide range of physical situations where QCD plays a primordial role; including the determination of Equations of State (EOS) that enables us to study the SQM in stable configurations. Static and dynamical models are both successfully used. An example of a static model is the MIT Bag Model, and a dynamical one is the Nambu-Jona-Lasinio Model (NJL). The first is static, in the sense that the quark masses have to be set in advance as input constants as well as the strong coupling constant $g$ and the Bag parameter $B_{\text{bag}}$, reproducing confinement as it will be shown later; while the second is dynamic, in the sense that the quark masses are determined self-consistently. The main shortcoming of this second model is that it does not reproduce the confinement of quarks; although it is noteworthy that recently has an extension to the Nambu-Jona-Polyakov-Lasinio Model(PNJL), where such effect is reproduced \cite{Ratti:2006wg}. From now on, we will focus on the study of the MIT Bag Model, which will be used in this thesis.

\subsection{MIT Bag Model}\label{epig1.1.1}

The phenomenological MIT Bag Model, introduced in the mid $'70s$ \cite{CH1, CH2, DG}, at the beginning was used to describe the hadronic matter; however, due to its versatility, its uses have spread to the search of EOS for the SQM, as well as the description of strangelets \cite{Farhi:1984qu, Madsen:1998uh, Chao:1995bk, Paulucci:2008jd, Gilson:1993zs}. This model describes free quarks confined in a limited region of space of closed surface $\mathcal{S}$ called the Bag, characterized by a $\delta$-Dirac potential.

Mathematically it is formulated through the Lagrangian density ~\cite{CH1, CH2, DG}:
\begin{equation}\label{LMIT}
\mathcal{L}_{\text{MIT}}=\left(\bar\psi(i\gamma^\mu D_\mu-m)\psi-\frac{1}{4}G^{a}_{\mu\nu}G_{a}^{\mu\nu}-B_{\text{bag}}\right)\theta_{s}-\dfrac{1}{2}\bar\psi\psi\delta_{s}\;.
\end{equation}
\noindent The factor $\theta_{s}$, represents a step function, taking the value 1 inside the volume enclosed by the surface $\mathcal{S}$ and 0 outside, reproducing phenomenologically the color confinement of QCD with the aid of the Bag constant $B_{\text{bag}}$. The last term of Eq.(\ref{LMIT}), represents a surface density term, whose interpretation is the following: quarks are distributed within the volume always running away from the surface.

Eq.\eqref{LMIT} by construction is invariant under Poincaré transformations, therefore, energy and momentum are conserved quantities and the equality $\partial\mathcal{T}^{\mu\nu}/\partial x^{\mu}=0$ holds, being $\mathcal{T}^{\mu\nu}$ the components of the energy-momentum tensor, which are given as
\begin{equation}\label{EMTensor}
\mathcal{T}^{\mu\nu}=\dfrac{\partial\psi}{\partial x^{\mu}}\dfrac{\delta\mathcal{L}_{\text{MIT}}}{\delta(\partial_{\nu}\psi)}-\delta^{\mu\nu}\mathcal{L}_{\text{MIT}},
\end{equation}
\noindent denoting by $\delta/\delta(\psi)$ the functional derivative. Taking this into account, one can impose the boundary conditions on the enclosing surface $\mathcal{S}$ for the quark and gluon fields $\psi(x)$ and $G^a_{\mu\nu}(x)$ respectively ~\cite{CH1, CH2}:
\begin{equation}\label{BagCurrentMIT}
\begin{cases}
B_{\text{bag}}=-\dfrac{1}{2}\eta_{\mu}\dfrac{\partial}{\partial
x^{\mu}}(\bar\psi\psi)-\frac{1}{4}G^{a}_{\mu\nu}G_{a}^{\mu\nu},\;\;\vec{x}\in\mathcal{S},& \\
&\\
\eta_{\mu}j^{\mu}=\eta_{\mu}\bar\psi\gamma^{\mu}\psi=0,\;\;\vec{x}\in\mathcal{S}, &\\
&\\
\eta_{\mu}G^{\mu\nu}_{a}=0,\;\;\vec{x}\in\mathcal{S}, &
\end{cases}
\end{equation}
\noindent where $\eta_{\mu}$ is an unitary normal vector to $\mathcal{S}$. The first boundary condition expresses the conservation of the total energy of the system, while the second and third, forbid the formation of quark and gluon color currents across the boundary surface.

\bigskip
\bigskip

This model was used in Refs \cite{CH1, CH2} to study the properties of hadrons, where the surface $\mathcal{S}$ plays an important role. Its uses have been also extended to situations where these surface effects are less important compared to the volumetric ones. In these cases are the QSs, which contain a high number of particles $\sim10^{56}$ and thus a thermodynamical formalism of the MIT Bag model is used by macroscopically averaging all physical quantities. It is then interpreted $B_{\text{bag}}$ as a negative pressure $P_{\text{vac}}=B_{\text{bag}}$ exerted on the system by the vacuum that surrounds it, mimicking the confinement \footnote{To be noted that in all the computations I will use the Natural System of Units, where $\hbar=1$, $c=1$ and $k=1$, where $k$ is the Boltzmann constant.}

\begin{figure}[h!t]
\centering
\includegraphics[width=0.5\textwidth]{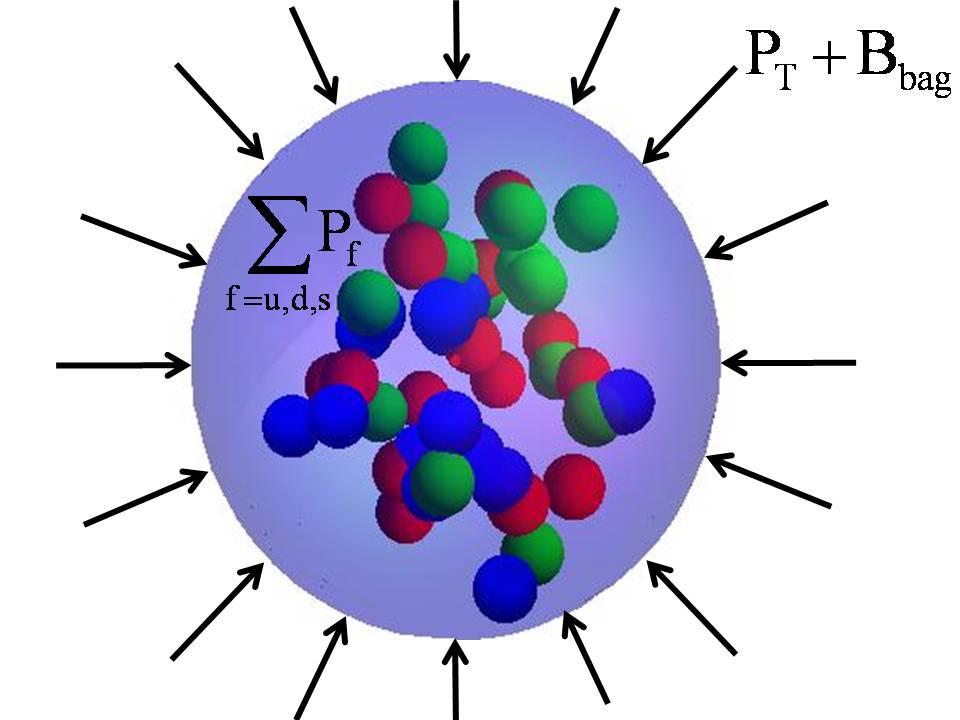}\\
\caption{Recreative illustration of how it is realized the color confinement in the MIT Bag Model for systems with high densities.}\label{MITBAG}
\end{figure}

The macroscopic average of the energy-momentum tensor given by Eq.\eqref{EMTensor}, for a perfect fluid (in co-moving coordinates) in a static and spherically symmetric space-time, it is diagonal and can be written as:
\begin{equation}\label{IdealEMTensor}
\langle{\cal T}_{\mu\nu}\rangle=\left(\begin{array}{llll}E & 0&0&0 \\0& P&0&0\\0&0&P&0\\ 0&0&0&P
\end{array}\right)
\end{equation}
\noindent being $E$ the total energy density and $P$ is the total pressure of the system. These two expressions, for an ideal gas made of quarks $u,d$ and $s$, can be written as
\begin{eqnarray}
E=\sum_{f=u,d,s}E_{f}+B_{\text{bag}},\;\;\;\;P=\sum_{f=u,d,s}P_{f}-B_{\text{bag}}\label{energypressureBAG},\\
E_{f}=\Omega_{f}+\mu_{f}N_{f}+T S_{f},\;\;\;\;P_{f}=-\Omega_{f}\;\;\;\;\;\;\;
\end{eqnarray}
\noindent where the sum is extended through the quarks flavor number; the terms $E_{f}$ and $P_{f}$ correspond to the kinetic contributions of quarks to the energy density and pressure given by $\Omega_{f}$, $\mu_{f}$, $N_{f}$, $T$, $S_{f}$, which are: the thermodynamical grand potential, the chemical potential, the particle density, temperature and entropy respectively; this is illustrated in Fig.(\ref{MITBAG}).

Despite its simplicity, it is a model widely used in the description of the SQM at high densities, which takes account of confinement and asymptotic freedom, which is essential for stars. Among its main deficiencies are: not being a dynamical model as the masses of quarks must be fixed in advance, like as well as $B_{\text{bag}}$. It also violates the chiral symmetry and confinement are described in terms of $ B_{\text{bag}}$ as a free parameter ~\cite{CH1, CH2}.

\section{The Bodmer-Witten-Terazawa hypothesis}\label{sec1.2}

The enigma of disintegration (at human accessible energy scales) of most of the hadrons (except the proton and its antiparticle) and the radioactivity of the atomic nuclei, makes us question about the stability of matter we know. This is one of the many puzzles of the millennium, as stated by David Gross. It is based on the knowledge of the fundamental state of matter and in which kind of environments this state could be found.

The hypothesis of Strange Quark Matter, idea proposed by A. R. Bodmer \cite{Bodmer:1971we} in 1971, gives a first answer to this question. This hypothesis states that the SQM is more stable than the $^{56}$Fe: the most stable isotope know in Nature. According to this hypothesis, then reinforced by E. Witten ~\cite{Witten:1984rs}, H. Terazawa ~\cite{Terazawa}, E. Farhi and Jaffe RL ~\cite{Farhi:1984qu}, the SQM is the ground state of the matter; one of the most intriguing speculations of Modern Physics. Its formation would require a rich environment of $s$ quarks or the direct formation of a QGP. It is important to note that this hypothesis introduces an upper bound for the values that can take the energy per baryon $E/A$ of the SQM, i.e., to preserve stability, it must always meet the relationship
\begin{equation}
\left.\dfrac{E}{A}\right|_{\text{SQM}}\leq\left.\dfrac{E}{A}\right|_{\;^{56}\text{Fe}}.
\end{equation}
\noindent This has been proven in Ref.\cite{Farhi:1984qu} for a wide range of values of the MIT Bag Model parameters. If true, it would be natural then to ask oneself: why if the SQM is the fundamental state of matter, are we not composed by it?

As stated in the \emph{Bodmer-Witten-Terazawa conjecture}, as is also known in the scientific literature, the fact that the SQM more stable than ordinary atomic nuclei, does not contradict at all what we see in everyday life. Nuclei do not decay into ``strange nuclei''. We know that the $s$ quarks decay into $u$ and $d$ quarks, electrons and neutrinos (antineutrinos) through Weak Interaction mechanisms; however, the inverse process to occur, requires a weak transition that makes $u$ and $d$ quarks, which make up the nucleon ($n$ and $p$), to transform into $s$ quarks, as shown in the chain reaction:
\begin{equation}\label{betadesint}
d \leftrightarrow u+e^{-}+\nu_{e},\;\;u+e^{-}+\overline{\nu_{e}}\leftrightarrow\; s,\;\;s+u\leftrightarrow d+u.
\end{equation}

\noindent In this case, the time that would take to materialize this transition is an eternity; times comparable to $10^{60}$ years, which is longer than the estimated age of the Universe. That's the reason why the matter we observe nowadays does not exhibit strangeness features; however, we know about the existence of forms of matter as old as the Universe itself: stars, where such mechanisms may take place ~\cite{Itoh:1970uw}. Therefore, one of the scenarios in Nature where the SQM might be present is reserved to the nuclei of NSs, where gravity is in charge of compressing matter, reaching densities of subnuclear orders and remain in that state for ages.

\subsubsection{Equation of State for the SQM}\label{acap1.2.1}

If Bodmer-Witten-Terazawa conjecture is true, then, to find an EOS for the SQM in high density and low temperature environments, under $\beta-$equilibrium and electric charge neutrality, which is supposed to be the natural habitat of this form of matter, would be of great importance to estimate characteristic parameters of QSs. In this subsection I show a first simple application of the MIT Bag Model to find an EOS for the SQM in stellar equilibrium.

\bigskip

The MIT Bag model, considers in this case free quarks moving inside the star, forming a gas of fermions; color confinement is ensured through the parameter $B_{\text{bag}}$, which has to be fixed in advance or calculated from some specific data from the star as the matter density, etc. For simplicity and understanding of what will be derived, I assume that the quarks $u, d$ and $s$ are in the ultra-relativistic regime, that is to say that their kinetic energies far exceed their rest energies or in a more simple way, assume $m_{f}\to0$, which is a great simplification since the $s$ quarks are very massive compared with $u$ and $d$. One then can obtain, under such assumptions, as it was already done in Ref.\cite{AuErMilRich} a very simple EOS for the quark gas; it would serve to study the mass-radius relations of stars, and the stability of the SQM itself. In this example, I do not take into account the presence of gluons, electrons and neutrinos, which could arise from the chain of reactions given by Eq.\eqref{betadesint}; in stellar equilibrium, it is supposed that time has passed enough such that neutrinos are no longer present \cite{AuErMilRich}.

From the degenerate limit\footnote{We use this approximation since in stars $T/T_{f}\sim10^{-4},$ where $T_{f}$ is the Fermi temperature.} $T=0$ and considering ultra-relativistic quarks, one can obtain the kinetic pressure for each quark gas:
\begin{equation}\label{ppotencial}
P_{f}=\dfrac{d_{f}}{24\pi^2}\,\mu_{f}^{4},\
\end{equation}
\noindent being $\mu_{f}$ the chemical potentials per quark flavor $f$ (Fermi momenta); the factor $d_{f}=6=3\times2$ counts the three color degrees of freedom times the two spin projections onto the $z-$axis.

The particle density and the energy density for each flavor are:
\begin{equation}\label{Npotencial}
N_{f}=\dfrac{d_{f}}{6\pi^2}\,\mu_{f}^{3},\;\;\;\;\;
\end{equation}
\noindent and
\begin{eqnarray}\label{Epotencial}
E_{f}=\dfrac{d_{f}}{8\pi^2}\, \mu_{f}^{4}\,
\end{eqnarray}
\noindent respectively. Notice that neglecting the presence of electrons and neutrinos, and taking into account the $\beta-$equilibrium condition, the chemical potentials satisfy the equations
\begin{equation}\label{betaequil}
\mu_{u}=\mu_{d}=\mu_{s}\equiv\mu,
\end{equation}
\noindent which imposes automatically the global electric charge neutrality due to the equality of the densities $N_{f}$, since $q_{u}+q_{d}+q_{s}=(2-1-1)/3=0$.

From Eqs.(\ref{ppotencial}) and (\ref{Epotencial}) one can obtain the relation $E_{f}=3P_{f}$, which substituting into Eqs.(\ref{energypressureBAG}), lead to the EOS we wanted to obtain:
\begin{equation}
P=\dfrac{1}{3}(E-4B_{\text{bag}})\,.
\end{equation}
The hydrostatic equilibrium condition $P=0$, allows the computation of the Bag constant as:
\begin{equation}\label{ep}
E=4 B_{\text{bag}},
\end{equation}
\noindent which means that the QSs are self bounded objects; the color confinement, through the negative pressure $B_{\text{bag}}$, is the responsible mechanism of the cohesion of matter inside the star.

\bigskip

Now we can proceed to estimate the value of $B_{\text{bag}}$ taking into account the typical density of a NS, equals to the normal nuclear density: $n_b=3.93\times10^{14}$~g cm$^{-3}\simeq0.16\;\text{fm}^{-3}$. Since we have ignored the presence of electrons and neutrinos, from Eq.\eqref{betaequil} one obtains:
\begin{equation}
\sum_{f=u,d,s} P_{f}=\dfrac{d_{f}}{8\pi^2}\,{\mu}^{4}\,.\label{presioni}
\end{equation}
\noindent Since the total pressure $P$ is zero, Eq.(\ref{energypressureBAG}) leads to the relation between the common chemical potential $\mu$ and $B_{\text{bag}}$:
\begin{equation}
\mu=\left({\dfrac{8 \pi^2 B_{\text{bag}}}{d_{f}}}\right)^{1/4}\,.\label{muBag}
\end{equation}

The star density coincides with the total baryon density, this relation is given by
\begin{equation}
n_{b}=\frac{1}{3}\sum_{f=u,d,s}N_{f}.\,\label{barion}
\end{equation}
\noindent Taking this into account, one can write $n_{b}=N_{f}=d_{f}\mu^{3}/(6\pi^2)\simeq 0.63$~fm$^{-3}$, which gives $\mu^{3}/\pi^2=0.63$~fm$^{-3}$. Finally, using Eq.(\ref{muBag}), one obtains, $B_{\text{bag}}^{1/4}\simeq145$~MeV.

\subsection{Stability of normal quark matter in stellar equilibrium}\label{epig1.2.1}

Now consider a gas of ultra-relativist quarks at $T=0$ composed only by $u$ and $d$ quarks; this state of the quark matter is called in the literature as \emph{normal quark matter}. The stellar equilibrium conditions imposes the charge neutrality, which in this case can be written as $$2N_u=N_d,\,$$
\noindent where for simplicity, no electrons and neutrinos have been taken into account. Defining $\mu_2$ in terms of the chemical potentials of the $u$ and $d$ quarks by $\mu_{2}=\mu_{u}=2^{-1/3}\mu_{d}$, then, the baryon density is
$$n_{b}=\frac{N_{u}+N_{d}}{3}=\frac{\mu_{2}^{3}}{\pi^{2}}\,.$$
\noindent The gas pressure is
$$ P_2=P_u+P_d=\frac{(1+2^{4/3})}{4\pi^2}\mu_2^4\,;$$
\noindent and substituting into Eq.\eqref{energypressureBAG} for the total pressure $P$, and imposing the hydrostatic equilibrium condition $P=0$, we get:
$$\mu_2=\left(\frac{4\pi^2}{1+2^{\frac{4}{3}}}\right)^{1/4}B_{\text{bag}}^{1/4}\,.$$
With this equation, one can express $n_{b}$ as a function of $B_{\text{bag}}$ and taking into account that $E=4B_{\text{bag}}$, the energy per baryon is
$$\frac{E}{A}=\frac{E}{n_{b}}=(2\pi)^{1/2}(1+2^{4/3})^{3/4}B_{\text{bag}}^{1/4}\,.$$
Evaluating for the value of $B_{\text{bag}}$ we found before: $B_{\text{bag}}^{1/4}=145~\text{MeV}$, one obtains
$$\left.\frac{E}{A}\right|_{u,d}\simeq 934~\text{MeV}.$$

Now let's explore one can very the values of $B_{\text{bag}}$, such that the energy per baryon is less than the corresponding to the normal nuclear matter. For a neutron gas, this energy is equal to the neutron mass $m_n=939,6\;\text{MeV}$, while for the isotope $^{56}$Fe, the energy per baryon can be computed as: $\left.E/A\right|_{^{56}Fe}=(56m_{n}-56\times8,8)/56=930$ MeV. The stability of normal matter with respect to the neutron gas, requires that $\left.\frac{E}{A}\right|_{u,d}<m_{n}$, which is always valid as long as $B_{\text{bag}}^{1/4}<145$ MeV. With respect to the isotope $^{56}$Fe, one has instead $\left.\frac{E}{A}\right|_{u,d}<\left.\frac{E}{A}\right|_{^{56}Fe}$, which puts a bound on $B_{\text{bag}}$ of $B_{\text{bag}}^{1/4}<144$ MeV. In Nature one can observe neutrons and the isotope $^{56}$Fe, but not normal matter, therefore, we can conclude that the values of $B_{\text{bag}}$ should be higher than those previously obtained. Usually in literature one takes the value $B_{\text{bag}}^{1/4}=144$ MeV as the lower bound of this parameter.

\subsection{Stability of strange quark matter in stellar equilibrium}\label{epig1.2.2}

In the case of a quark gas with the three flavors $u$, $d$ and $s$, one writes again the electric charge neutrality condition as,
$$2N_u=N_d+N_s\,.$$
\noindent Defining now $\mu_3=\mu_u=\mu_d=\mu_s,\,$ the expression for the pressure of the three flavor quark gas, given by Eq.(\ref{presioni}),  becomes
$$
P_3=P_u+P_d+P_s=\frac{3\mu_3^4}{4\pi^2}\,.
$$

The equivalent relation to Eq.(\ref{muBag}), between the chemical potential $\mu_{3}$ and $B_{\text{bag}}$, in this case is:
$$ \mu_3=\left(\frac{4\pi^2}{3}\right)^{1/4}
B_{\text{bag}}^{1/4}\,.
$$
Using the condition given by Eq.(\ref{ep}), and writing $n_{b}$ as a function of $B_{\text{bag}}$, one obtains the energy per baryon evaluated for the value of $B_{\text{bag}}^{1/4}=145$~MeV
$$
\left.\frac{E}{A}\right|_{u,d,s}=\frac{E}{n_{b}}=(2\pi)^{1/2}
3^{3/4} B_{\text{bag}}^{1/4}\simeq 829~\text{MeV}.
$$

Recall that the energy per baryon of a neutron gas is given by the neutron mass $m_n=939.6$~MeV, while for the isotope $^{56}$Fe it is $E/A=930$~MeV. One can conclude that
$$
\left.\frac{E}{A}\right|_{u,d,s}<\left.\frac{E}{A}\right|_{^{56}\text{Fe}}
< \left.\frac{E}{A}\right|_{u,d} < m_n\,,
$$
\noindent for values of $B_{\text{bag}}^{1/4}<163$ MeV, therefore, the SQM is the most stable.

\begin{figure}[h!t]
\centering
\includegraphics[width=0.5\textwidth]{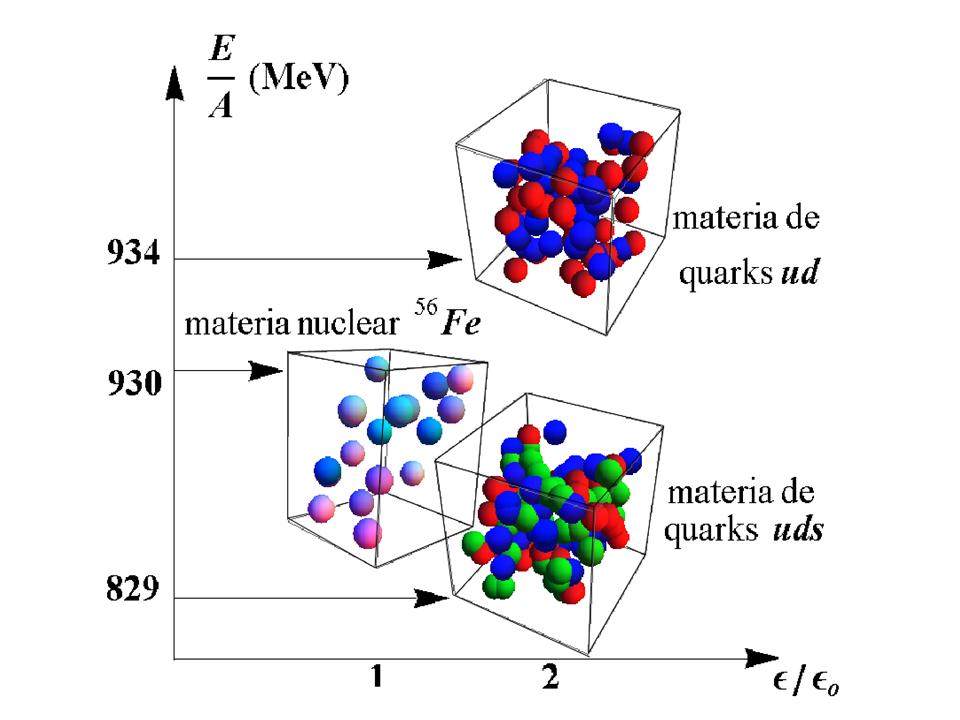}
\caption{Comparison of the energy per baryon $E/A$, for the isotope $^{56}$Fe,
normal and strange quark matter as a function of the baryon density ~\cite{Weber}.} \label{stable}
\end{figure}

It worths to emphasize that one earns approximately $100$ MeV per baryon when a new quark flavor is introduced in the description. This fact proves the Bodmer's conjecture using the MIT Bag Model: the SQM is more stable than nuclear matter and in particular, than the most stable nucleus that exists in Nature: the iron isotope $^{56}$Fe, see Fig.(\ref{stable}). We can understand, intuitively, that this result is due to the Pauli exclusion principle. In the case of the SQM, the baryon number is divided into three Fermi energy seas instead of two, as in the case of normal matter. Thus, in a regime of high densities, the state with unconfined quarks $u$, $d$ and $s$ will have a lower energy than the state with only two flavors $u$ and $d$.

\section{QCD phase diagram and color superconductivity}

The analysis performed in the preceding section, has taken into account the zero temperature ultra-relativistic limits; however, there are certain environments where those limits are not appropriate, they don't describe adequately the overall behavior of the system under study; for example, depending on the temperature $T$ and chemical potential $\mu$, quark matter in general, may appear essentially two regimes. The first would be a ``hot phase'' when $T\gg\mu$, forming the QGP. The Universe could have been probably in this stage the first seconds after the Big Bang, when the temperature was extremely high and very low baryon density. Another environment where one could find the QGP in the hot phase are heavy ion colliders, where the plasma temperature is very high, compared with the density.

As the QGP temperature keeps decreasing, the QGP undergo through a series of different stages, beginning with the nucleation of hadronic bubbles; mechanisms like this are the basis of the hadronization of the Universe. Strangelets are considered as possible remains of this process and they are  supposed to have survived till nowadays, in equilibrium, minimizing its free energy. Once formed hadrons, the first nuclear reactions and the synthesis of the light elements such as Deuterium and Helium\footnote{The QGP is considered the starting point for the synthesis of all the chemical elements.}. The latter, then they occupy the centers of stellar objects capturing other elements to form those super-dense bodies that register the astrophysicists.

On the other hand, the QGP can also appear in a low temperature and high density $T\ll\mu$ environment such as inside the NSs to later transform into QSs. This phase transition, would be occurring in the Universe each time a massive star explodes as a Supernova, with the consequent emergence of a NS. Once reached the equilibrium state in the interior of the NS, the QGP moves toward the SQM.

\begin{figure}[h!t]
\centering
\includegraphics[width=0.45\textwidth]{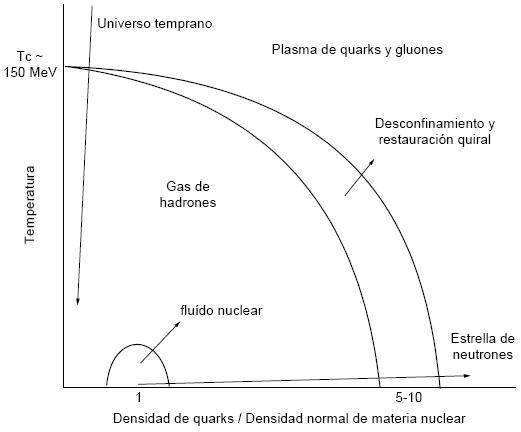}
\includegraphics[width=0.45\textwidth]{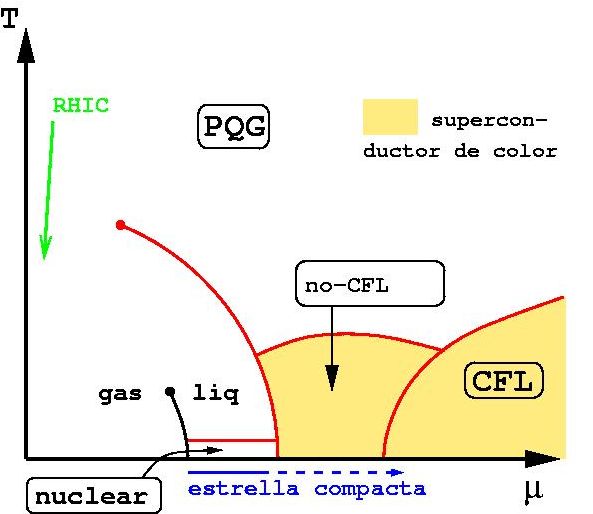}
\caption{\textbf{\small{Phase diagram of QCD:}}
\footnotesize{The figure shows the temperature and density regions at which matter exists as a nuclear fluid, hadron gas, or plasma of quarks and gluons. The line on the left shows the path that will follow RHIC experiments, and try to reproduce the conditions in the early Universe. It starts from the stage of the PQG until normal nuclear matter. The line at the bottom of the diagram, traces the path taken by a NS. Heavy ion collisions follow a path between these two cases, simultaneously increasing the temperature and density. In the right panel is shown an extension of the phase diagram where are also shown the regions where the SQM can be found in the color superconducting phase.}}\label{phased}
\end{figure}

A qualitative scheme of the situations described in the preceding paragraphs are shown in Fig.(\ref{phased}). At low temperatures and densities, the system can be described in terms of hadrons nucleons and internal excitation states as mesons, etc. In the regions of high temperatures $T\geq T_{c}$ and higher densities, $\sim5-10\;n_{b}$, the proper description is in terms of quarks and gluons. The transition between these regions can be abrupt, as in boiling water with a latent heat (\emph{first-order transition}), no latent heat (\emph{second order}), or just a gentle pace but fast (\emph{crossover}). In any case, the physics changes drastically between the two regimes of low and high temperatures. In Fig.(\ref{phased}) is shown also a second phase transition, from normal nuclei, which are in a liquid state to a gas of nucleons. These two phases can coexist at temperatures below $15-20\,$ MeV, and densities in the range of the normal nuclear matter $n_{b}$.

\bigskip

At relatively low temperatures and high densities, a new phase transition is conjectured to appear, from the SQM towards a more symmetric state: the color superconductivity state \cite{Alford:2001zr, Rajagopal:2000ff, Alford:2007xm}. This idea, first proposed in Ref.\cite{Ivanenko1969}, suggests that the color superconducting phase of the SQM, consists of diquark condensates, analogous to the electron Cooper pairs existing in ordinary superconductors. This latter phenomenon arises from the coupling of electrons whose fundamental interaction is repulsive. The attractive interaction needed for pairing electrons and form Cooper pairs, comes from the phonons or lattice vibrations. The difference with color superconductivity is that in this case, the attractive interaction arises itself naturally from the fundamental interactions of the theory, mediated by gluons.

Quarks are spin $\frac{1}{2}$ fermions, hence, they obey the Pauli exclusion principle. At zero temperature and high densities, asymptotic freedom keeps them free of interactions, filling all the energy levels $E_{p}$ till the Fermi energy $\mu_{F}$, which is represented by the Fermi-Dirac distribution function at $T=0:$ $f(p)=\theta(\mu_{F}-E_{p}).$ Quarks with energies near the Fermi level interact via the exchange of gluons; such interaction is attractive if both quarks are in a color antisymmetric state, which lead to the formation of quark pairs breaking locally the symmetry. As a consequence, there appears at least one energy gap $\Delta$ that characterizes the condensate \cite{alemanes}:

\begin{eqnarray}\label{DeltaCSC}
\Delta\propto\langle\psi^{t}\mathfrak{O}\psi\rangle\\
\mathfrak{O}=\mathfrak{O}_{\text{color}}\otimes\mathfrak{O}_{\text{sabor}}\otimes\mathfrak{O}_{\text{spin}},
\end{eqnarray}
\noindent where the operator $\mathfrak{O}$ acts on the color, flavor and spin states $\psi$.

Depending on the flavors and colors the pairs are formed, the color superconducting state is classified. The two most important patterns are: the pairs are formed between only two flavors $(u,d)$ or $(u,s),$ or even $(d,s)$, know as 2SC phase; and the most symmetrical pattern, where the three quark flavors $u,d$ and $s$ may form pairs $(u,d),$ $(u,s),$ and $(d,s)$. This last pairing pattern is known as Color Flavor Locked phase or CFL phase.

The 2SC phase is called like this because of the pairing between only two quark flavors, conventionally it always points in the blue-antiblue direction, i.e., quarks $u-$reds pair with quarks $d-$greens, while quarks $d-$reds pair with quarks $u-$greens, forming antiblue pairs; blue quarks do not pair \cite{alemanes} and therefore do not contribute to the energy gap. But for the CFL phase, such quark pairings are uniform, all contribute to this energy gap.

In astrophysics, color superconductivity became more popular when it was discovered that the value of the energy gap of the color superconductor, could reach $100\;\text{MeV}$, for baryon densities typical of the NSs. Theoretical results suggest that inside those stars may be a superconducting phase of the SQM, of CFL-type, which would be a more stable state than SQM ~\cite{Alford:2001zr, Rajagopal:2000ff, Alford:2007xm}.


%

\chapter{Strangelets and properties}\label{chap2}

\section{Strangelets}\label{sec2.1}

Among the biggest challenges of this century in stellar astrophysics, figure the proof if the SQM can actually exist inside dense compact objects. If the SQM or preferably the CFL phase is the fundamental state of matter, then we can expect to find inside NSs, forming Hybrid Stars or QSs ~\cite{Itoh:1970uw, Ivanenko:1965dg}. To check this, one needs indications that clearly justify the presence of quarks, for example, with the emission of particles from such celestial bodies.

\bigskip

One possible solution would be to detect strangelets from the interstellar medium. The formation of these may be due to the collision of compact objects, supernovae explosions or even from the first glimmerings caused by the conversion of NSs to QSs. Other studies also suggest that strangelets could be produced by collisions between heavy ions in the ultra-relativistic accelerators such as the LHC and RHIC ~\cite{RHICII}; experiments on cosmic rays are also in search of these particles ~\cite{Klingenberg:2001qs, Finch:2006pq}.

\bigskip
But, what are the strangelets? In Ref.\cite{Farhi:1984qu} were defined as small lumps of the SQM, whose dimensions are comparable with atomic nuclei, a few femtometres (fermi) $\text{fm}=10^{-13}\text{cm}$, and baryon numbers in the range $1\leq A <10^{7}$. As systems with a few quarks, one might think on Lattice QCD to study their main properties, however, the MIT Bag model has more applications, because the calculations are computationally less difficult.

\bigskip

\begin{figure}[h!t]
\centering
\includegraphics[width=0.3\textwidth]{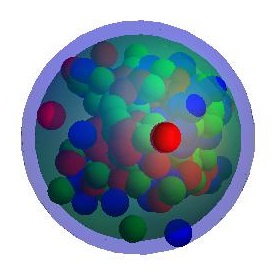}\vspace{0.5cm}
\includegraphics[width=0.5\textwidth]{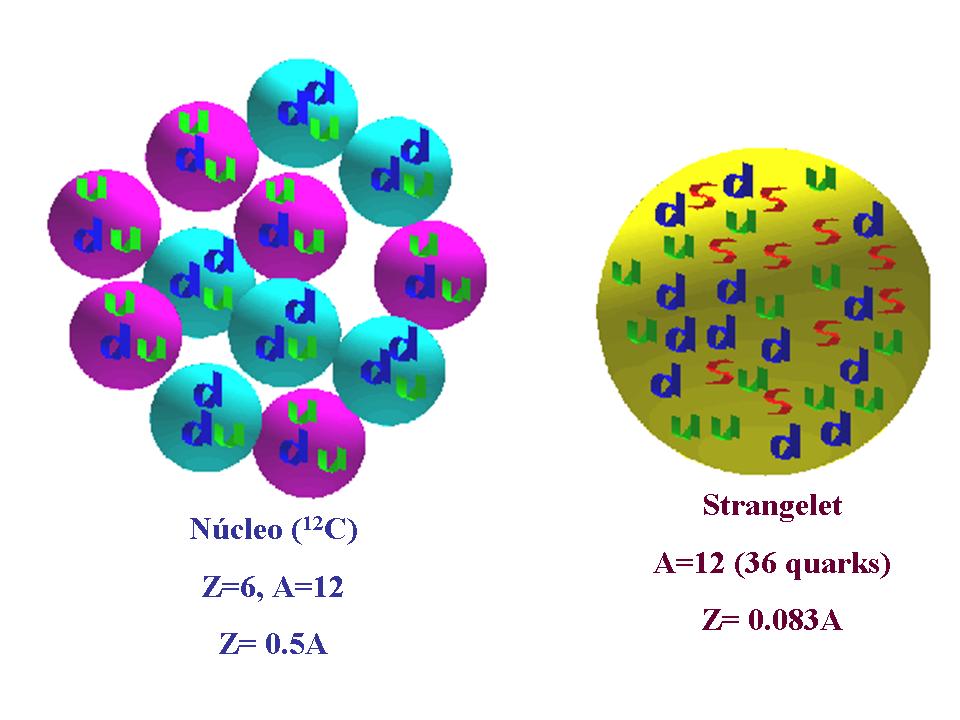}
\includegraphics[width=0.5\textwidth]{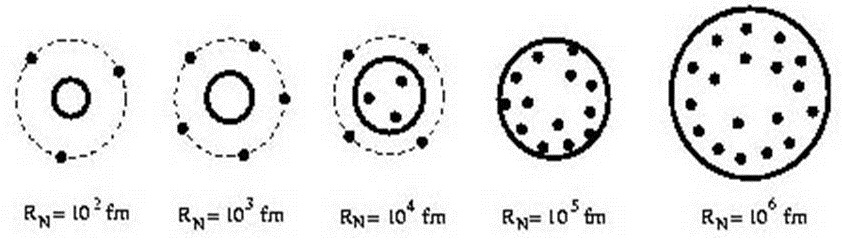}\\
\caption{In the left panel is shown an artistic recreation of a strangelet. The right panel shows a comparison between a stangelet and the Carbon isotope $^{12}$C with the same baryon number. The inferior panel shows a strangelet and an orbiting electron cloud; as the strangelet's radius increases, electrons can go inside the strangelet.}\label{strangelet}
\end{figure}

Recall that in the MIT Bag model, it is necessary to introduce an auxiliary surface $\mathcal{S}$ to reproduce the confinement of quarks. In the works of \cite{Farhi:1984qu, Madsen:1998uh, Chao:1995bk, Gilson:1993zs}, have considered a spherical surface and they have estimated their radii, i.e., strangelets would be like packed spheres with no more than $10^{7}$ quarks inside, see Fig.(\ref{strangelet}). Furthermore, these studies suggest that the radius $R$ depends on the baryon number similar to that of spherical atomic nuclei $R=r_{0}A^{1/3}$, where $r_{0}$ is a critical radius, which for nuclei is estimated to be equal to 1.12 fm \cite{Chao:1995bk}. These estimates may be of great interest to the astrophysics community, who seek strangelets upcoming from cosmic rays and could reach the Earth's atmosphere.

One of the main features of strangelets is their electric charge, which is due to their small sizes. Their radii are estimated to be not greater than the Compton's wavelength of electrons: $\lambda_{e}\simeq2.43\times10^{3}\;\text{fm}$ and therefore the presence of electrons inside strangelets is forbidden, analogous to what happens in the atomic nucleus, see Fig.(\ref{strangelet}). This leads to electrically charged strangelets, result which is in contrast to what happens in stars, where the overall charge neutrality is a prerequisite.

The sign of electric charge would also be a very important factor; many authors speculate cataclysmic situations depending on this. If strangelets have positive charge, they will repel nuclei who are in their way; however, if they are negatively charged, they would attract nuclei producing internal excitation states, becoming strange nuclei and fastly decaying and leading to a natural disaster ~\cite{Madsen:1998uh}.

\bigskip

Regarding the stability of strangelets; in the works of ~\cite{Farhi:1984qu, Madsen:1998uh, Chao:1995bk, Wen:2005uf, Gilson:1993zs} they studied strangelets at $T=0,$ showing that their energy per baryon is lower than the $^{56}$Fe. At finite temperature $T\neq0$, strangelets can be found in meta-stable states, but if they survive at least $0.01$ milliseconds, their presence can be detected by heavy ion collision experiments ~\cite{Madsen:1998uh}, by the traces they leave in the particle detectors.

\section{Shell model}\label{sec2.2}

As a first approach to the study of strangelets, one can use the MIT Bag model described in the previous chapter; this model has already been used to study the EOS and the energy by baryon of a quark gas at high densities. In this and the next sections, I will consider two equivalent formalisms of the MIT Bag model to study the main properties of strangelets, which keep close analogy with two of the best known models used in Nuclear Physics. The first of these is similar to the Nuclear Shell Model.
\bigskip

In Refs.\cite{Madsen:1998uh, Gilson:1993zs} were studied spherical strangelets considering quarks as noninteracting fermions. When solved the Dirac equation and imposing the appropriate boundary conditions on the sphere $\mathcal{S}$ of radius $R$, given in Eqs.\eqref{BagCurrentMIT} (neglecting the gluon fields, i.e. $G_{\mu}^{a}=0$) a quark distribution within strangelets is obtained analogous to the distribution of protons and neutrons in
the Nuclear Shell Model. Protons and neutrons are distributed in shells thanks to the nuclear potential, while in the case for quarks quark, there's no such potential, but the shell distribution comes from the boundary conditions.

\bigskip

The total energy of the strangelet is given by the sum of the kinetic energy of quarks plus the Bag contribution; this is expressed through the following expression:
\begin{equation}\label{ShellEB0}
E=\sum_{f=u,d,s}\sum_{i}\left[N_{f,i}\sqrt{p_{f,i}^{2}+m_{f}^{2}}\;\right]+\dfrac{4\pi}{3} B_{\text{bag}}R^{3}
\end{equation}
\noindent where the sums is performed over the three quark flavors and total angular momentum quantum number $j,$ related to the index $i$ by: $i=\pm(j+\frac{1}{2}).$ The factor $N_{f,i}=3(2j+1)$ corresponds to the number of quark flavors per energy levels (the number 3 corresponds to the three color degrees of freedom), $m_{f}$ is the quark mass, their momenta $p_{f,i}$ and the radius of the sphere $R$. After solving the Dirac equation one obtains the recurrence relations:
\begin{equation}\label{recurrence}
f_{i}(p_{f,i}R)=\dfrac{-p_{f,i}}{\sqrt{p_{f,i}^{2}+m_{f}^{2}}+m_{f}}f_{i-1}(p_{f,i}R)
\end{equation}
\noindent and the equivalent to Eqs.\eqref{BagCurrentMIT} is given by $\dfrac{\partial E}{\partial R}=0.$ The functions $f_{i}(p_{f,i}R)$ are given in terms of
\begin{equation}
f_{i}(z)=\begin{cases}
j_{i}(z),& \text{si}\; i\geq0\\
(-1)^{i+1}j_{-i-1}(z),& \text{si}\; i<0
\end{cases}
\end{equation}
\noindent which contain the Bessel spherical functions $j_{i}(z)$; while the baryon number can be written as:
\begin{equation}\label{barionicnumber}
A=\dfrac{1}{3}\sum_{f=u,d,s}\sum_{i}N_{f,i},
\end{equation}
\noindent which is considered as a constant.

\begin{figure}[h!t]
\centering
\includegraphics[width=0.7\textwidth]{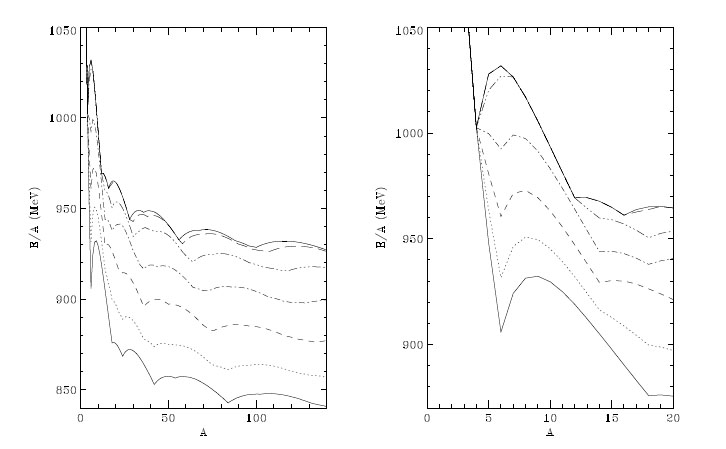}\\
\caption{Energy per baryon for strangelets with a Bag constant of $B_{\text{bag}}^{1/4}=145\;\text{MeV}$. It is also shown the curves corresponding to different masses of the $s$ quarks $m_{s}=0-300\;\text{MeV}$, in steps of 50 MeV and increasing order, with the corresponding magic numbers\cite{Madsen:1998uh}.}\label{shell}
\end{figure}

In Fig.(\ref{shell}) is shown the energy per baryon $E/A$ for strangelets made up by $u, d$ and $s$ free quarks, zero temperature and spherical surface. For small values of $A$, $E/A$ takes higher values, overcoming the energy per baryon of the isotope $^{56}\text{Fe}$: $E/A=930\;\text{MeV}$ and therefore are vulnerable to radioactive disintegration. On the other hand, in the limit $A\rightarrow\infty,$ the energy per baryon approaches decreasingly a constant value which is lower than the energy per baryon of $^{56}\text{Fe}$ for the used value of $B_{\text{bag}}.$

The quark distribution in strangelets follows a shell filling model, which are easily recognized in Fig.(\ref{shell}). The first shell is filled completely at $A=6$ (3 colors and 2 spin orientations for each quark flavors); the following complete shells are observed for
$A = 18, 24, 42, 54, 60, 84, 102,$ etc. (magic numbers). As the value of $m_{s}$ increases, the distribution of the magic numbers gets shifted to the left, filling the first shell at $A=4.$

\section{Liquid drop model}\label{sec2.3}

Another model used in the study of strangelets, is an extrapolation of the Liquid Drop Model, widely used in Nuclear Physics. This model predicts the spherical shape of some nuclei, the relationship between the electric charge and the baryon number, some of the mechanisms of nuclear disintegration and the energy levels of small deformed nuclei.

Extrapolated to the study of strangelets, as thermodynamic formalism of the MIT Bag model, it has the advantage that the effects of finite temperature $T$, surface corrections and Coulomb interaction can be taken into account and the magnitudes that depend on $T$ are relatively easy to handle numerically, unlike the shell model \cite{Madsen:1998uh}, which requires working with functions of mathematical physics such as Bessel functions. Another advantage is that the surface corrections are considered as a perturbation to the density of states of free quarks
and gluons using the Multiple Reflection-Expansion Method; these in turn can be interpreted as corrections to the thermodynamic potential and related thermodynamical functions such as the particle density, entropy, etc. The Multiple Reflection-Expansion Method was developed by R. ~Balian and C. ~Bloch in Ref.~\cite{Balian:1970fw}, and extended to the study of strangelets in Ref.\cite{Farhi:1984qu, Madsen:1998uh}, whereas the surface and curvature terms are identified with a correction to the density of states of gas of non-interacting particles
\begin{equation}\label{MERM}
\dfrac{dN_{f}}{dp}=\dfrac{d_{f}p^{2}V}{2\pi^{2}}+d_{f}G_{f,s}\left(\dfrac{m_{f}}{p}\right)pS+d_{f}G_{f,c}\left(\dfrac{m_{f}}{p}\right)C,
\end{equation}
\noindent being $V=\frac{4\pi}{3}R^{3}$, $S=4\pi R^{2},$ $C=8\pi R,$ and $R,$ the volume, surface area, curvature and radius of the spherical surface enclosing the gas. The structure of the factors $G_{f,s}\left(\dfrac{m_{f}}{p}\right)$ and $G_{f,c}\left(\dfrac{m_{f}}{p}\right)$ will be described in the following subsection.

\subsection{Thermodynamical potential}\label{granpotencial}

In the Liquid Drop Model formalism of the MIT Bag model, strangelets are described in terms of a gas of quarks $u, d$ and $s$ and gluons in thermodynamical equilibrium with surface and curvature corrections, and for simplicity it considers the surface enclosing the gas is a static closed surface, such that the expansion pressure generated by the gas is counteracted by a negative pressure exerted from the Bag surface.

\bigskip

To balance both pressures, one must know the expression of the thermodynamic potential\footnote{In this work, the thermodynamic potential $\Omega$ will be computed at the tree level.} $\Omega,$ which in Ref.\cite{Madsen:1998uh} is written as:
\begin{align}
\Omega\;\;&=\Omega_{g}+\Omega_{q\overline{q}},\\
\Omega_{g}\;&=\Omega_{g,v}\,V+\Omega_{g,c}\,C,\label{GluonO}\\
\Omega_{q\overline{q}}&=B_\text{bag}\,V+\sum_{f=u,d,s}\left[\Omega_{f,v}\,V+\Omega_{f,s}\,S+\Omega_{f,c}\,C\right].\label{OTotal}\,
\end{align}

The contribution of gluons to the thermodynamic potential is represented by the term $\Omega_{g}$, while $\Omega_{q\overline{q}}$ corresponds to quarks $q$ (antiquarks $\overline{q}$). As gluons are massless spin 1 bosons, their contributions to the surface thermodynamic potential is zero ~\cite{Farhi:1984qu, Madsen:1998uh, Madsen:2001fu}, that is why in Eq.(\ref{GluonO}) only appears the volumetric and curvature contributions
\begin{align}\label{Ogluon}
\Omega_{g,v}(T)=-\frac{d_{g}\pi^{2}}{90}\,T^{4},\quad \Omega_{g,c}(T)=\frac{d_{g}}{36}\,T^{2},
\end{align}
\noindent being $d_{g}=16$ the degeneracy factor (8 gluons times 2 spin projections).

\bigskip

On the other hand, the quarks-antiquarks contribution to the bulk thermodynamical potential, present in Eq.~\eqref{OTotal}, is written as
\begin{equation}\label{OVolumen}
\Omega_{f,v}(\mu_{f},T)=-\dfrac{d_{f}T}{(2\pi)^{3}}\int \ln\bigl(f_{p}^{+}f_{p}^{-}\bigr)\,d^{3}p,
\end{equation}
\noindent where
\begin{equation}\label{FermiDiracDistrib}
f_{p}^{\pm}=1+e^{-(E_{p,f}\mp\mu_{f})/T},
\end{equation}
\noindent contains the details of the particle ($f_{p}^{+}$) and antiparticle ($f_{p}^{-}$) distribution functions, $T$ denotes the temperature, $E_{p,f}$ the energy of each quark given by the spectrum, $d_{f}=6$ is the degeneracy factor, and $\mu_{f}$ are the chemical potentials.

\bigskip

The factor $B_\text{bag}\,V $ takes into account the Bag's energy, interpreting $B_\text{bag} $ as a vacuum pressure, which reproduces the color confinement of quarks as has been already explained in the previous chapter. It is important to note that in this case, the appropriate units for $B_{\text{bag}}$ in this context are in MeV fm$^{-3}$ and not MeV$^{4}$ as we used to write in the previous chapter. To convert MeV$^{4}$ to MeV fm$^{-3}$, simply divide by $(\hbar c)^{3}$ and extract the fourth root \footnote{The value of $\hbar c $ is $197.326968$ MeV fm.}. Thus, a value of $B_{\text{bag}}^{1/4}=145$ MeV equals to $B_{\text{bag}}\simeq57.8~\text{MeV}\,\text{fm}^{-3}$. From now on, this conversion will be used for the values of $B_{\text{bag}}.$
\bigskip

The surface contribution to the thermodynamical potential for the quark-antiquark gas present in Eq.~\eqref{OTotal} is written as:
\begin{align}\label{OSuperficie}
\begin{split}
\Omega_{f,s}&=\dfrac{d_{f}T}{16\pi^{3}}\int G_{s}\ln
\left(f_{p}^{+}f_{p}^{-}\right)\dfrac{d^{3}p}{|\vec{p}|},\\
G_{s}\;\,&=\arctan \left(m_{f}/|\vec{p}|\right).
\end{split}
\end{align}

\noindent The factor $G_{s}$ has into account the modification to the density of states according to the Multiple Reflection-Expansion Method, see Eq.\eqref{MERM}. For massless particles this factor vanishes even at finite temperature; but for massive particles, the density of particles derived from it is always negative, which implicitly contains the boundary conditions of the MIT Bag Model.

\begin{figure}[h!t]
\centering
\includegraphics[width=0.45\textwidth]{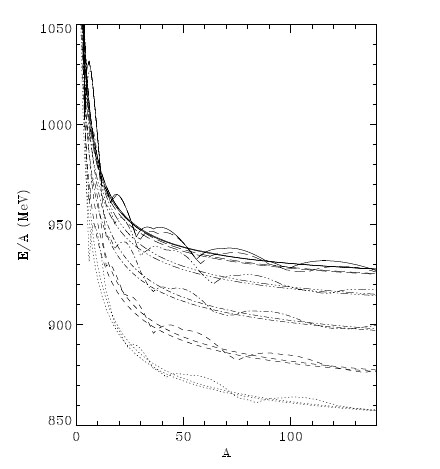}\\
\caption{Energy per baryon for strangelets at $T=0\,\text{MeV}$ and Bag constant $B_{bag}=57\;\text{MeV\;fm}^{-3}$. It is shown the curves corresponding to the Shell Model and the Liquid Drop Model (continuous lines) for different values of $m_{s}=0-300\;\text{MeV}$
(in steps of 50 MeV).}\label{equivalence}
\end{figure}

In an analogous fashion, one can write the curvature corrections to the thermodynamical potential for quarks and antiquarks~\cite{Madsen:1998uh}:
\begin{align}\label{OCurvatura}
\begin{split}
\Omega_{f,c}&=-\dfrac{d_{f}T}{48\pi^{3}}\int G_{c}\ln
\left(f_{p}^{+}f_{p}^{-}\right)\dfrac{d^{3}p}{|\vec{p}|^{2}}\,,\\
G_{c}\;\,&=1-\dfrac{3}{2}\dfrac{|\vec{p}|}{m_{f}}\arctan\left(m_{f}/|\vec{p}|\right).
\end{split}
\end{align}
In this case, the factor $G_{c}$ was derived in Ref.\cite{Madsen:1998uh} to fit the computations of the energy per baryon in the Shell Model for large values of the baryon number as it is shown in Fig.(\ref{equivalence}).

Once know the thermodynamical potential containing the bulk Eq.\eqref{OVolumen}, surface Eqs.\eqref{OSuperficie} and curvature \eqref{OCurvatura} corrections, one can compute the particle densities by
\begin{equation}\label{PDens}
N_{f,v}=-\frac{\partial\Omega_{f,v}}{\partial\mu_{f}},\;
N_{f,s}=-\frac{\partial\Omega_{f,s}}{\partial\mu_{f}}, \;
N_{f,c}=-\frac{\partial\Omega_{f,c}}{\partial\mu_{f}},
\end{equation}
\noindent and later compose the total number of particles of each flavor
\begin{equation}
N_{f}=N_{f,v}\,V+N_{f,s}\,S+N_{f,c}\,C.
\end{equation}
\noindent A third of the sum flavor of the total number of particles is the baryon number $A$ given in Eq.\eqref{barionicnumber}, which is considered as a constant for strangelets.

In the same fashion, one can compute the entropies as:
\begin{eqnarray}
S_{f,v}=-\frac{\partial\Omega_{f,v}}{\partial T},\;
S_{f,s}=-\frac{\partial\Omega_{f,s}}{\partial T}, \;
S_{f,c}=-\frac{\partial\Omega_{f,c}}{\partial T},\\
S=\sum_{f=u,d,s}\left[S_{f,v}V+S_{f,s}S+S_{f,c}C\right].\;\;\;\;\;\;\;\;\;\;\;\;\label{entropy}
\end{eqnarray}

It is noteworthy that this formalism of the MIT Bag Model has a purely thermodynamic features, so it doesn't describe with precision the physical situation for strangelets with a few quarks; however, using the surface and curvature terms, they reproduce similar results as the Shell formalism (except the magic numbers), for small values of baryon number; this is shown in Fig.(\ref{equivalence}), where certain equivalence is observed between the two descriptions for large values of baryon number $A$ \cite{Madsen:1998uh}. This allows us to use the Liquid Drop formalism to study the main properties of strangelets, such as: energy per baryon, radius and electric charge.

\subsection{Electric charge and Coulomb interaction}\label{zapantallada}

As I already argued in Sec.(\ref{sec2.1}) of this chapter, the electric charge of the strangelets is due to its small size. For $A<10^{7}$, their radii are estimated not to exceed the Compton wavelength of electrons; therefore, the presence of these within strangelets can be completely neglected. This leads to strangelets having a net electric charge, which is not homogeneously distributed within the strangelets Ref.\cite{Heiselberg:1993dc}.

The free electric charge distribution inside strangelets is given by:
\begin{eqnarray}\label{charge}
Z=Z_{V}+Z_{S},\;\;\;\;\;\;\;\;\;\;\;\;\\
Z_{V}=\rho_{V}V,\;\;\; Z_{s}=\rho_{S}S+\rho_{C}C
\end{eqnarray}
\noindent being
\begin{align}
\rho_{V}=\sum_{f=u,d,s}q_{f}N_{f,v},\;\;\; \rho_{S}=\sum_{f=u,d,s}q_{f}N_{f,s},\;\;\; \rho_{C}=\sum_{f=u,d,s}q_{f}N_{f,c}
\end{align}

\noindent the corresponding bulk, surface and curvature charge density distributions\footnote{The magnitudes $\rho_{S}S$ and $\rho_{C}C$ contribute to  the total surface electric charge.}; however, when the strangelet radius overcome the Debye length $R\geq\lambda_D\simeq 5.1\;\text{fm}$
\begin{equation}
(\lambda_D)^{-2}= 4\pi \sum_{f=u,d,s} q_f^2 \left(\frac{\partial N_{f,v}}{\partial \mu_f}\right),
\end{equation}
\noindent the charge screening effects and the coulombian interactions are of great importance and have to be included ~\cite{Heiselberg:1993dc,Endo:2005zt,Alford:2006bx}. This effect predicts that for $R\geq\lambda_D$, strangelets behave as conductors, being electrically neutral in the interior and the excess of the electric charge is distributed on the surface, precisely within a layer of width $\lambda_{D}.$

\bigskip

I will treat the screening of the electric charge in the framework of the Thomas-Fermi approximation, following Ref.\cite{Heiselberg:1993dc}. In this case we have three charge carriers, the quarks $u$, $d$ and $s$ respectively. The chemical potentials of the gases formed by these carriers vary spatially as a result of the electrostatic potential $V (r)$ generated by the quarks electric charges; however, the chemical potentials must always satisfy the conditions of $\beta-$ equilibrium
\begin{equation}
\mu_{u}(r)+q_{u}V(r)=\mu_{d}(r)+q_{d}V(r)=\mu_{s}(r)+q_{s}V(r)=\mu,
\end{equation}
\noindent where the chemical potential $\mu$ is constant due to the Thomas-Fermi approximation.

Considering $V(r)\ll\mu,$ the bulk charge density, depending on the radial coordinate $r$, is given by
\begin{equation}\label{rhor}
\rho_{V}(r)=\rho_{V}-\dfrac{V(r)}{4\pi\lambda_{D}^{2}},
\end{equation}
\noindent from where one can write the associated Poisson equation for the electrostatic potential and the boundary conditions it must satisfy:
\begin{eqnarray}
\nabla^{2}V(r)=-4\pi\rho_{V}(r),\label{Poisson}\\
\lim_{r\to0}V(r)<\infty,\\
V(R)=\dfrac{eZ_{\text{apant}}}{R}\label{PotZscreen}.
\end{eqnarray}
\noindent Integrating in spherical coordinates Eq.(\ref{Poisson}), one gets:
\begin{eqnarray}
V(r)=4\pi\rho_{V}\lambda_{D}^{2}\left[1-\dfrac{1}{\cosh\left(\frac{R}{\lambda_{D}}\right)}\dfrac{\sinh\left(\frac{r}{\lambda_{D}}\right)}{\frac{r}{\lambda_{D}}}\right]\\
\rho_{V}(r)=\dfrac{\rho_{V}}{\cosh\left(\frac{R}{\lambda_{D}}\right)}\dfrac{\sinh\left(\frac{r}{\lambda_{D}}\right)}{\frac{r}{\lambda_{D}}};
\end{eqnarray}
\noindent hence, from Eq.(\ref{PotZscreen}) one obtains finally
\begin{equation}\label{Zscreen}
Z_{\text{apant}}=\frac{4\pi}{e} R \lambda_D^2\, \rho_V \left[1-\frac{\lambda_D}{R}\tanh\left(\frac{R}{\lambda_D}\right)\right].
\end{equation}

The Coulomb energy is given by:
\begin{align}
E_{C}=2\pi\int_{0}^{R}\left[V(r)\rho_{V}(r)r^{2}dr\right],\;\;\;\;\;\;\;\;\;\;\;\;\;\;\;\\
E_C=4\pi^2R\rho_{V}^{2} \lambda_D^4 \left[2-\frac{3\lambda_D}{R}\tanh\left(\frac{R}{\lambda_D}\right) +\cosh^{-2} \left(\frac{R}{\lambda_D}\right)\right]\label{Ecoulomb},
\end{align}
\noindent which should be added to the total energy.

\bigskip
\bigskip


\subsection{Free energy and hydrostatic equilibrium condition}\label{energialibre}

To obtain hydrostatically stable configurations of strangelets, we need that the free energy attains a minimum with respect to the volume. This minimum represents the most stable configuration the strangelet can reach and it is found by solving the equation
\begin{align} \label{equilconf}
\left.\frac{\partial F}{\partial V}\right|_{N,T}=0,
\end{align}
\noindent keeping fixed the baryon number $A$ given by Eq.\eqref{barionicnumber}. This equilibrium state is achieved when the total expanding pressure exerted by the gas of quarks and gluons compensates the negative pressure exerted by the Bag surface, which means to ask that $P=0$, where $P=-\Omega/V$, and $\Omega$ is given by Eq.\eqref{OTotal}.
Globally, strangelets may be considered as closed statistical systems where the temperature and generalized forces, including the volume ~\cite{CRMaite} and the total number of particles is constant.

\bigskip

The Helmholtz Free Energy or simply Free Energy, is written as
\begin{align}\label{freeenergy}
\begin{split}
F\;\;&=F_{g}+F_{q\overline{q}}+E_{C},\\
F_{g}\;\,&=\Omega_{g,v}\,V+\Omega_{g,c}\,C,\\
F_{q\overline{q}}\;&=B_\text{bag}\,V+\sum_{f}\left[F_{f,v}\,V+F_{f,s}\,S+F_{f,c}\,C\right],
\end{split}
\end{align}
\noindent where $F_{g}$ and $F_{q\overline{q}}$ represent the contributions of gluons and quarks(antiquarks) respectively, while $E_{C}$ is the Coulomb energy given by Eq.(\ref{Ecoulomb}). The bulk, surface and curvature contributions of quarks and antiquarks to the free energy, are given by the following expressions:
\begin{align}\label{FVSC}
\begin{split}
F_{f,v}&=\Omega_{f,v}+\mu_{f}N_{f,v},\\
F_{f,s}&=\Omega_{f,s}+\mu_{f}N_{f,s},\\
F_{f,c}&=\Omega_{f,c}+\mu_{f}N_{f,c}.
\end{split}
\end{align}

The total energy of the strangelets is obtained from the free energy by adding the contribution from the entropy $TS$, i.e $E=F+T\,S$, begin $S$ the total entropy of the system, given by Eq.\eqref{entropy}.

\section{Stability of strangelets}\label{sletstability}

Similarly as we proceeded in the sections (\ref{epig1.2.1}) and (\ref{epig1.2.2}), I will study now the conditions under which strangelets are stable, considering quarks $u, d$ and $s$ in the ultra-relativistic limit, $T=0$ and neglecting the contribution of gluons. The set of thermodynamical quantities to begin the study are given by
\begin{eqnarray}
\Omega_{f,v}=-\dfrac{\mu_{f}^{4}}{4\pi^{2}},\;\;\;N_{f,v}=\dfrac{\mu_{f}^{3}}{\pi^{2}},\;\;\;E_{f,v}=\dfrac{3\mu_{f}^{4}}{4\pi^{2}},\;\;\;\\
\Omega_{f,c}=\dfrac{\mu_{f}^{2}}{8\pi^{2}},\;\;\;N_{f,c}=-\dfrac{\mu_{f}}{4\pi^{2}},\;\;\;E_{f,c}=-\dfrac{\mu_{f}^{2}}{8\pi^{2}}.\;
\end{eqnarray}
\noindent Notice that I haven written the surface contributions $\Omega_{f,s}$ since they become negligible for ultra-relativistic particles $m_{f}\sim0.$ Since electrons cannot coexist with quarks within strangelets, I take the three chemical potentials as equals, which means that I use Eq.\eqref{betaequil}. Under these considerations, one obtains electrically neutral strangelet configurations due to the fact that the sum of the quarks electric charges are zero; reason why I ignore also the screening effects and the Coulomb energy contribution, and the free energy can be written as:
\begin{equation}\label{masslesstrangfreee}
F=\dfrac{4\pi}{3}R^{3}B_{\text{bag}}+\dfrac{3\mu^{4}}{\pi}R^{3}-\dfrac{3\mu^{2}}{\pi}R.
\end{equation}
\noindent Using Eq.\eqref{equilconf}, which is equivalent to minimize the free energy $F$ given in Eq.\eqref{masslesstrangfreee} with respect to the radius $R,$ one obtains the following relationship:
\begin{equation}\label{masslesstrangradius}
R=\dfrac{\mu}{\sqrt{3\mu^{4}+\frac{4\pi^{2}}{3}B_{\text{bag}}}}.
\end{equation}
\noindent If the chemical potential increases, the radius decreases and viceversa; the same happens with the Bag constant $B_{\text{bag}}.$ The baryon number is given by
\begin{equation}\label{masslesstrangbnumber}
A=\dfrac{4\mu^{3}}{3\pi}R^{3}-\dfrac{2\mu}{\pi}R;
\end{equation}
\noindent therefore, now we can compute the chemical potential and estimate $B_{\text{bag}}$ for a baryon number of $A=100$ for example.

Indeed, using $R=0.89 A^{1/3}\; \text{fm},$ which will be plenty justified later, one obtains
$R=4.13\; \text{fm},$ giving $\mu\simeq300\;\text{MeV}.$ Substituting these results in Eq.\eqref{masslesstrangfreee}, one has
$E/A=F/A=681.52\;\text{MeV}+2.95B_{\text{bag}}\;\text{fm}^{3}$.

\bigskip

The stability, relative to the isotope $^{56}$Fe, requires $E/A<930$ MeV, from where we obtain that $B_{\text{bag}}$ is allowed to vary between: $0\leq B_{\text{bag}}<84$ MeV fm$^{-3};$ which means that for strangelets with baryon number $A=100$, which corresponds to approximately 300 quarks, stability requires that $B_{\text{bag}}<84$ MeV fm$^{-3}.$ In \textbf{Cap IV} we generalize these results for strangelets with massive quarks, finite temperature, the presence of gluons, strong magnetic fields and the effects of the color superconductivity.

\section{Production of strangelets and ongoing experiments}

In the astrophysics literature, one can find a number of phenomena attributable to strangelets; among them they are: collisions of EQs, anomalies of  cosmic ray bursts in Cygnus X-3, the extraordinarily high luminosity of the Gamma Ray Bursts in the N49 remnant of the Large Magellanic Cloud and even the events such as Centaury \cite{Klingenberg:2001qs}. Therefore, the most obvious places for detecting strangelets are those regions of the Universe where involves cosmic ray events. In this context, it may be mentioned that there is an extensive literature on the production of exotic cosmic ray events with an unusually small charge-to-mass ($Z/A$) \cite{Kasuya1993, Ichimura1993, Saito1995, Capdeville1996}. It seems natural to identify these events with strangelets; however, it has not yet reached a consensus due to ambiguities of the mechanisms of propagation through the Earth's atmosphere. For example, as cited in Ref.\cite{Banerjee2006}, if a strangelet with baryon number of $A\sim1000$ reaches the atmosphere, it encounter serious problems to penetrate, because the mass of the strangelet will rapidly decrease by the constant collisions with the air molecules. It would reach a critical mass, below which, the strangelet simply evaporates forming neutrons.

\bigskip

Among the most cited experiments seeking strangelets from astrophysical sources, are the Alpha Magnetic Spectrometer (AMS) on the International Space Station, the muon detector ALEPH in the experiment CosmoLEP (Cosmic-Ray Experiment by Low Energy Physic ) and experiments SLIM symbolic circuit compactor in Chacaltaya (Bolivia).

\bigskip
\bigskip

Despite the efforts of astrophysicists looking for strangelets, this is not the only way to try to detect them; many scientists support the idea that they could be also produced in particle accelerators experiments \cite{Lourenco:2002c, Wiener2006}. Repeatedly BNL-AGS (Brookhaven National Lab Experiments) and SPS (Super Proton Synchrotron) experiments are cited, and the detector system CASTOR (calorimeter), proposed the latter as a subsystem of the ALICE (A Large Ion Collider Experiment) in the LHC belonging to CERN \cite{Angelis:2001c}. The BNL-AGS, experiments are studying heavy ion collisions in an open geometry spectrometer at energies of the order of $11$ GeV Au + Pb, but so far no one has found evidences of strangelets \cite{Armstrong:1997l}. Another experiment that is being carried out at CERN is the NA52, where they collide Sulfur-Tungsten ions and Lead-Lead \cite{Klingenberg:1996a}, to look out for strange massive particles with relatively short lifetimes.
%

%

\chapter{Strange quark matter and magnetic fields}\label{chap3}

This chapter will be devoted to study the main properties of the quarks gas that makes up strangelets in the presence of an external constant and homogeneous magnetic field in the direction of the $z$ axis; for this, I use again Liquid Drop Model formalism of the MIT Bag Model described in the previous chapter. Due to the interaction of the magnetic field with electrically charged particles, the $x-y$ components of the momentum are quantized according to Landau levels, therefore, the thermodynamic properties differ from those at zero magnetic field. I will investigate how the properties of strangelets vary in the presence of strong magnetic fields.

\section{Energy spectrum of quarks in the presence of a magnetic field}

\bigskip

To compute the thermodynamical quantities as in Sec.(\ref{sec2.3}), one needs to know first the energy spectrum of the quarks $u,d$ and $s$. Such spectrum, in the presence of an external magnetic field $\cal B$, and described by the vector potential $A^{\mu},$ is obtained by solving the Pauli-Dirac equation
\begin{equation}
 \left[\gamma^{\mu}(\partial_{\mu}+iq_{f}A_{\mu})+m_{f}\mathds{I}_{4}\right]\psi_{f}=0.
\end{equation}
Considering that the magnetic field in homogeneous, constant and oriented along the $z-$axis, the spectrum is given by \cite{Felipe:2007vb}:
\begin{equation}\label{espectroquarks}
E_{p,f}^{\nu,\eta}=\sqrt{p_{z}^{2}+p_{f\;\perp}^{2}+m_{f}^{2}},\;\;p_{f\;\perp}=\sqrt{q_{f}\mathcal{B}(2\nu+1-\eta)}.
\end{equation}

\noindent As one should expect, the presence of the field produces a spatial symmetry breaking, with the subsequent separation of the three components of the linear momentum of each charged particle: one longitudinal $p_{z}$ along the field's direction, and the two others are equals and both perpendicular to the direction of the magnetic field, i.e., transversal component $p_{f\;\perp}$. The transversal component is quantized in discrete Landau levels $\nu$. The index $f=u,d,s$, runs over the quark flavors as usual, the numbers $\eta=\pm1$ are the eigenvalues of the spin operator; $q_{f}$, and $m_{f}$ represent the electric charge and the rest mass of each particle respectively. In Tab.\ref{tabla1} are shown the values of such magnitudes used in this thesis:
\bigskip
\begin{table}[!ht]
\begin{center}
\label{tabla1}
\begin{tabular}{|c|c|c|c|}
\hline
\textbf{Quarks} & \emph{u} & \emph{d} & \emph{s} \\
\hline
$q_{f}\;(\emph{e})$&$+2/3$&$-1/3$&$-1/3$\\
\hline
$m_{f}\;(\text{MeV})$&$\sim5$&$\sim5$&$\sim150$\\
\hline
\end{tabular}\caption{Values of the electric charge and rest mass of the particles involved in the study.}
\end{center}
\end{table}

\noindent Notice that Eq.(\ref{espectroquarks}) can be written equivalently as
\begin{equation}
E_{p,f}^{\nu,\eta}=\sqrt{p_{z}^{2}+M_{\nu,f}^{\eta\;2}},\;\;\;\;M_{\nu f}^{\eta}=\sqrt{q_{f}\mathcal{B}(2\nu+1-\eta)+m_{f}^{2}}
\end{equation}
\noindent compatible with the spectrum of a particle with mass $M_{\nu,f}^{\eta}$ which depends on the magnetic field and the states $\nu,\eta$, which means a kind of ``quantized magnetic mass''. The hypothetical particle moves freely with linear momentum oriented along the field\cite{Felipe:2007vb}.

\bigskip

\section{Pressure anisotropy}

Already know the energy spectrum of quarks, one can compute the contributions of each of them to the bulk thermodynamical potential, using Eq.(\ref{OVolumen}). Due to the anisotropy in the linear momentum, imposed by the presence of the magnetic field, the integration over the transversal components $dp_{x}dp_{y}$ of the linear momentum in Eqs.~\eqref{OVolumen}, \eqref{OSuperficie} and \eqref{OCurvatura} should be replaced by the rule
\begin{equation}
\int_{-\infty}^{+\infty}\int_{-\infty}^{+\infty}\;[\;\;] dp_{x}\,dp_{y}
\rightarrow 2\pi q_{f} \mathcal{B} \sum_{\eta=\pm1}
\sum_{\nu=0}^{\nu_{\mbox{\tiny{max}}}^{f}}\;[\;\;],
\end{equation}
\noindent where the sum over the Landau levels is finite till the maximum level
\begin{equation}\label{Landaulevels}
\nu_{\mbox{\tiny{max}}}^{f}=I\left[\dfrac{\mu_{f}^{2}-m_{f}^{2}}{2q_{f}\mathcal{B}}\right],
\end{equation}
\noindent due to the Fermi momenta of each quark gas $p^{f}_{F}=\sqrt{\mu_{f}^{2}-M_{f\,\nu}^{\pm\;2}}$ have to be real quantities ~\cite{Felipe:2007vb}. The function $I[x]$ represents the integer part of the real number $x$. Taking this into account, the thermodynamical potential can be written as
\begin{eqnarray}\label{BOmega}
\Omega_{f,v}=-\dfrac{d_{f}q_{f}\mathcal{B}}{2\pi^{2}}\sum_{\nu=0}^{\nu^{f}_{\text{max}}}[\omega_{f,\nu}^{+}(q\overline{q})+\omega_{f,\nu}^{-}(q\overline{q})],\;\;\;\;\;\;\;\;\;\;\;\;\;\;\;\;\;\;\;\;\;\;\;\;\\
\omega_{f,\nu}^{\pm}(q\overline{q})=\dfrac{1}{\beta}\int_{0}^{+\infty}\left[\ln\left[1+e^{-\beta(E_{p,f}^{\nu,\pm}- \mu_{f})}\right]+\ln\left[1+e^{-\beta(E_{p,f}^{\nu,\pm}+\mu_{f})}\right]\right]dp_{z},
\end{eqnarray}
\noindent where I denoted by $\omega_{f,\nu}^{\pm}(q\overline{q})$ the quark and antiquark contributions respectively, with spin projections $\pm1$ and being $\beta=T^{-1}$ the inverse temperature. In an analogous way, one can write the bulk particle densities
\begin{eqnarray}\label{BNden}
N_{f,v}=\dfrac{d_{f}q_{f}\mathcal{B}}{2\pi^{2}}\sum_{\nu=0}^{\nu^{f}_{\text{max}}}[N_{f,\nu}^{+}(q\overline{q})+N_{f,\nu}^{-}(q\overline{q})],\;\;\;\;\;\;\;\;\;\;\;\;\;\;\;\\
N_{f,\nu}^{\pm}(q\overline{q})=\int_{0}^{+\infty}\left[\dfrac{1}{1+e^{\beta(E_{p,f}^{\nu,\pm}-\mu_{f})}}-\dfrac{1}{1+e^{\beta(E_{p,f}^{\nu,\pm}+\mu_{f})}}\right]dp_{z},\;\;\;  \end{eqnarray}
\noindent energy
\begin{eqnarray}\label{BEnergy}
E_{f,v}=\dfrac{d_{f}q_{f}\mathcal{B}}{2\pi^{2}}\sum_{\nu=0}^{\nu^{f}_{\text{max}}}[E_{f,\nu}^{+}(q\overline{q})+E_{f,\nu}^{-}(q\overline{q})],\;\;\;\;\;\;\;\;\;\;\;\;\;\;\;\;\;\;\\
E_{f,\nu}^{\pm}(q\overline{q})=\int_{0}^{+\infty}\left[E_{p,f}^{\nu,\pm}\left(\dfrac{1}{1+e^{\beta(E_{p,f}^{\nu,\pm}-\mu_{f})}}+\dfrac{1}{1+e^{\beta(E_{p,f}^{\nu,\pm}+\mu_{f})}}\right)\right]dp_{z},\;\;\;        \end{eqnarray}
\noindent and the magnetization
\begin{eqnarray}\label{BMag}
M_{f,v}=-\dfrac{\partial\Omega_{f,v}}{\partial\mathcal{B}},\;\;\;\;
M_{f,v}=\dfrac{d_{f}q_{f}\mathcal{B}}{2\pi^{2}}\sum_{\nu=0}^{\nu^{f}_{\text{max}}}[M_{f,\nu}^{+}(q\overline{q})+M_{f,\nu}^{-}(q\overline{q})],\;\;\;\;\;\;\;\;\;\;\;\;\;\;\;\;\;\;\;\;\\
M_{f,\nu}^{\pm}(q\overline{q})=\dfrac{1}{\mathcal{B}}\int_{0}^{+\infty}\left[\left(p_{z}\dfrac{\partial E_{p,f}^{\nu,\pm}}{\partial p_{z}}-\mathcal{B}\dfrac{\partial E_{p,f}^{\nu,\pm}}{\partial\mathcal{B}} \right)\left(\dfrac{1}{1+e^{\beta(E_{p,f}^{\nu,\pm}-\mu_{f})}}+\dfrac{1}{1+e^{\beta(E_{p,f}^{\nu,\pm}+\mu_{f})}}\right)\right]dp_{z}.\;\;\;
\end{eqnarray}

\bigskip

The macroscopic limit of the energy-momentum tensor for a gas made of quarks $u, d$ and $s$ (without considering surface effects) is diagonal in the presence of a strong magnetic field as in Eq.(\ref{IdealEMTensor}), but in this case there appear two pressures due to the anisotropy induced by the magnetic field  \cite{Felipe:2007vb,Felipe:2008cm,Martinez:2010sf,Ferrer:2010wz,Boligan}, which is observed in the macroscopic energy-momentum tensor:
\begin{eqnarray} \langle{\cal T}_{\mu\nu}\rangle = \left(
\begin{array}{llll}E & 0&0&0 \\0& P_{\perp}&0&0\\0&0&P_{\perp}&0\\ 0&0&0&P_{\parallel}
\end{array}\right) , \hspace{0.3 cm} P_{\perp}=P_{\parallel}-M\mathcal{B},\\
P_{\parallel}=-\sum_{f=u,d,s}\Omega_{f,v}-B_{\text{bag}},\;\;\;E=\sum_{f=u,d,s}E_{f,v}+B_{\text{bag}},\;\;\;M=\sum_{f=u,d,s}M_{f,v} , \label{presiones}
\end{eqnarray}
\noindent where $P_{\perp}$ corresponds to the transversal and $P_{\parallel}$ to the parallel components of the pressure; $E$ and $M$ are the energy and magnetization densities respectively. Notice that always $P_{\perp}\leq P_{\parallel}$; such inequality is more evident as the magnetic field becomes even stronger. Therefore, strong magnetic fields produce an anisotropy, which convey to an equatorial deformation of the quark gas, even producing a collapse condition if the magnetic field strength reaches a critical value \cite{Boligan,Martinez:2003dz}. However, in this study, I use a magnetic field of the order of $\mathcal{B}=5\times10^{18}$~G, and consider that the regime is still isotropic ($P_{\perp}\sim P_{\parallel}$), and I  can still use the spherical model for strangelets.

Evaluating numerically the difference $P_{\parallel}-P_{\perp}$, for $\mathcal{B}=5\times10^{18}\;\text{G},$ $A=100,$ $T=0$ and $T=15\;\text{MeV}$ and $B_{\text{bag}}=75$ MeV fm$^{-3}$ one gets:
\begin{equation}
\left.(P_{\parallel}-P_{\perp})\right|_{T=0}=2.8\;\text{MeV\;fm$^{-3}$} ,\;\;\;\;\left.(P_{\parallel}-P_{\perp})\right|_{T=15}=5.9\;\text{MeV\;fm$^{-3}$}.
\end{equation}

\begin{figure}[h!t]
\centering
\includegraphics[width=0.7\textwidth]{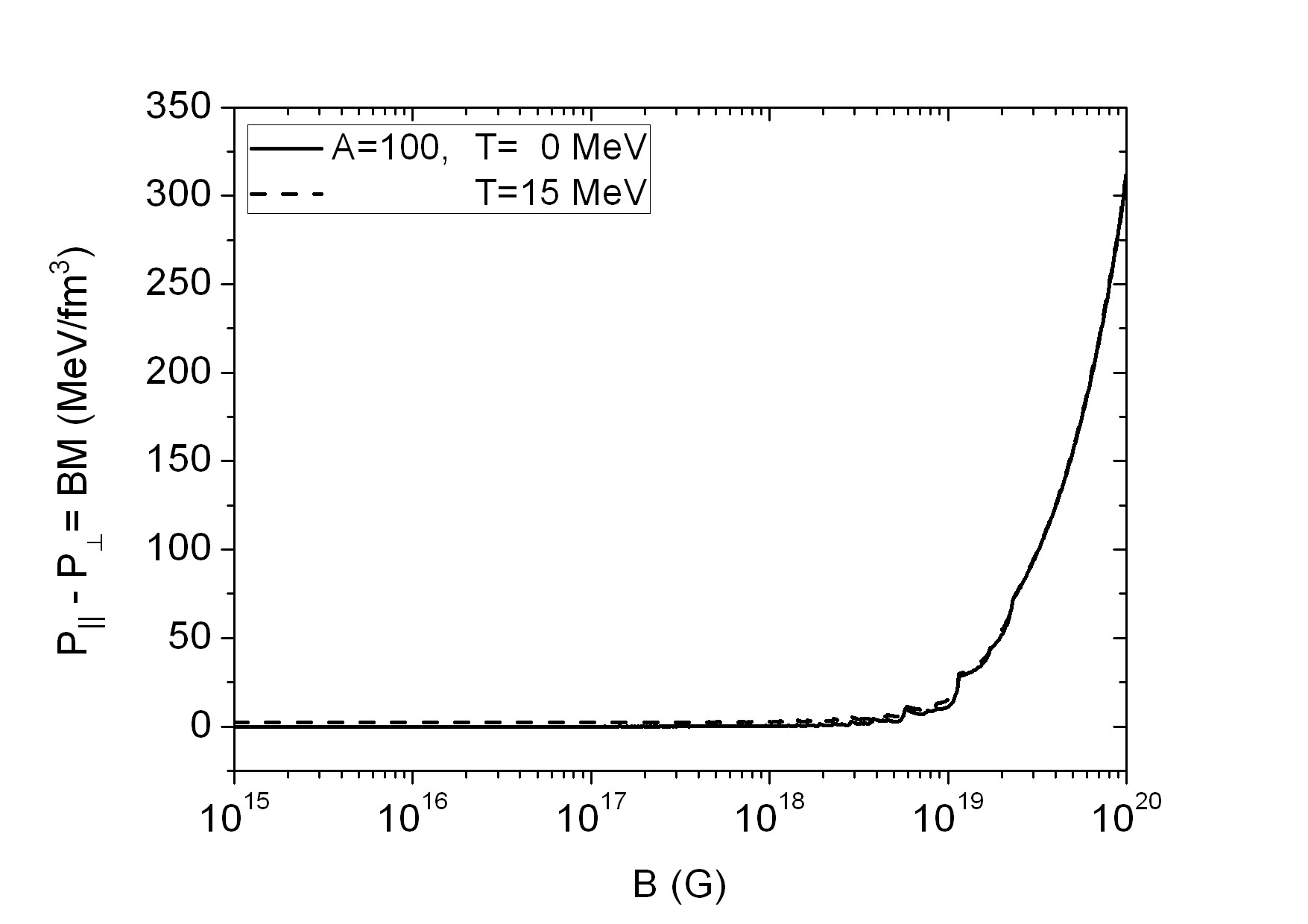}\\
\caption{Behavior of the anisotropy of the pressures $P_{\parallel}-P_{\perp}=\mathcal{B}M$ with the magnetic field strength, for $A=100,$ $T=0, 15\;\text{MeV}$ and $B_{\text{bag}}=75$\;MeV\;fm$^{-3}$.}\label{BM}
\end{figure}

\noindent In Fig.(\ref{BM}) is shown this difference $P_{\parallel}-P_{\perp}$ as a function of the magnetic field $\mathcal{B}$, taking $A=100,$ $T=0$ and $T=15\;\text{MeV}$ and $B_{\text{bag}}=75$ MeV fm$^{-3}$. The effects of the anisotropy become relevant precisely for fields stronger than $\mathcal{B}=5\times10^{18}\;\text{G}.$

\bigskip

\section{Surface terms in the presence of a magnetic field}

Now I compute the surface $\Omega_{f,s}$ and curvature $\Omega_{f,c}$ contributions to the thermodynamical potential using Eqs.(\ref{OSuperficie}) and (\ref{OCurvatura}), where I only modify the energy spectrum of quarks, which contains the effects of the magnetic field. Starting with $\Omega_{f,s}$, magnitude identified with the surface tension of the gas; for which one gets
\begin{eqnarray}\label{OSuperficieMagn}
\Omega_{f,s}=\dfrac{d_{f}q_{f}\mathcal{B}}{4\pi^{2}}\sum_{\nu=0}^{\nu_{\text{max}}^{f}}[\sigma_{f,\nu}^{+}(q\overline{q})+\sigma_{f,\nu}^{-}(q\overline{q})],\;\;\;\;\;\;\;\;\;\;\;\;\;\;\;\;\;\;\;\;\;\;\;\;\;\;\;\;\\
\sigma_{f,\nu}^{\pm}(q\overline{q})=\dfrac{1}{\beta}\int_{0}^{+\infty}\left[G^{\pm}_{f,\nu,s}(p)\left(\ln\left[1+e^{-\beta(E_{p,f}^{\nu,\pm}- \mu_{f})}\right]+\ln\left[1+e^{-\beta(E_{p,f}^{\nu,\pm}+\mu_{f})}\right]\right)\right]dp_{z},\;\;\;\\
G^{\pm}_{f,\nu,s}(p)=\dfrac{1}{\sqrt{p_{z}^{2}+p_{f\;\perp}^{2}}}\arctan\left(\dfrac{m_{f}}{p_{z}}\right).\;\;\;\;\;\;\;\;\;\;\;\;\;\;\;\;\;\;\;\;\;\;\;\;
\end{eqnarray}

There appears a singularity for $\nu=0,$ at $p_{z}=0$ due to the factor $G^{\pm}_{f,0}(p).$ The integration of $\sigma_{f,\nu}^{\pm}(q\overline{q})$ at $T=0$ leads to the following result:
\begin{equation}\label{divergence}
\sigma_{f,\nu}^{\pm}(q\overline{q})\propto-\ln[p_{f\;\perp}/m_{f}](\mu_{f}-M_{f\;\nu}^{\eta}).
\end{equation}

\bigskip

Similarly one gets for $\Omega_{f,c}:$
\begin{eqnarray}\label{OCurvatMagn}
\Omega_{f,c}=-\dfrac{d_{f}q_{f}\mathcal{B}}{12\pi^{2}}\sum_{\nu=0}^{\nu_{\text{max}}^{f}}[\gamma_{f,\nu}^{+}(q\overline{q})+\gamma_{f,\nu}^{-}(q\overline{q})],\;\;\;\;\;\;\;\;\;\;\;\;\;\;\;\;\;\;\;\;\;\;\;\;\;\\
\gamma_{f,\nu}^{\pm}(q\overline{q})=\dfrac{1}{\beta}\int_{0}^{+\infty}\left[G^{\pm}_{f,\nu,c}(p)\left(\ln\left[1+e^{-\beta(E_{p,f}^{\nu,\pm}- \mu_{f})}\right]+\ln\left[1+e^{-\beta(E_{p,f}^{\nu,\pm}+\mu_{f})}\right]\right)\right]dp_{z},\;\;\;\\
G^{\pm}_{f,\nu,c}(p)=\dfrac{1}{\sqrt{p_{z}^{2}+p_{f\;\perp}^{2}}}\left(1-\dfrac{3}{2}\dfrac{p_{z}}{m_{f}}\arctan\left(\dfrac{m_{f}}{p_{z}}\right)\right).\;\;\;\;\;\;\;\;\;\;\;\;\;\;
\end{eqnarray}

Analogously to $\sigma_{f,\nu}^{\pm}(q\overline{q}),$ for $\gamma_{f,\nu}^{\pm}(q\overline{q})$ one takes the limit $T=0$ and one gets a divergent term as:
\begin{equation}\label{gammadiv}
\gamma_{f,\nu}^{\pm}(q\overline{q})\propto\dfrac{\pi m_{f}}{2p_{f\;\perp}}+\ln[p_{f\;\perp}/m_{f}].
\end{equation}

Eqs.~\eqref{OSuperficieMagn} and \eqref{OCurvatMagn}, at $\nu=0$ show an infrared singularity. For example, the surface contribution at $T=0$, diverges logarithmically as Eq.~\eqref{divergence} to $+\infty$, while the curvature contribution diverges as Eq.~\eqref{gammadiv}, on one side to $+\infty$ ($1/p_{f\;\perp}$) and logarithmically on the other side to $+\infty$. The second divergence is dominant against the first one, therefore, $\gamma_{f,\nu}^{\pm}(q\overline{q})\to+\infty$. This leads that the total surface pressure for each quark gas diverges as $P_{f,S}=-(\Omega_{f,s}S+\Omega_{f,c}C)\to-\infty$, which become infinitely negative. These divergences can be avoided by introducing a momentum cutoff $p_{f\;\perp}$ of each particle. The lower integration limit in Eqs.(\ref{OSuperficieMagn}) and (\ref{OCurvatMagn}) will be augmented by this cutoff, and beyond it, one can safely integrate both expressions. The divergences are a consequence of considering a spherical system when obviously the magnetic field breaks this symmetry and produces such divergences.
A rigourous treatment of this phenomenon will be in deducing the expression of the surface and curvature contributions considering the effects of a magnetic field.

\bigskip

%


\chapter{Magnetized strangelets at finite temperature}\label{chap4}

In this chapter, I will solve numerically the equations governing the hydrostatic equilibrium of strangelets formed by MSQM and MCFL quark matter phases. I will study the behavior of the energy per baryon $E/A$, the radius $R$ and electric charge $Z$, as a function of the baryon number $A$, fixing in the first case the parameters ($B_{\text{bag}}, \mathcal{B}, T$), while in the second case I fix $(B_{\text{bag}}, \mathcal{B}, T, \Delta).$ In all the computations, I will compare to the already obtained results with $\mathcal{B}=0$, to the see how, strong magnetic fields modify the stability, radio and electric charge. The quark masses, the electric charge screening and the Coulomb interaction will be also taken into account.

\section{Strangelets from Magnetized Strange Quark Matter}\label{MQEM}

\subsubsection{Hydrostatic equilibrium conditions}

Knowing a priori the energy spectrum of quarks, leads directly to compute all thermodynamic expressions described in the previous two chapters, in particular, the free energy. The hydrostatic equilibrium condition for strangelets, is obtained by minimizing the total free energy with respect to the volume, which is represented by Eq.\eqref{equilconf}, and it is written as
\begin{eqnarray}\label{equilibrio0}
\left.\frac{\partial F_{g}}{\partial V}\right|_{N,T,\mathcal{B}}+
\left.\frac{\partial F_{q\overline{q}}}{\partial
V}\right|_{N,T,\mathcal{B}}+\left.\dfrac{\partial E_C}{\partial V}\right|_{N,T,\mathcal{B}}=0,\label{equilibrio}
\end{eqnarray}
\noindent being
\begin{eqnarray}\label{equilibrio1}
\left.\frac{\partial F_{g}}{\partial
V}\right|_{N,T,\mathcal{B}}=\Omega_{g,v}+\dfrac{2}{R^{2}}\Omega_{g,c},\;\;\;\;\;\;\;\;\;\;\;\;\;\;\;\;\;\;\;\;\;\;\;\;\;\;\;\;\;\;\\
\left.\frac{\partial F_{q\overline{q}}}{\partial
V}\right|_{N,T,\mathcal{B}}=B_{\rm bag}+\Omega_{V}+ \dfrac{2}{R}\Omega_{S}+\dfrac{2}{R^{2}}\Omega_{C},\;\;\;\;\;\;\;\;\;\;\;\;\label{equilibrio2}
\end{eqnarray}
\noindent and
\begin{align}\label{equilibrio3}
\Omega_{V}=\sum_{f=u,d,s}\Omega_{f,v},\quad
\Omega_{S}=\sum_{f=u,d,s}\Omega_{f,s},\quad
\Omega_{C}=\sum_{f=u,d,s}\Omega_{f,c}.
\end{align}

Given the values of $B_{\text{bag}},$ the temperature and magnetic field, the system of equations written above together with Eq.\eqref{barionicnumber}, become a complete set of equations to determine the common chemical potential $\mu$ and the radius $R$ of strangelets; this fact allows to evaluate the thermodynamical quantities, such as the particle density, total energy and electric charge. To compute this last magnitude, the Debye screening effects will be included in the bulk contributions while on the surface I will take the free electric charge. The sum of both contributions will give the total electric charge:

\begin{equation}\label{Ztotal}
Z=Z_{\text{screen}}+Z_{s}.
\end{equation}

Due to the analytical complexity of all the quantities mentioned above, which are mostly introduced by the temperature, the chemical potential and the magnetic field, all the results below will be obtained via numerical methods.

\subsubsection{Energy per baryon}

\begin{figure}[h!t]
\centering
\includegraphics[width=0.7\textwidth]{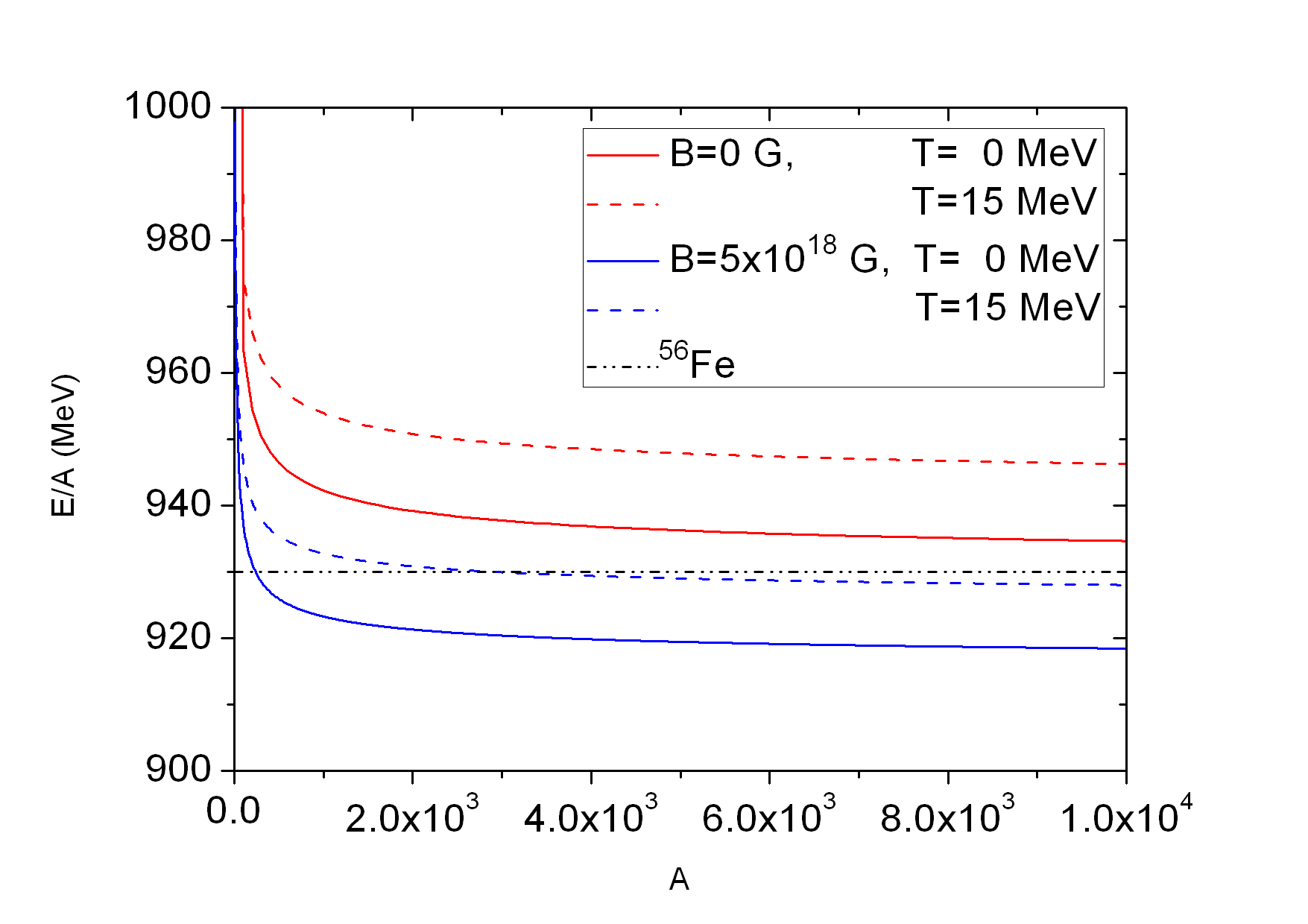}\\
\caption{Energy per baryon for MSQM and SQM strangelets at $T=0,15$~MeV, considering $\mathcal{B}=5\times10^{18}$~G and $B_\text{bag}=75$~MeV fm$^{-3}$.} \label{AEA}
\end{figure}

In Fig.(\ref{AEA}) are shown the energy per baryon curves for MSQM and SQM strangelets at zero and finite temperature $T=15$ MeV, assuming $m_{u}=m_{d}=5$~MeV, $m_{s}=150$~MeV, for $B_\text{bag}=75$~MeV fm$^{-3}$ and $\mathcal{B}=5\times10^{18}$~G. It is observed that for strangelets of MSQM, the energy per baryon lines always lie below the ones of SQM strangelets, therefore, the magnetic field contributes to their stability for the same value of $B_\text{bag}=75$~MeV fm$^{-3}$. The horizontal line corresponds to the energy per baryon of the isotope $^{56}$Fe; for a better comparison one should compare with the energy per baryon of the same isotope in the presence of an external magnetic field, but the energy difference is irrelevant for the magnetic field used in this study \cite{Martinez:2003dz,Paulucci:2010uj}, whic means that  $$\left.\dfrac{E}{A}\right|_{^{56}\text{Fe}}^{\mathcal{B}=0}\simeq\left.\dfrac{E}{A}\right|_{^{56}\text{Fe}}^{\mathcal{B}=5\times10^{18}~G}\simeq 930\;\text{MeV},$$
\noindent and therefore, it is enough to compare with the energy per baryon of $^{56}$Fe at zero magnetic field.

\bigskip

The presence of a magnetic field reduces the values of the $E/A$, even at finite temperature, for the used quark masses and Bag constant. The bulk contribution contains only a finite number of terms due to the Landau level quantization produced by the magnetic field. As the field becomes stronger, the maximum number of Landau levels decreases, contributing less to the bulk energy; these effects predominate for large values of $A$. Magnetized strangelets with relatively large baryon numbers could be absolutely stable even up to $T=15$ ~MeV, indicating the possibility of finding strangelets in ``warm environments'', for example, in particle accelerators. On the other hand, the effects of the temperature tend to increase the values of the $E/A$ as expected, due to the thermal motion of quarks and gluons.

\bigskip

One obtains in the first case, baryon numbers $A_\text{crit}$, where $E/A_{\text{crit}}=930$ ~MeV. Strangelets with baryon numbers less than this critical value, could be found in a meta-stable state, while they could be absolutely stable with baryon numbers over $A_\text{crit}.$ Such ranges of meta-stability, for strangelets of MSQM are: $A\leq 241\,(N\leq723)$ at $T=0$, and $A\leq2891\,(N\leq8673)$ at $T=15$ ~MeV; above these values, strangelets of MSQM are all stable for the value of $B_{\text{bag}}$ used; this doesn't happen to SQM strangelets, which are in meta-stable states for the same $B_{\text{bag}}$.

\begin{figure}[h!t]
\centering
\includegraphics[width=0.7\textwidth]{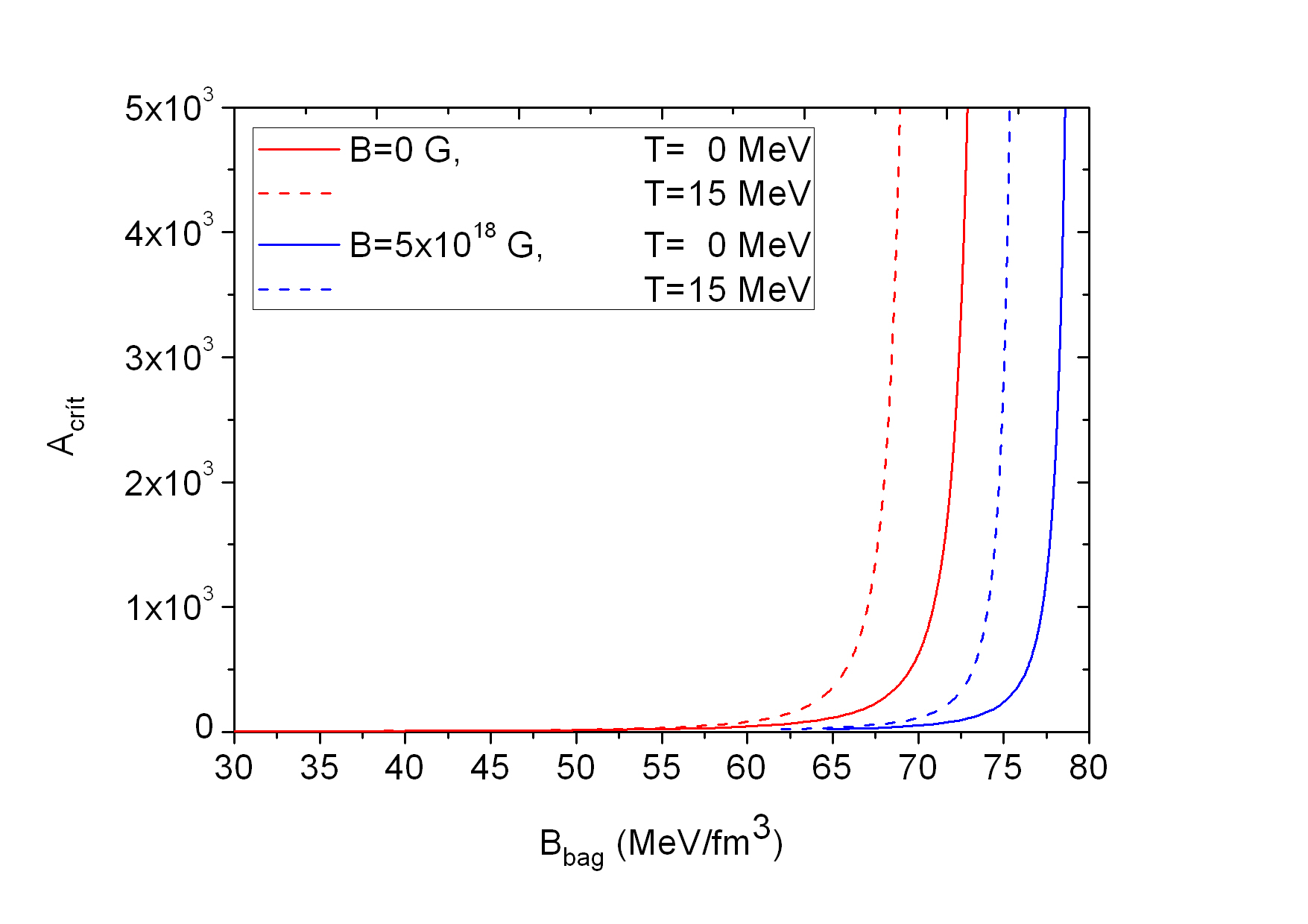}\\
\caption{Dependence of the critical baryon number with $B_{\text{bag}}$ for MSQM and SQM strangelets at $T=0,15$~MeV and $\mathcal{B}=5\times10^{18}$~G.} \label{BACRIT}
\end{figure}

Worth to mention that under this description, the strangelet with $A=56$ corresponding to the ``strange iron nucleus'', in both cases: $\mathcal{B}=0$ and $\mathcal{B}=5\times10^{18}$ G, lies inside the meta-stability range analyzed before; this is due to the fact that the value of $B_{\text{bag}}$ chosen is higher than the cohesion energy of the isotope $^{56}$Fe; therefore, depending on $B_{\text{bag}}$, the temperature and the magnetic field, one can compute a $A_\text{crit}$, for which the strangelet energy per baryon equals to the one of $^{56}\text{Fe}$. This new restriction could be added to the set of equations Eqs.\eqref{barionicnumber} and \eqref{equilibrio}, allowing us to compute $B_{\text{bag}}$ as a function of this $A_\text{crit}$; this is shown in Fig.(\ref{BACRIT}) and constitutes a generalization of the computations done in Sec.(\ref{sletstability}).

The curves correspond to MSQM strangelets configurations with baryon number $A_{\text{crit}}$, such that  $E/A$ coincides with the one of the isotope $^{56}$Fe and the corresponding value of $B_{\text{bag}}.$ For a constant $B_{\text{bag}}$ there's a fixed value of $A_{\text{crit}}$ in each case. The straight lines of constant $B_{\text{bag}}$ they cross two regions: $A<A_{\text{crít}}$, which corresponds to the metastability regions; and $A\geq A_{\text{crit}}$ to the absolute stability relative to $^{56}$Fe. The absolute stability range allowed by $B_{\text{bag}}$, is large in the presence of the magnetic field as shown in Fig.(\ref{BACRIT}).

Notice also that in Fig.(\ref{BACRIT}), for large values of $B_{\text{bag}}$, one obtains an abrupt increase of $A_\text{crit}$, reaching an asymptotic value of $B_{\text{bag}}$, above which, there are no more solutions to the equations. If $\mathcal{B}=0$, the stability requires that $B_{\text{bag}}<73\;\text{MeV}\; \text{fm}^{-3}$ at $T=0$ and $B_{\text{bag}}<70\;\text{MeV}\; \text{fm}^{-3}$ at $T=15$~MeV; when $\mathcal{B}=5\times10^{15}$~G, such limits correspond to $B_{\text{bag}}<79\;\text{MeV}\; \text{fm}^{-3}$ at $T=0$ and $B_{\text{bag}}<76\;\text{MeV}\; \text{fm}^{-3}$ at $T=15$~MeV.

\bigskip

Finally, for $A_\text{crit}=56,$ which corresponds to the strange iron nucleus, one obtains $B_{\text{bag}}\simeq61.4\;\text{MeV}\; \text{fm}^{-3}$ at $T=0$ and $B_{\text{bag}}\simeq58.1\;\text{MeV}\; \text{fm}^{-3}$ at $T=15$~MeV; similar, when $\mathcal{B}=5\times10^{18}$~G, the ranges are $B_{\text{bag}}\simeq70.3\;\text{MeV}\; \text{fm}^{-3}$ at $T=0$ and $B_{\text{bag}}\simeq67.4\;\text{MeV}\; \text{fm}^{-3}$ at $T=15$~MeV respectively.

\subsubsection{Strangelets radii}

\begin{figure}[h!t]
\begin{center}
\includegraphics[width=0.7\textwidth]{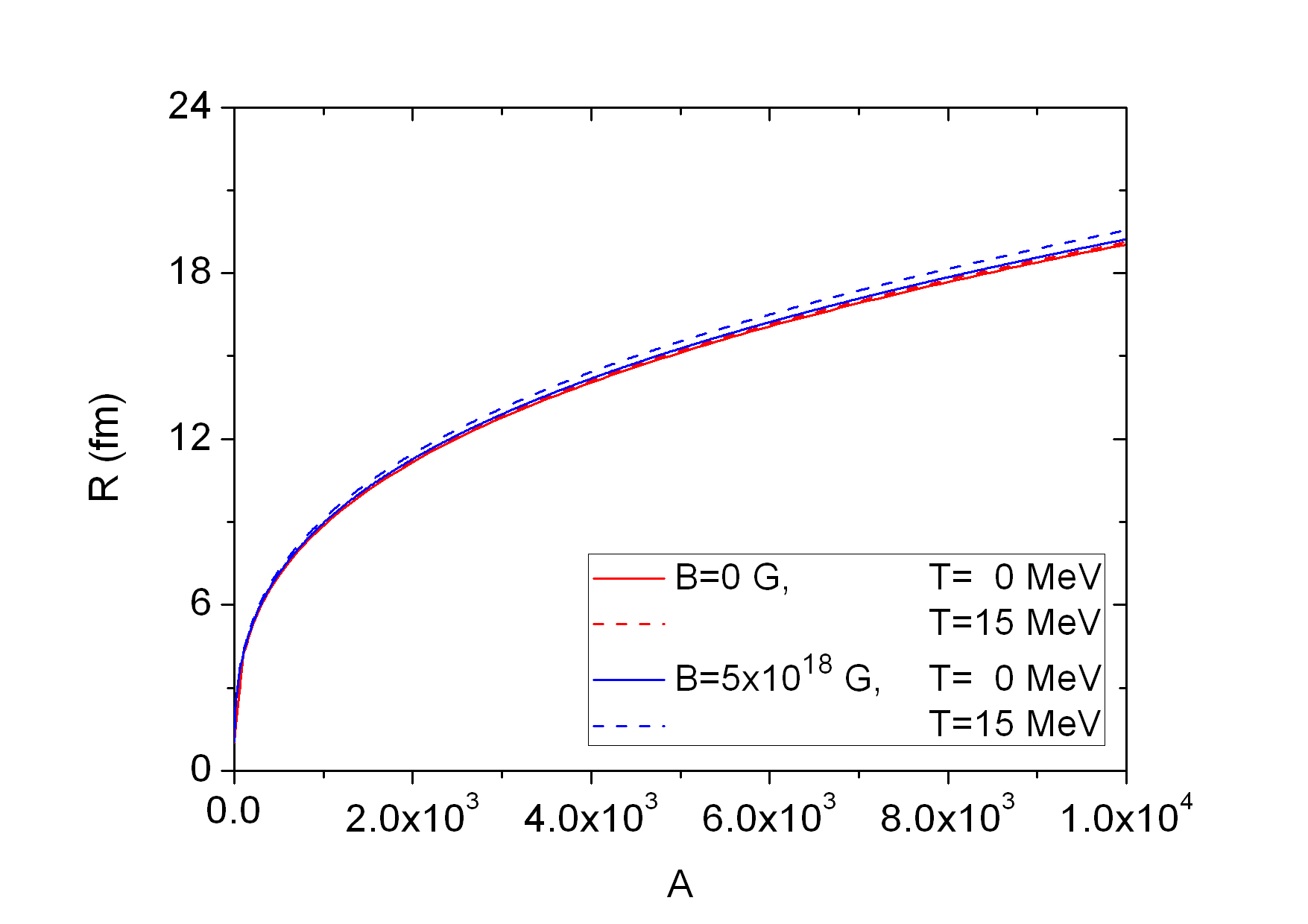}
\end{center}
\caption{Dependence of the radii with the baryon number for $B_\text{bag}=75$~MeV fm$^{-3}$ and $T=0, 15$~MeV. The figure shows the radii of MSQM ($\mathcal{B}=5.0\times10^{18}$~G) and SQM strangelets ($\mathcal{B}=0$) respectively.}\label{ARA}
\end{figure}

The dependence of the radii of MSQM and SQM strangelets with the baryon number $A$, is shown in Fig.(\ref{ARA}). Due to the quarks and gluons thermal motion, the radius increases with the temperature for a fixed baryon number. The same behavior is observed with the magnetic field, which is a direct result of the relaxation produced on the surface. The energy preserving the stability is diminished by the effects of the magnetic field, which tends to stabilize strangelets, but also to increase their radii.

Analogous to atomic nuclei, between the radius $R$ and the baryon number, one finds a relationship like $R=r_{0}A^{\frac{1}{3}}$, for large values of $A$; which for the chosen parameters is:
\begin{equation}\label{RAMSQM}
\left.R\right|_{0}=0.89\, A^{\frac{1}{3}}~\text{fm}, \quad
\left.R\right|_{15}=0.91\, A^{\frac{1}{3}}~\text{fm},
\end{equation}
\noindent which correspond to MSQM strangelets, meanwhile
\begin{equation}\label{RASQM}
\left.R\right|_{0}=0.88\, A^{\frac{1}{3}}~\text{fm}, \quad
\left.R\right|_{15}=0.89\, A^{\frac{1}{3}}~\text{fm},
\end{equation}
\noindent for SQM strangelets. Notice that, the temperature and magnetic field modify the values of $r_{0}$ around $1\%$, being $r_{0}$ greater in the case of MSQM strangelets, but always lower to the nuclear critical radius $1.12$~fm.

\bigskip
In Fig.(\ref{ARA}), as well as in the coefficients $r_{0}$ of Eqs.\eqref{RAMSQM} and \eqref{RASQM} respectively, one can see a small deviation of the radius as a function of the magnetic field and the temperature, for a fixed baryon number. This is due to the fact that the $s$ quarks have a great mass  compared to $u$ and $d$ quarks. For $\mathcal{B}=5\times10^{18}\;\text{G}$ and $T=15$ MeV, the effects of the magnetic field and temperature are almost irrelevant for the $s$ quarks, as can be noticed from:
\begin{align}
M_{0,u}^{\pm}\simeq11\;\text{MeV}>m_{u},&\;\;\;M_{0,d}^{\pm}\simeq20\;\text{MeV}>m_{d},&M_{0,s}^{\pm}\simeq150.02\; \text{MeV}\simeq m_{s}\label{Magmasa}\\
\;\;\;\;\;\;\;\;\;\;\;T=15\;\text{MeV}>m_{u},&\;\;\;\;\;\;\;T=15\;\text{MeV}>m_{d},&T=15\;\text{MeV}<m_{s},
\end{align}
\noindent where $M_{\nu,f}$ is the magnetic mass of each quark and $T$ is the thermal energy.

\begin{figure}[h!t]
\includegraphics[width=0.5\textwidth]{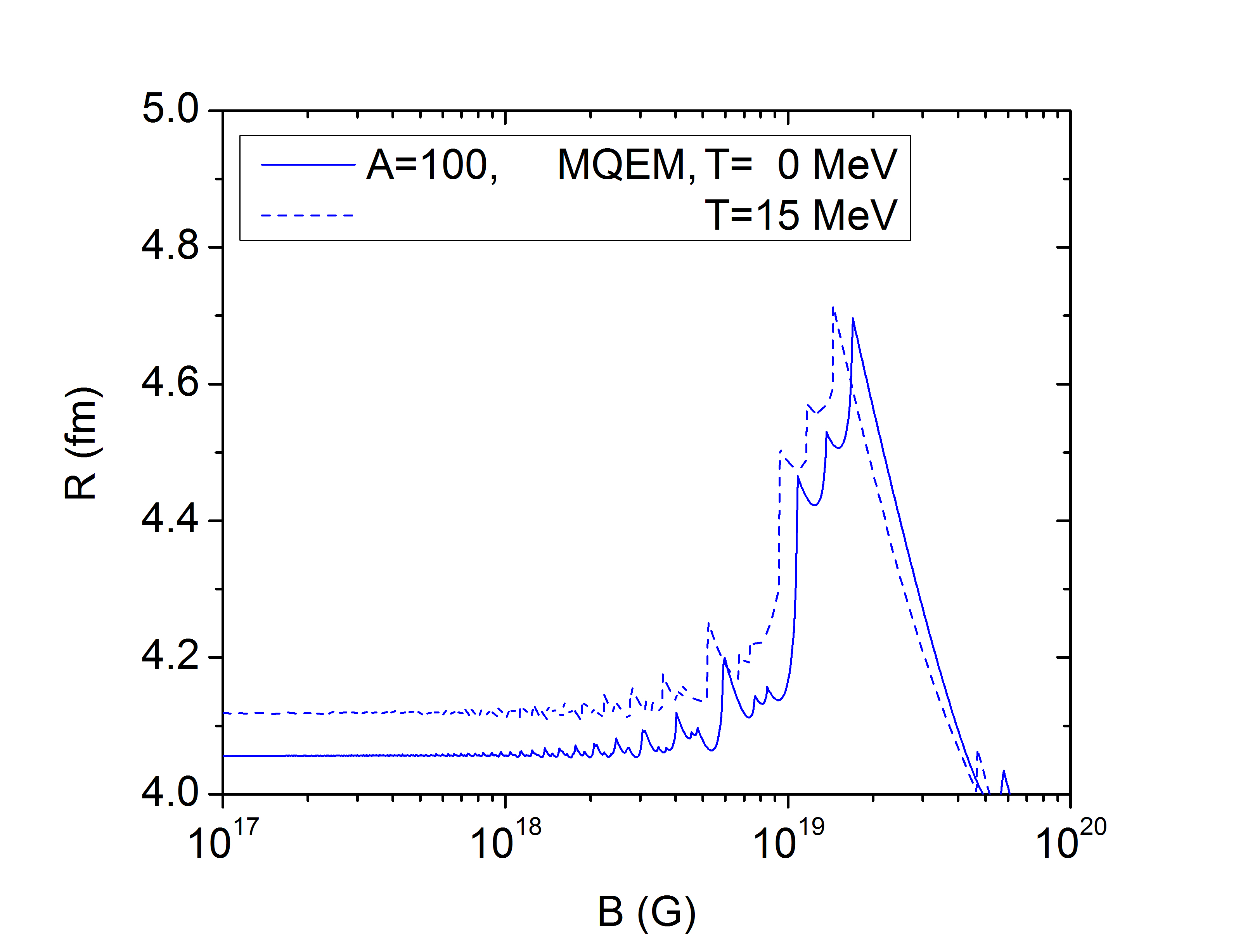}
\includegraphics[width=0.5\textwidth]{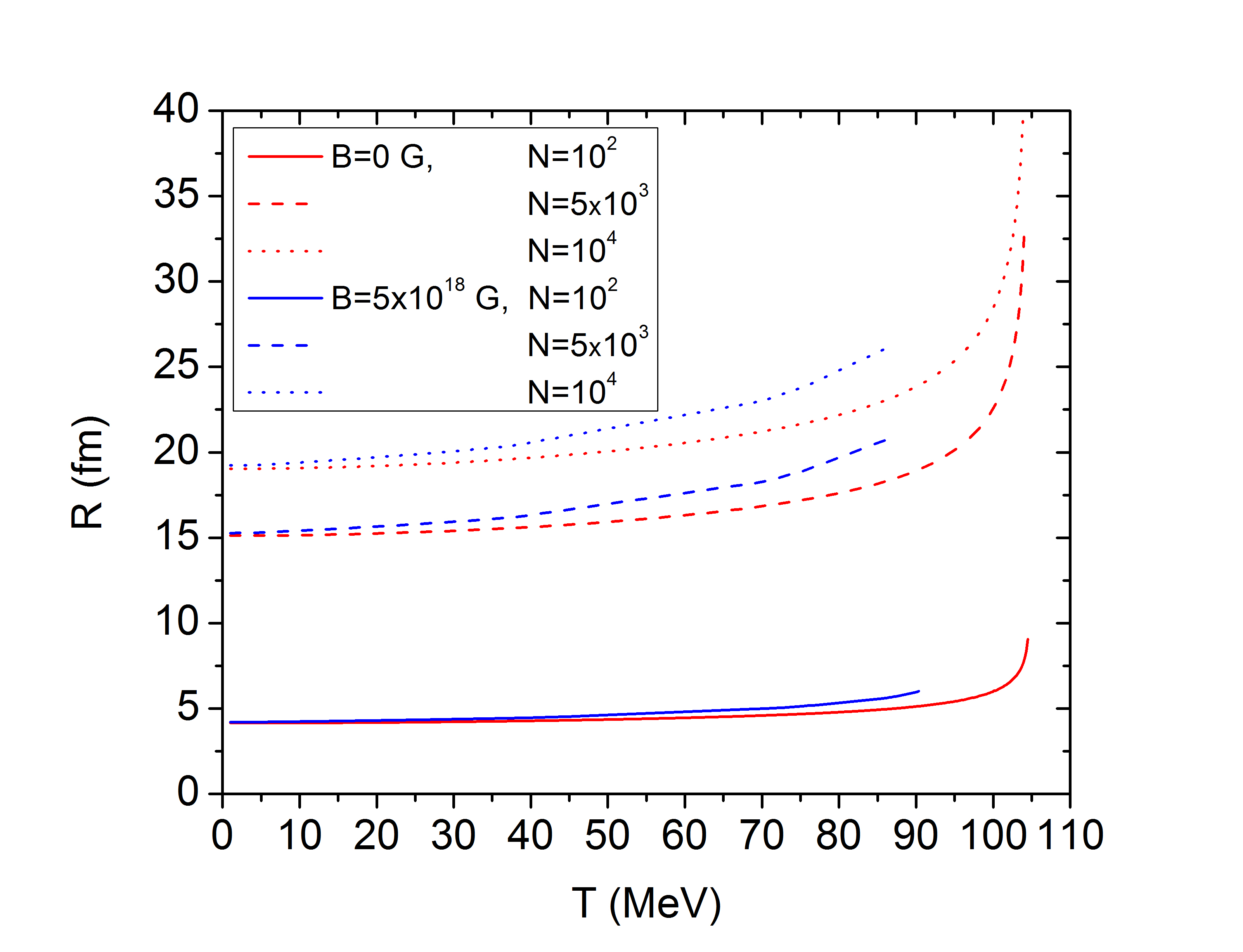}
\caption{Dependence of the radius of MSQM and SQM strangelets with the magnetic field and temperature for a fixed baryon number and $B_\text{bag}=75$~MeV fm$^{-3}$.}\label{BRAT}
\end{figure}

When computing the radius $R$ through Eqs.\eqref{equilibrio0} and \eqref{equilibrio1}, massive particles contribute relevantly to $\Omega_{S},$ (in this case the $s$ quarks); classically one has the behavior $R\sim2\left|\dfrac{\Omega_{S}}{\Omega_{V}}\right|,$ which makes the curves of $R$ as a function of the magnetic field and the temperature vary little with the values taken of $\mathcal{B}$\; and \; $T$. On the other hand, if $\mathcal{B}$\; and/or\; $T$ increases such that they reach the values $\mathcal{B}\sim m_{s}^{2}/q_{s}$\; and \; $T\sim m_{s}$, then one can notice suddenly that $R$ starts to vary considerably, as shown in Figs.(\ref{BRAT}).

In the left panel of Figs.(\ref{BRAT}) is shown the behavior of the radius of MSQM strangelets with $A=100$ at $T=0, 15$~MeV respectively, as a function of the magnetic field. In the right panel is shown the behavior of the radius with the temperature, for $A=100, 5000$ and $A=10000$, with $\mathcal{B}=5.0\times10^{18}$~G, and in both cases $B_\text{bag}=75$~MeV fm$^{-3}$. One can observe that for fields below $\mathcal{B}=5\times10^{18}$ G and temperatures below $40$ MeV, the strangelets radius behaves almost constant; noticeable variations are observed for fields and temperatures above the one mentioned before, i.e. when their values become comparable to the $s$ quarks mass.

\noindent

\subsubsection{Electric charge}

\begin{figure}[h!t]
\begin{center}
\includegraphics[width=0.7\textwidth]{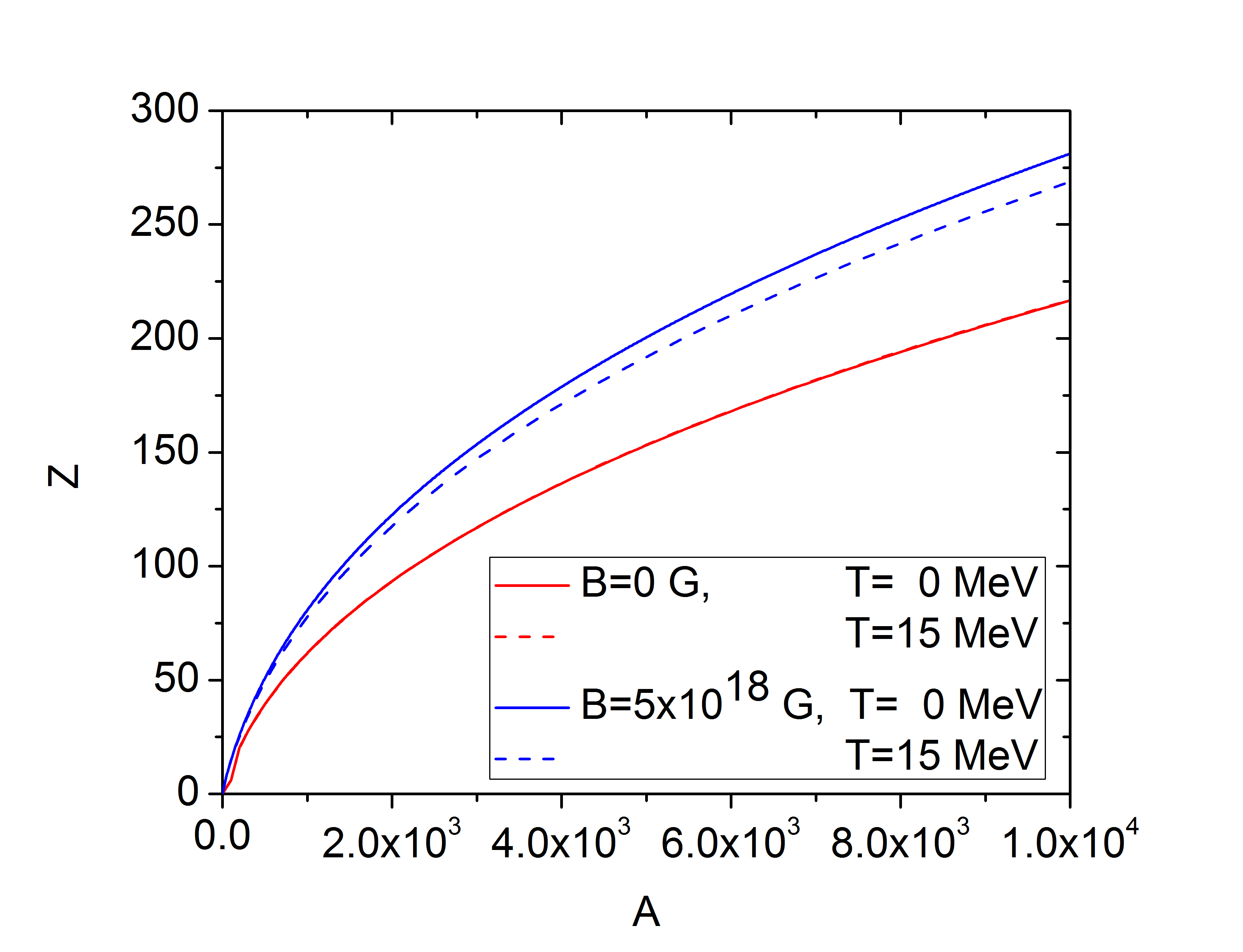}
\end{center}
\caption{MSQM and SQM strangelets electric charge for $B_\text{bag}=75$~MeV fm$^{-3}$, $T=0, 15$~MeV and $\mathcal{B}=5.0\times10^{18}$~G.}\label{ZA}
\end{figure}

Finally, the dependence of the total electric charge given by Eq.\eqref{Ztotal}, with respect to the baryon number is shown in Fig.(\ref{ZA}). The electric charge increases with $A$ as expected; it also increases with the magnetic field at a fixed baryon number; however, with the temperature, it decrease since antiparticles begin to populate states outside the ``Dirac sea''.

The relationship between electric charge and baryon number is a topic well discussed in the scientific literature related to strangelets. At small radii, which do not exceed $\lambda_{D},$ this dependence should be linear, because the screening effects play no role (contribute little) and the  surface contribution is negligible. As the radius increases, and consequently the baryon number, the screening effects of the electric charge become of relevant importance and therefore, the relation $Z-A$ is not longer linear.

\bigskip

In Ref.\cite{Heiselberg:1993dc} is obtained a relation like $Z\sim A^{\frac{1}{3}}$. In addition to the bulk charge screening, which would produce a behavior like Ref.\cite{Heiselberg:1993dc}, I have included the effects of the free surface charge through Eq.\eqref{Ztotal}, which introduces a factor  proportional to $A^{\frac{1}{3}}$ due to the curvature term, and one of the form $A^{\frac{2}{3}}$ due to the surface area. Obviously, for very large values of $A$, the behavior is $Z\sim A^{\frac{2}{3}} $. However, in the range of baryon numbers I used, one can find a relationship like $Z=Z_{0}A^{\alpha}$ with $Z_{0}$ and $\alpha$ given by 
\begin{equation}\label{Z0B}
\left.Z\right|_{0}\;\,=2.21\,A^{0.53},\;\;\;\;\left.Z\right|_{15}=2.16\,A^{0.53},
\end{equation}
\noindent for MSQM strangelets, and
\begin{equation}\label{Z0B0}
\left.Z\right|_{0}\;\,=1.62\,A^{0.53},\;\;\;\;\left.Z\right|_{15}=1.61\,A^{0.53},
\end{equation}
\noindent for SQM strangelets.

\bigskip

Eqs.\eqref{Z0B} and \eqref{Z0B0} always show positively-charged strangelets configurations; as the temperature increases, the coefficients $Z_{0}$ decrease, in both cases, as a consequence of the antiparticles contribution. With the magnetic field, a contrary effect is produced since the coefficients $Z_{0}$ increase with the field.

\bigskip

At zero field, the bulk densities of quarks $u$ and $d$ coincide because of the equality of their rest masses, while the bulk density of quarks $s$ is different and lower than the previous two because of the high mass, and because the three gasses have the same Fermi energy. The effects of the magnetic field on the electrical charge of strangelets, start to become noticeable when the two densities of quarks $u$ and $d$, initially equal, they are split into two separate and unequal densities due to the difference between the electric charges thereof; the density of quarks $s$ decreases little with the field in the range studied. Since the effects of the magnetic field affect more the quark $d$, see Eqs.\eqref{Magmasa}, the bulk density of these decreases and consequently also the density of negative charges.

\begin{figure}[h!t]
\begin{center}
\includegraphics[width=0.7\textwidth]{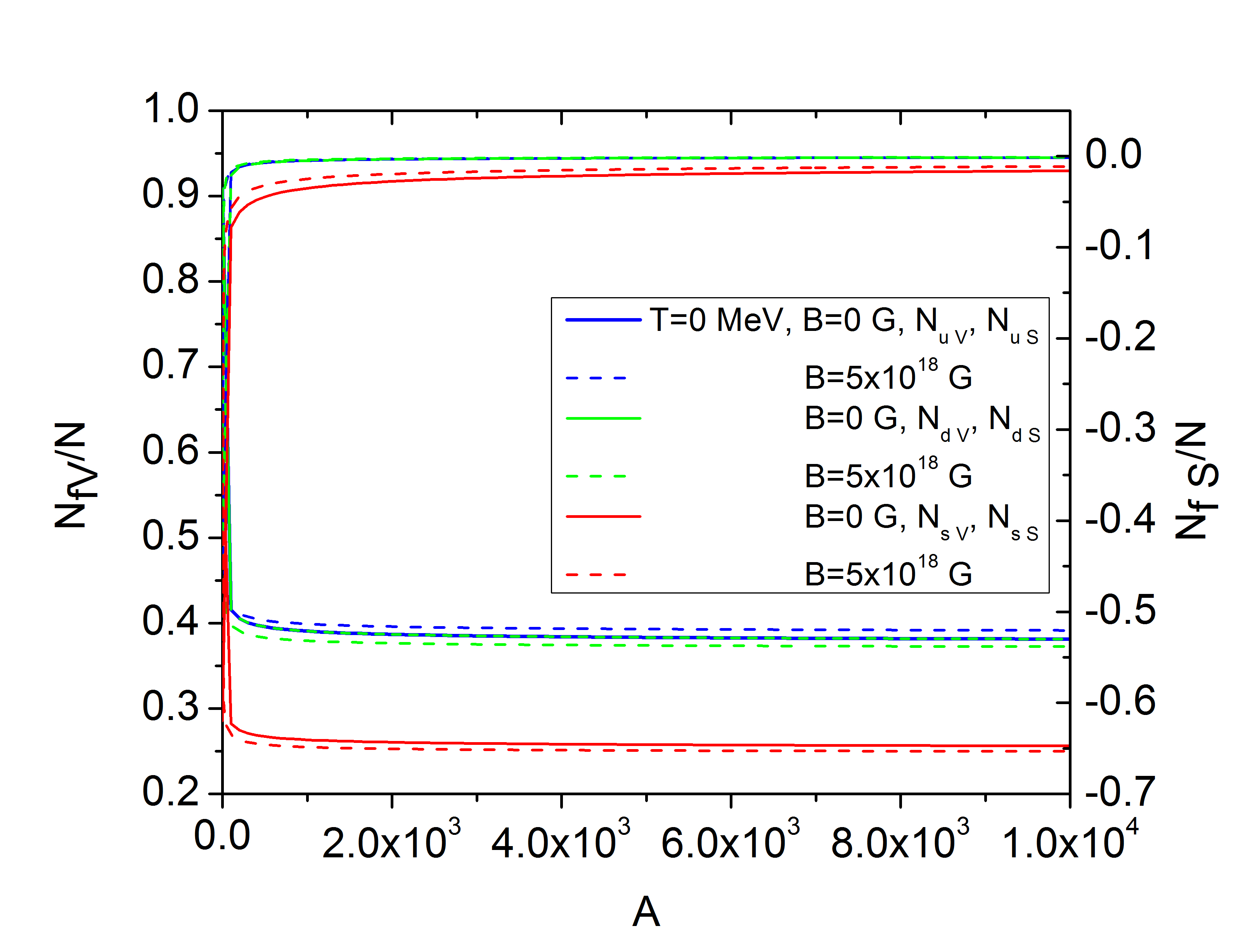}
\end{center}
\caption{Bulk and surface fraction of $u$, $d$ and $s$  quarks for MSQM and SQM strangelets with $B_\text{bag}=75$~MeV fm$^{-3}$, $T=0, 15$~MeV and $\mathcal{B}=5.0\times10^{18}$~G.}\label{PDEN}
\end{figure}

On the other hand, as the field affects little the $s$ quarks, and as the baryon number is fixed, the density of $u$ quarks have to increase, thereby producing also an increase in the bulk density of positive charges.

\section{Strangelets of magnetized strange quark matter in the color superconductor phase CFL}\label{MCFL}

Similarly to the case of MSQM strangelets, I study in this section the relevant properties MSQM strangelets in the magnetized CFL phase (MCFL), at finite temperature. The model used for the study of this color superconducting phase of the MSQM, which is the most symmetric phase, will be briefly explained. In the limit $\mathcal{B}=0$ and $T=0$, the results obtained in this section are consistent with those obtained in Refs.\cite{Paulucci:2008jd, Madsen:2001fu}. The parameter space in this case is ($B_{\text{bag}}, \mathcal{B}, T, \Delta$).

\subsection{Color superconductor in the presence of a magnetic field}\label{sec3.3}

To study strangelets in the MCFL phase, first I will discuss some of the main features of this phase of the SQM. The main difference between this phase and the ordinary superconductivity is that in the last, the local symmetry $\text{U}(1)_{\gamma}$ is broken with the consequent appearance of a photon mass term; this leads to the well known Meissner effect, which basically consists in ejection of the magnetic field lines inside the superconductor. In the case of the color superconductivity, the broken local symmetry is of $\text{SU}(3)_{\text{color}}$ and therefore the eight gluons become massive.

Quarks not only interact via gluon exchange, but also with photons since they are electrically charged; this leads to the appearance of a massless $\widetilde{A_{\mu}}=A_{\mu}\cos\theta-G_{\mu}^{a}\sin\theta$ and a massive $\widetilde{G_{\mu}^{a}}=A_{\mu}\sin\theta+G_{\mu}^{a}\cos\theta$ linear combinations between the photon $A_{\mu}$ and the gluon $G^{a}_{\mu}$ components. As a result, there appear a rotated magnetic field $\tilde{B}$ and a rotated electric charge $\tilde{e}=e\cos\theta $, such that the mixing angle $\theta$ depends on the gap structure, and in the CFL phase is given by $\cos\theta=g/\sqrt{e^{2}/3+g^{2}}$ ($g$ is the QCD coupling constant) ~\cite{Alford:1999pb, Gorbar:2000ms}. Since the the rotated photon remains massless, the rotated magnetic field within the color superconductor $\tilde{B}$ is not screened and therefore there's no Meissner effect.

In the region of interest for Astrophysics, $e\ll g,$ so $\cos\theta\sim1,$ then the strength of the magnetic field inside and outside the system in the CFL phase will be approximately equal, i.e. $\tilde{e}\tilde{B}\simeq e\mathcal{B} $ ~\cite{Fukushima:2007fc}. This fact is relevant for Astrophysics, since one of the results is that the field inside the color superconductor could be ``frozen'' and even strengthened in the nuclei of compact objects with strong magnetic fields, such as magnetars \cite{Fukushima:2007fc, Felipe:2010vr, Ferrer:2005vd, Cristinam:2007, Noronha:2007wg}. Studies on rotating of pulsars could confirm if their interiors are formed by a color superconducting phase of the MSQM.

Discussed the effect of the field on the CFL phase of the SQM, this model will be used also to study strangelets. I will take into account a single and common value of $\Delta$ based on Ref.\cite{Felipe:2010vr} for the predominant color pairing pattern ($ud,us,ds$) ~\cite{Fukushima:2007fc, Noronha:2007wg}, and any dependence of the energy gap with the magnetic field is neglected; however, a temperature dependence will be taken into account. A more robust model would require the determination of $\Delta$ from the equations of motion, and where appears an explicit dependence on the magnetic field \cite{Ferrer:2005vd, Paulucci:2010uj}. In that case, one obtains two energies gaps, one depends on the field and the other is independent; this is a direct consequence of the appearance of a rotated charge in the MCFL phase \cite{Ferrer:2005vd, Paulucci:2010uj}.

\subsubsection{Hydrostatic equilibrium condition}

The dependence of $\Delta$ on the temperature $T$ ~\cite{Paulucci:2008jd,Alford:2007xm,Schmitt:2002sc} that I will use in this thesis is given by:
\begin{equation}\label{Delta}
\Delta=2^{-1/3} \Delta_{0}\left[1-\left(\frac{T}{T_c}\right)^{2}\right]^{1/2},
\end{equation}
\noindent where $T_{c}=2^{1/3} e^\gamma \Delta_0/\pi \simeq 0.71 \Delta_{0}$ is the critical density, above which the color superconductor phase disappears, forbidding the quark pairings. I will consider the effects of the color superconductivity only in the bulk terms, therefore, one needs to add a new contribution to the thermodynamical potential, which depend on $\Delta$ as:
\begin{equation}\label{CFL}
\Omega_{V}=\sum_{f=u,d,s}\Omega_{f,v}-\dfrac{3\Delta^2\mu_{B}^{2}}{\pi^2},
\end{equation}
\noindent where the second term of the right-hand-side, takes into account the variation of the free energy, proportional to the square of the baryon chemical potential $\mu_{B}=(\mu_{u}+\mu_{d}+\mu_{s})/3$, and it is compensated with the quark pairs formation ~\cite{Felipe:2010vr}.

\bigskip

The color superconducting phase, is also characterized by the equality of the quarks bulk densities, which minimizes the free energy, forcing the system to be electrically neutral \cite{Alford:2001zr, Rajagopal:2000ff, Alford:2007xm, Alford:2002kj, Alford:2004pf}. This last requirement, leads to the following set of equations
\begin{equation}
N_{u,v}+\dfrac{2\Delta^2\mu_B}{\pi^2}=N_{d,v}+\dfrac{2\Delta^2\mu_B}{\pi^2}
=N_{s,v}+\dfrac{2\Delta^2\mu_B}{\pi^2}. \label{Nequality}
\end{equation}

\noindent From Eqs.\eqref{Nequality} is clear that:

\begin{itemize}
\item each chemical potential is different $\mu_{u}\neq\mu_{d}\neq\mu_{s},$
\item the bulk electric charge of MCFL strangelets vanishes, as well as the Coulomb contribution to the total energy, therefore, the electric charge of MCFL strangelets comes from the surface charge distribution.
\item The screening of the electric charge is negligible, in contrast with MSQM strangelets.
\end{itemize}

The hydrostatic equilibrium equations for MCFL strangelets can be written as:
\begin{eqnarray}
\left.\dfrac{\partial F_{g}}{\partial V}\right|_{N,T,\mathcal{B}}^{MCFL}+
\left.\dfrac{\partial F_{q\overline{q}}}{\partial
V}\right|_{N,T,\mathcal{B}}^{MCFL}=0,\label{equilibrioCFL}
\end{eqnarray}
\noindent where
\begin{eqnarray}\label{equilibrio1CFL}
\left.\dfrac{\partial F_{g}}{\partial
V}\right|_{N,T,\mathcal{B}}^{MCFL}=\Omega_{g,v}+\dfrac{2}{R^{2}}\Omega_{g,c},\;\;\;\;\;\;\;\;\;\;\;\;\;\;\;\;\;\;\\
\left.\dfrac{\partial F_{q\overline{q}}}{\partial
V}\right|_{N,T,\mathcal{B}}^{MCFL}=B_{\rm bag}+\Omega_{V}+ \dfrac{2}{R}\Omega_{S}+\dfrac{2}{R^{2}}\Omega_{C},\label{equilibrio2CFL}
\end{eqnarray}
\noindent and
\begin{align}\label{equilibrio3CFL}
\Omega_{V}=\sum_{f=u,d,s}\Omega_{f,v}-\dfrac{3\Delta^2\mu_{B}^{2}}{\pi^2},\quad
\Omega_{S}=\sum_{f=u,d,s}\Omega_{f,s},\quad
\Omega_{C}=\sum_{f=u,d,s}\Omega_{f,c}.
\end{align}

As in MSQM strangelets, given the values of $B_{\text{bag}},$ the magnetic field, temperature and the gap energy $\Delta,$ one obtains a system of equations: Eq.\eqref{barionicnumber} and Eqs.\eqref{Nequality}, which allows us to determine the three chemical potentials $\mu_{u}, \mu_{d}, \mu_{s}$ and the radius $R$, to later evaluate the other thermodynamical parameters and the electric charge.

\subsubsection{Energy per baryon}

\begin{figure}[h!t]
\begin{center}
\includegraphics[width=0.7\textwidth]{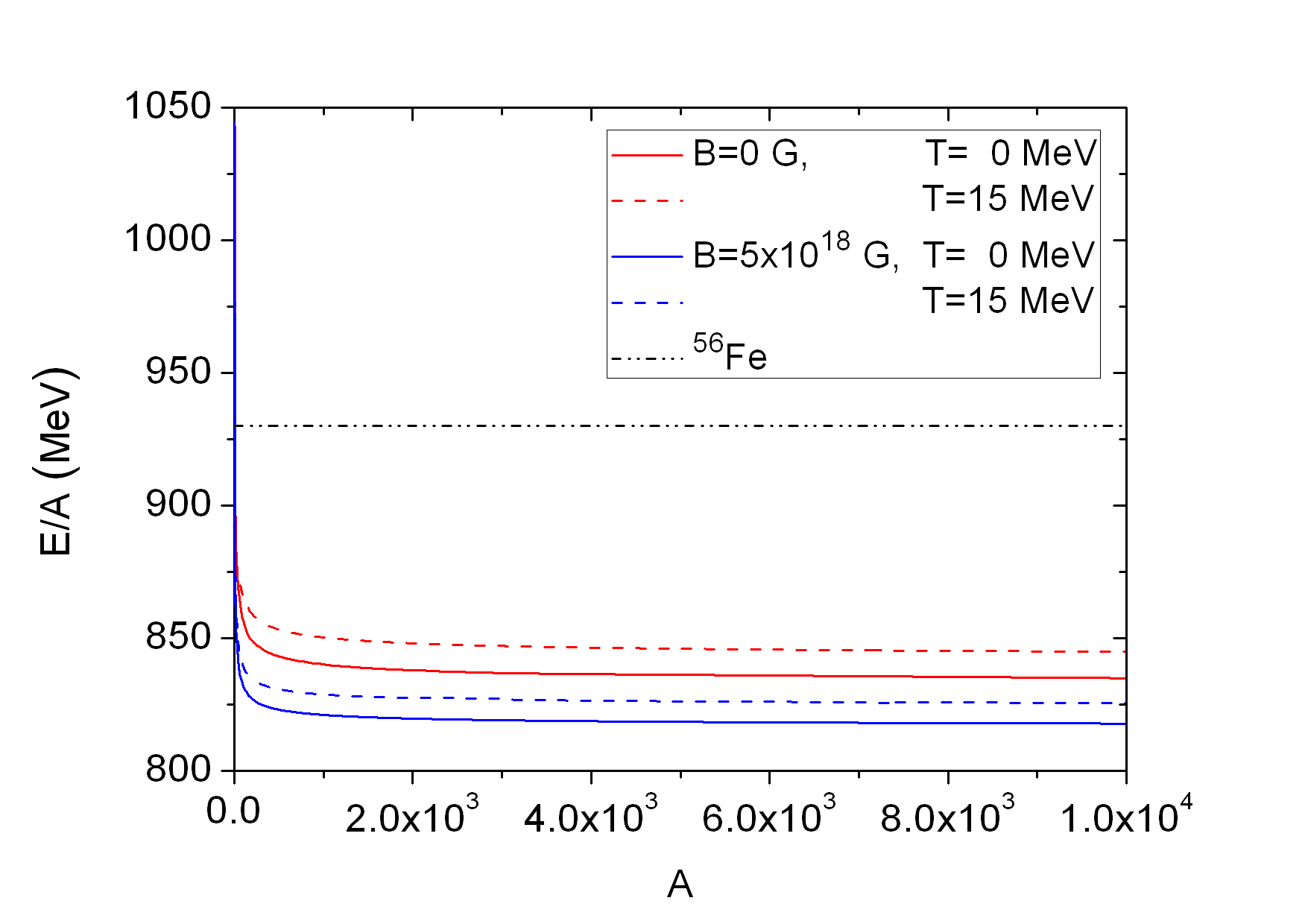}
\end{center}
\caption{\label{AEACFL} Energy per baryon of MCFL and CFL strangelets at $T=0, 15$~MeV, $B_\text{bag}=75$~MeV fm$^{-3}$, gap energy $\Delta=100$~MeV and $\mathcal{B}=5\times 10^{18}$~G, $\mathcal{B}=0$.}
\end{figure}

In Fig.(\ref{AEACFL}) is shown the energy per baryon for MCFL and CFL strangelets , depending on the baryon number $A$. In this case, the parameter space is the same as the previous section, but with the addition of gap energy $\Delta=100$ MeV. Notice that for the chosen $\Delta$, the $E/A$ is lower than in the MSQM case, even at zero magnetic field. The formation of pairs of quarks decreases the values of the energy per baryon; a similar result is obtained in atomic nuclei, where isotopes with equal numbers of protons and neutrons are more stable. Similarly to what happens for strangelets of MSQM (previous section), the surface curvature terms and play an essential role in the stability at small baryon numbers.

\bigskip

Fixed the values of $B_\text{bag}$, the temperature and the magnetic field, one can always find a $A_\text{crit}$ such that $E/A_\text{crit}=930$~MeV. This $A_\text{crit}$ depends strongly on the gap energy $\Delta$. For $B_\text{bag}=75$~MeV fm$^{-3}$, MCLF strangelets with $\Delta=100$~MeV are meta-stables for $A\leq3.6$ at $T=0$, and $A\leq11.6$ at $T=15$~MeV, which corresponds to $N \leq 11$ and $N\leq 12$ particles respectively. However, for CFL strangelets the ranges are: $A\leq7.9$ at $T=0$ and $A\leq9.2$ at $T=15$~MeV, with $N\leq24$ and $N\leq28$ respectively. These computations support the idea that the color superconducting phase CFL, may be the most stable phase of matter in Nature, since it has a lower energy per baryon than that of the SQM.

\bigskip

\begin{figure}[h!t]
\begin{center}
\includegraphics[width=0.7\textwidth]{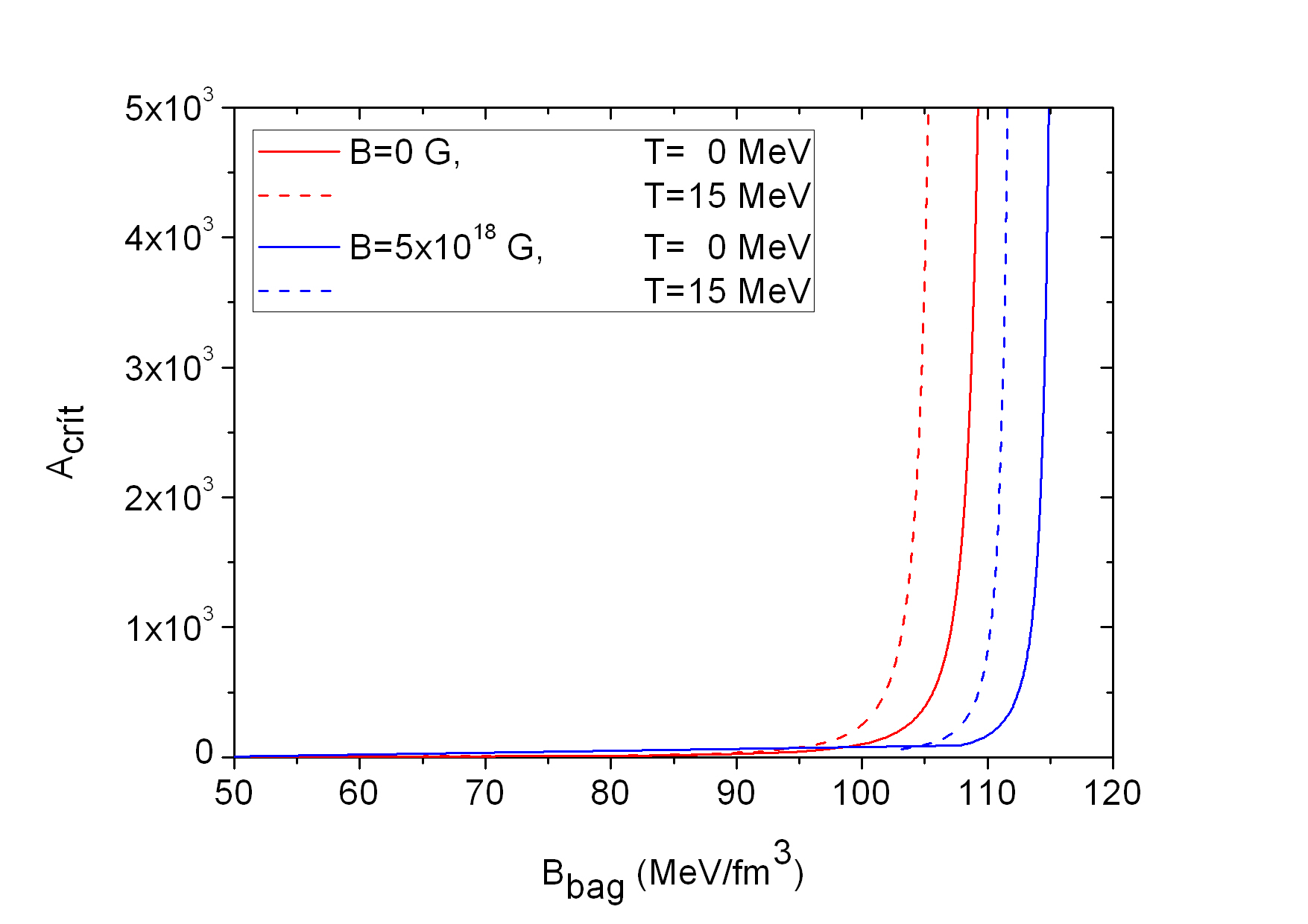}
\end{center}
\caption{Dependence of the critical baryon number $A_{\rm crit}$ with $B_{\rm bag}$ for MCFL and CFL strangelets. The values of $\mathcal{B}=5\times10^{18}$~G and $\mathcal{B}=0$, $T=0$ and 15~MeV, plus $\Delta = 100$~MeV have been fixed.}\label{AcritCFL}
\end{figure}

In Fig.(\ref{AcritCFL}) is shown the dependence of the critical baryon number on $B_{\text{bag}}$. The curves represent MCFL strangelet configurations with baryon number $A_{\text{crit}}$, such that their $E/A$ coincide with the one of the isotope $^{56}$Fe and the corresponding value of $B_{\text{bag}}$. For a constant value of $B_{\text{bag}}$ the critical baryon number $A_{\text{crit}}$ is fixed in each case. The lines of constant $B_{\text{bag}}$ divide two regions: $A<A_{\text{crit}}$, which corresponds to the meta-stability region; and $A\geq A_{\text{crit}}$ to the absolute stability region relative to the $^{56}$Fe. The absolute stability range allowed for $B_{\text{bag}}$, is bigger in the presence of a magnetic field, as observed in Fig.(\ref{AcritCFL}). Again, for values of $B_{\rm bag}$ greater than $115$~MeV fm$^{-3}$, there are no longer solutions to the hydrostatic equilibrium equations and therefore, no stable configurations of strangelets appear. At $\mathcal{B}=0$, this condition requires that $B_{\rm bag} \leq110$~MeV fm$^{-3}$ at $T=0$ and $B_{\rm bag} \leq 106$~MeV fm$^{-3}$ at $T=15$~MeV. On the other hand, when $\mathcal{B}=5\times 10^{18}$~G, these values correspond to $B_{\rm bag} \leq 115$~MeV fm$^{-3}$ at $T=0$, and $B_{\rm bag} \leq 112$~MeV fm$^{-3}$ at $T=15$~MeV respectively. Hence, the presence of a magnetic field and the color superconductivity, allow a greater stability range for strangelets.

\subsubsection{Strangelets radii}

\begin{figure}[h!t]
\begin{center}
\includegraphics[width=0.7\textwidth]{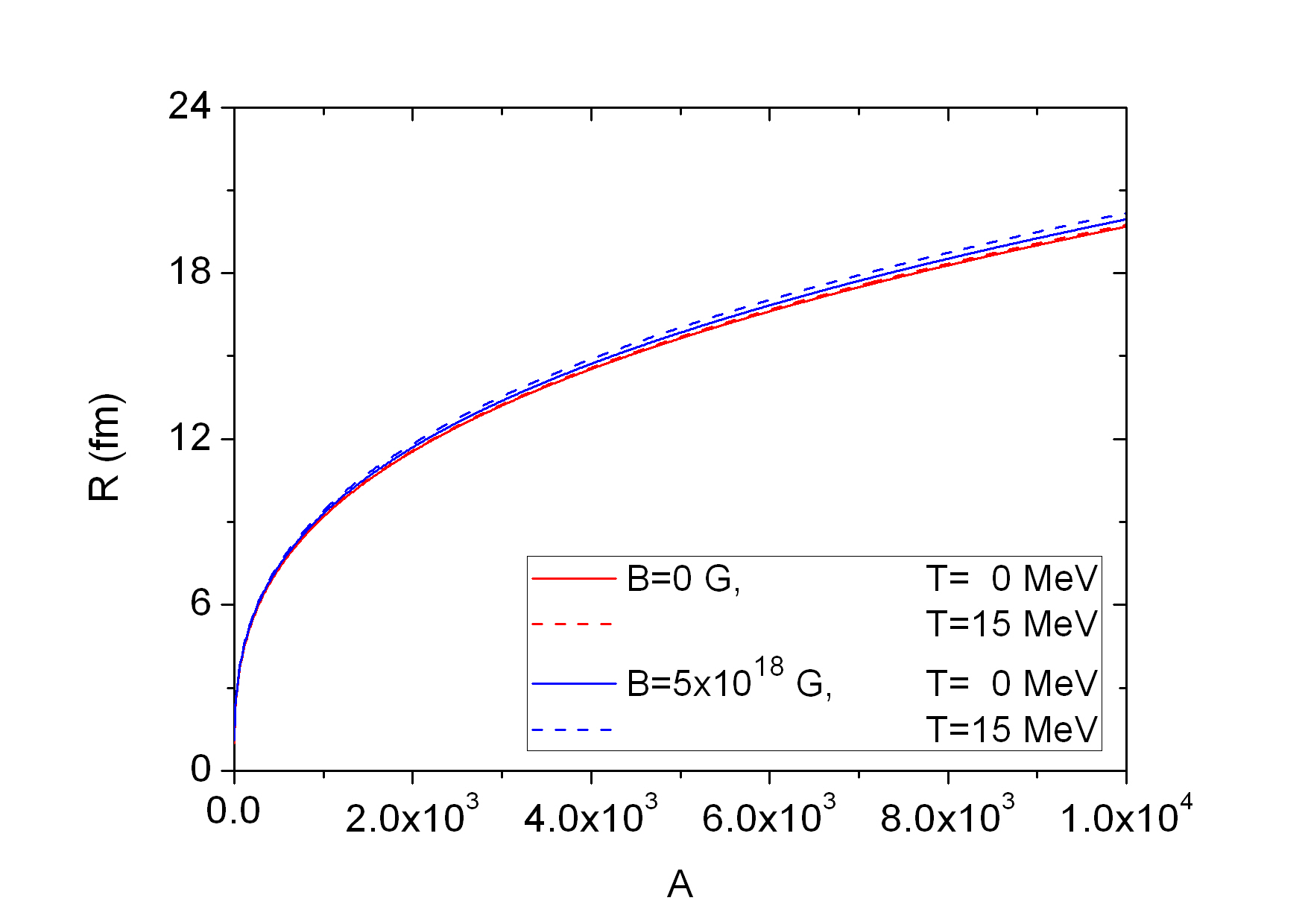}
\end{center}
\caption{Dependence of the radius of MCFL and CFL strangelets with the baryon number $A$, for $\mathcal{B}=5.0\times10^{18}$~G and $\mathcal{B}=0$, $T=0,15$~MeV, $B_\text{bag}=75$~MeV fm$^{-3}$ and $\Delta=100$~MeV.}\label{RACFL}
\end{figure}

The radius of MCFL and CFL strangelets at $B_\text{bag}=75$~MeV fm$^{-3}$, $\Delta=100$~MeV, $T=0,15$~MeV, $\mathcal{B}=5.0\times10^{18}$~G and $\mathcal{B}=0$, is shown in Fig.(\ref{RACFL}) as a function of the baryon number $A$. In this case, one observes the same behavior of $R$ with $A$, as in MSQM and SQM strangelets.

As expected, the radius of strangelets increases with the temperature, the magnetic field and the gap energy (similar to QSs ~\cite{Felipe:2010vr}). As in the previous section, one can look for an analytical dependence of $R$ with $A$ according to $R=r_{0}A^{\frac{1}{3}}$; from where one gets:

\begin{equation}
\left.R\right|_{0}=0.93\, A^{\frac{1}{3}}~\text{fm}, \quad
\left.R\right|_{15}=0.94\, A^{\frac{1}{3}}~\text{fm},
\end{equation}
\noindent for MCFL strangelets, and
\begin{equation}
\left.R\right|_{0}=0.91\, A^{\frac{1}{3}}~\text{fm}, \quad
\left.R\right|_{15}=0.92\, A^{\frac{1}{3}}~\text{fm},
\end{equation}
\noindent for CFL strangelets. There's an increase of $2\%$ in the values of the critical radius with the magnetic field, $1\%$ with the temperature, and around $3\%$ with the energy gap.

\begin{figure}[h!t]
\includegraphics[width=0.5\textwidth]{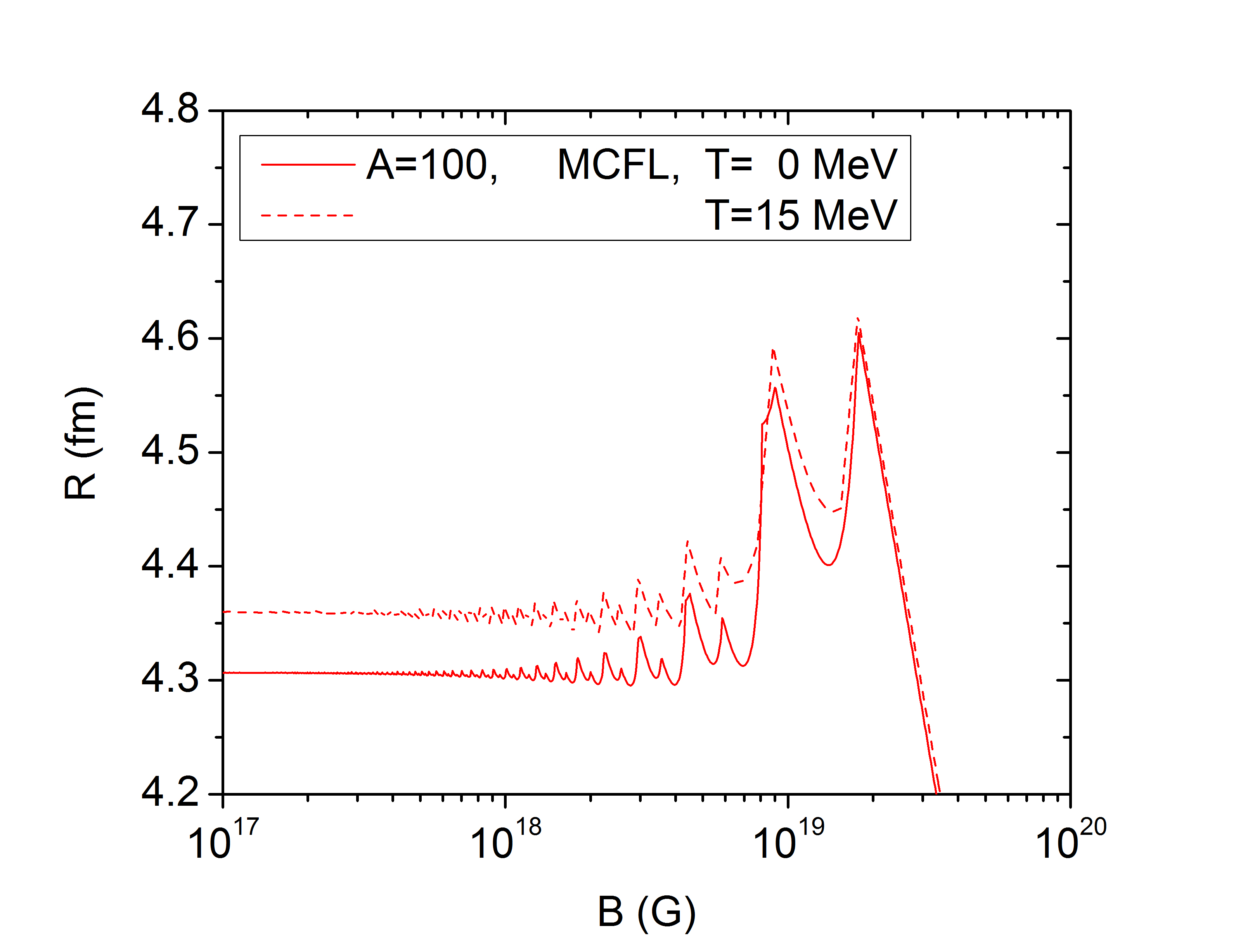}
\includegraphics[width=0.5\textwidth]{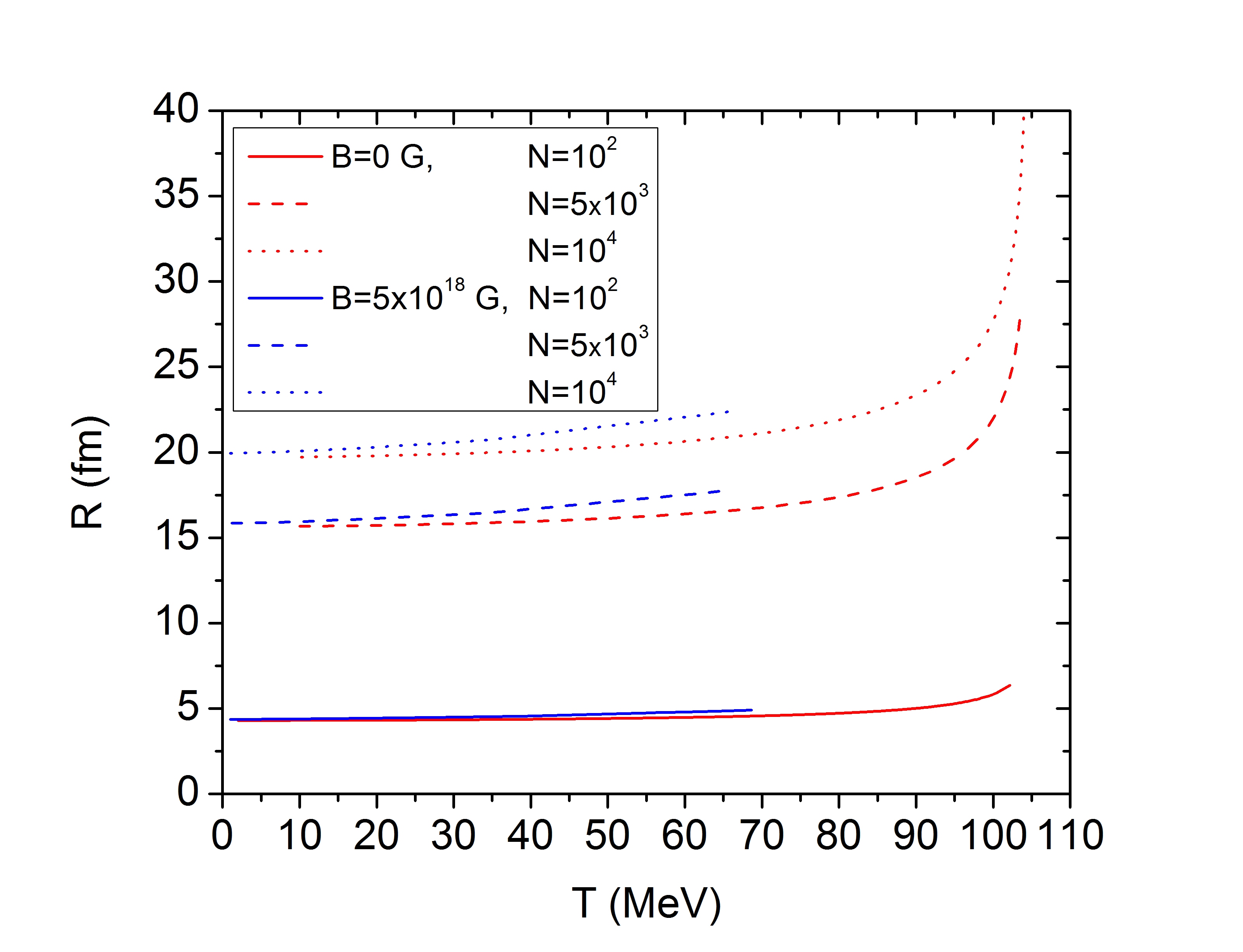}
\caption{Dependence of $R$ for MCFL strangelets on the magnetic field and temperature at a fixed baryon number, $B_\text{bag}=75$~MeV fm$^{-3}$ and $\Delta=100$ MeV.}\label{BRATMCFL}
\end{figure}

Analogously to MSQM strangelets, the little variations that suffers the radius of MCFL strangelets with respect to the magnetic field and temperature, for a fixed baryon number, is due to the mass of the $s$ quarks. This can be seen in Figs.(\ref{BRATMCFL}), which in the left panel show the behavior of the MCFL strangelets radius with $A=100$ at $T=0,15$ ~MeV respectively, as a function of magnetic field. In the right panel is shown the behavior of $R$ with the temperature, for $ A=100, 5000$ and $A=10000$, with $\mathcal{B}=5.0\times10^{18}$ ~G, and in both cases $B_\text{bag}=75$ ~MeV fm$^{-3}$ and $\Delta=100$ MeV.

\subsubsection{Carga eléctrica}

\begin{figure}[h!t]
\begin{center}
\includegraphics[width=0.7\textwidth]{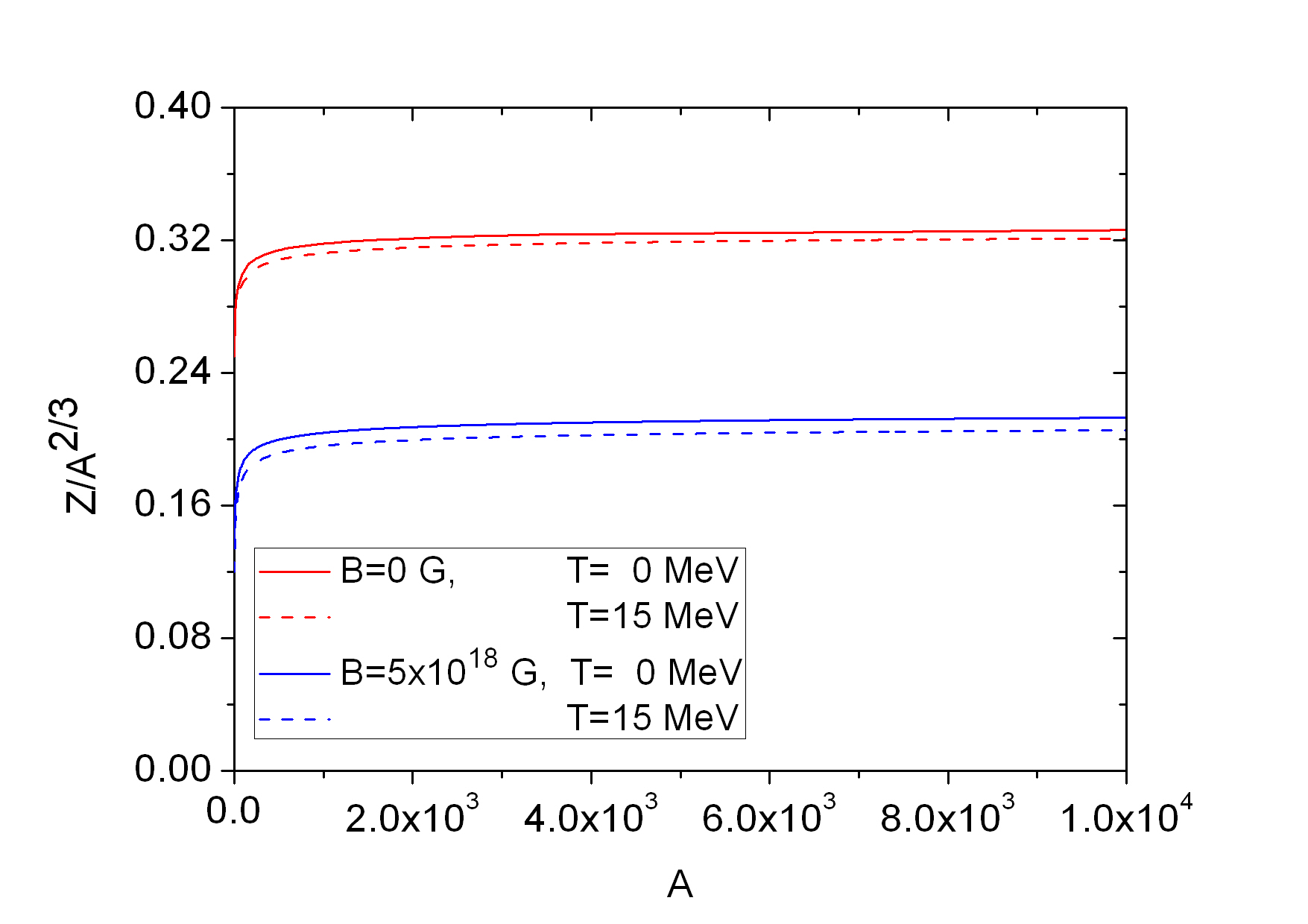}
\end{center}
\caption{Electric charge of MCFL and CFL strangelets as a function of the baryon number for $\mathcal{B}=5.0\times10^{18}$~G, $T=0,15$~MeV, $B_\text{bag}=75$~MeV fm$^{-3}$ and $\Delta=100$~MeV.}\label{ZACFL}
\end{figure}

Finally, in Fig.(\ref{ZACFL}) is shown the behavior of the electric charge with the baryon number, for MCFL and CFL strangelets. As already discussed, the effects of the temperature tend to decrease the values of the electric charge; the bulk contribution vanishes by the conditions given in Eq.\eqref{Nequality}; to the total charge, which decreases for the magnetic field chosen, only contribute quarks near the surface. In Fig.(\ref{ZACFL}), is shown the ratio $Z/A^{\frac{2}{3}}$ to observe better the constant behavior of $Z$ with the baryon number for large values of $A$.

\bigskip
For large baryon numbers, the electric charge behaves as $Z=Z_{0}A^{\frac{2}{3}}$:
\begin{equation}\label{Z0BCFL}
\left.Z\right|_{0}\;\,=0.21\,A^{\frac{2}{3}},\;\;\;\;\left.Z\right|_{15}=0.20\,A^{\frac{2}{3}},
\end{equation}
\noindent for MCFL strangelets, while 
\begin{equation}\label{Z0B0CFL}
\left.Z\right|_{0}\;\,=0.32\,A^{\frac{2}{3}},\;\;\;\;\left.Z\right|_{15}=0.31\,A^{\frac{2}{3}},
\end{equation}
\noindent for CFL strangelets. Notice that from Eqs.\eqref{Z0BCFL} and \eqref{Z0B0CFL}, the magnetic field chosen reduces the surface electric charge as well as the temperature.

%
%

%

\chapter*{Conclusions and recommendations}\label{chap5}
\addcontentsline{toc}{chapter}{Conclusions and recommendations}

\noindent

\subsubsection{Conclusions}

In this thesis we have studied bounded states of a gas of quarks and gluons: strangelets. Using the Liquid Drop Model formalism of the MIT Bag Model, it was found that the stability of these states depends strongly on the surface properties and other characteristic thermodynamic parameters of the system, to say, the magnetic field, temperature, Bag constant $B_{\text{bag}}$ and energy gap $\Delta$. Two different cases of magnetised strangelets were considered: those formed by quarks in the MSQM phase and in the MCFL phase. A comparison with the results obtained by other researchers in the field to zero magnetic field cases in both phases was also performed. We examined how the stability, size and electrical charge of strangelets are modified by the presence of strong magnetic fields and temperature.

\bigskip
\bigskip

The main results of this work were:

\begin{enumerate}

\item The presence of a strong magnetic field produces a lower $E/A$, contributing positively to the stability of strangelets in both phases of the quark matter. As one should expect, temperature effects do the opposite by increasing the $E/A$ reaching values such that there's still a range in baryon numbers where there's no stability compared to the $E/A$ of the isotope $^{56}$Fe. At $T=0$, strangelets achieve the minimum energy allowed by the space of parameters under consideration.

\item Depending on the chosen space of parameters, I find ranges of baryon number where still $E/A\leq930\,\text{MeV}.$ It is still proven that, this range is even greater when the magnetic field effects are included, and also the effects of the color superconductivity; obtaining a case of absolute stability for $\Delta=100\text{MeV}.$ These results do not constitute a decisive prove since the MIT Bag Model is just an approximation of QCD in a very special regime.

\item The radii of strangelets increase with the temperature and also with the magnetic field. The first produces an increase in the thermal energy of gas of quarks and gluons; while for the magnetic field, the confining energy exerted by the surface decreases. Theoretical studies on the Mass-Radius relations of magnetized QSs with color superconductivity effects included, reproduce a similar result. The gap energy also increases the radius of strangelets.

\item The electric charge of strangelets in both phases decreases with the temperature due to the presence of the antiparticles. The magnetic field contributes to the increase of the screening of the electric charge, while the surface charge decreases due to the boundary conditions in the Liquid Drop Model. The screening charge dominates in the phase MSQM, while only contributes the surface charge in the MCFL phase. This last could be an important property in detecting strangelets in particle colliders or coming from astrophysical sources.

\item The ratio $Z/A$ represents another important feature that would allow us to distinguish the phase of the strange quark matter present in strangelets. For large values of $A$, strangelets in MSQM phase exhibit a constant rate of $Z/A^{0.53}\to Z/A^{2/3}$, while for small baryon numbers would be constant the ratio $Z/A.$ For strangelets in the MCFL phase, the situation is similar: for large baryon number, the ratio $Z/a^{2/3}$ would be constant, while for small values of $A$ the ratio $Z/A^{1/3}$ would be constant.

\item Finally, in both cases studied, the solutions obtained always lead to positively charged configurations of strangelets.

\end{enumerate}


\subsubsection{Directions to future works}

From the results, many are lines of research can be addressed in the future and that are derived from the results of this thesis. They could cover the following topics:

\begin{itemize}

\item Repeat the same study of strangelets in the presence of strong magnetic fields using the shell model.

\item Extend the Multiple Reflection-Expansion Method to those cases where the magnetic field is strong enough, to study the expressions of $\Omega_{f,s},$ $\Omega_{f,c}$ in the presence of those strong fields.

\item Include the effects of the magnetic field in the gap energy $\Delta$.
\end{itemize}

%

%





\begin{thebibliography}{99}

\bibitem{Yndurain} F. J. Yndur\'{a}in, The theory of quark and gluon interactions (Springer-Verlag, Heilderberg, 1999).

\bibitem{GW} D. J. Gross, F. Wilczec, Phys. Rev. Lett. {\bf 30}, 1343 (1973).

\bibitem{Politzer} H. D. Politzer, Phys. Rev. Lett. {\bf 30}, 1346 (1973).

\bibitem{SB} S. Bethke, Prog. Part. Nucl. Phys. {\bf 58}, 351-386, (2007).

\bibitem{RHICI}I. V. Selyuzhenkov [STAR Collaboration], Rom. Rep. Phys. {\bf
58}, 049 (2006); D. E. Kharzeev, Phys. Lett. {\bf B633}, 260 (2006);
D. E. Kharzeev, L. D. MacLerran and H. J. Warriga, arXiv:0711.0950
[hep-ph].

\bibitem{Baym:2006rq}
  G.~Baym,
  AIP Conf.\ Proc.\  {\bf 892}, 8 (2007)

\bibitem{Bodmer:1971we}
  A.~R.~Bodmer,
  Phys.\ Rev.\  D {\bf 4}, 1601 (1971).

\bibitem{Bailin:1983bm}
  D.~Bailin and A.~Love,
  Phys.\ Rept.\  {\bf 107}, 325 (1984).


\bibitem{Alford:2001zr}
  M.~G.~Alford, K.~Rajagopal, S.~Reddy and F.~Wilczek,
  Phys.\ Rev.\  D {\bf 64}, 074017 (2001).


\bibitem{Farhi:1984qu}
  E.~Farhi and R.~L.~Jaffe,
  Phys.\ Rev.\  D {\bf 30}, 2379 (1984).

\bibitem{RHICII}
  M.~Gyulassy, L.~McLerran,
  Nucl.\ Phys.\ A {\bf 750}, 30 (2005);

\bibitem{Klingenberg:2001qs}
  R.~Klingenberg,
  J.\ Phys.\ G {\bf 27}, 475 (2001).


\bibitem{Finch:2006pq}
  E.~Finch,
  J.\ Phys.\ G {\bf 32}, S251 (2006).


\bibitem{Felipe:2007vb}
  R.~G.~Felipe, A.~P.~Mart\'{\i}nez, H.~P~Rojas and M.~G.~Orsaria,
  Phys.\ Rev.\  C {\bf 77}, 015807 (2008).


\bibitem{Felipe:2008cm}
  R.~G.~Felipe and A.~P.~Mart\'{\i}nez,
  J.\ Phys.\ G {\bf 36}, 075202 (2009).

\bibitem{Martinez:2010sf}
  A.~P.~Mart\'{\i}nez, R.~G.~Felipe, D.~M.~Paret,
  Int.\ J.\ Mod.\ Phys.\ D {\bf 19}, 1511 (2010).

\bibitem{Ferrer:2010wz}
  E.~J.~Ferrer, V.~de la Incera, J.~P.~Keith {\it et al.},
  Phys.\ Rev.\ C {\bf 82}, 065802 (2010).

\bibitem{Chakrabarty:1996te}
  S.~Chakrabarty,
  Phys.\ Rev.\  D {\bf 54}, 1306 (1996).


\bibitem{Duncan:1992hi}
  R.~C.~Duncan, C.~Thompson,
  Astrophys.\ J.\  {\bf 392}, L9 (1992).

\bibitem{Kouveliotou:1998ze}
  C.~Kouveliotou, S.~Dieters, T.~Strohmayer {\it et al.},
  Nature {\bf 393}, 235 (1998).

\bibitem{Madsen:1998uh}
  J.~Madsen,
  Lect.\ Notes Phys.\  {\bf 516}, 162 (1999).

\bibitem{Chao:1995bk}
  W.~Q.~Chao, C.~S.~Gao, Y.~B.~He {\it et al.},
  Phys.\ Rev.\  C {\bf 53}, 1903 (1996).


\bibitem{Paulucci:2008jd}
  L.~Paulucci and J.~E.~Horvath,
  Phys.\ Rev.\  C {\bf 78}, 064907 (2008).

\bibitem{Wen:2005uf}
  X.~J.~Wen, X.~H.~Zhong, G.~X.~Peng {\it et al.},
  Phys.\ Rev.\ C {\bf 72}, 015204 (2005).


\bibitem{Gilson:1993zs}
  E.~P.~Gilson, R.~L.~Jaffe,
  Phys.\ Rev.\ Lett.\  {\bf 71}, 332 (1993).


\bibitem{Madsen:2001fu}
  J.~Madsen,
  Phys.\ Rev.\ Lett.\  {\bf 87}, 172003 (2001).


\bibitem{MBuballa2005}
  M.~Buballa,
  Phys.\ Rept.\  {\bf 407}, 205-376 (2005).


\bibitem{CH1}
  A.~Chodos, R.~L.~Jaffe, K.~Johnson, C.~B.~Thorn and V.~F.~Weisskopf,
  Phys.\ Rev.\  D {\bf 9}, 3471 (1974).

\bibitem{CH2}
  A.~Chodos, R.~L.~Jaffe, K.~Johnson and C.~B.~Thorn,
  Phys.\ Rev.\  D {\bf 10}, 2599 (1974).



\bibitem{Lepage2003}
  G.~Peter Lepage,
  Phys.\ Rev.\ Lett.\ {\bf 92}, 022001 (2004).


\bibitem{Heller}
U.~M.~Heller
Eur. Phys. Journal A \textbf{31}, (4) (2007).

\bibitem{Ratti:2006wg}
  C.~Ratti, S. Roessner, M. A. Thaler, W. Weise
  Eur.\ Phys.\ J.\ {\bf C49}, 213-217 (2007).


\bibitem{DG}
  T.~A.~DeGrand, R.~L.~Jaffe, K.~Johnson and J.~E.~Kiskis,
  Phys.\ Rev.\  D {\bf 12}, 2060 (1975).


\bibitem{Witten:1984rs}
  E.~Witten,
  Phys.\ Rev.\  D {\bf 30}, 272 (1984).



\bibitem{Terazawa} H. Terazawa, INS-Report 336, Univ. of Tokyo (1979).

\bibitem{Itoh:1970uw}
  N.~Itoh,
  Prog.\ Theor.\ Phys.\  {\bf 44}, 291 (1970).

\bibitem{AuErMilRich} A. Pérez Martínez, M. Orsaria, R. González Felipe,  E. López
Fune. Materia extraña en el universo. Rev. Mex. Fís. E 54 (2008)175.

\bibitem{Weber} F. Weber, Prog. Part. Nucl. Phys. {\bf 54}, 193 (2005).

\bibitem{alemanes}
  S. Bernhard R\"{u}ster,
  arXiv:nucl-th/0612090v1 (2006).


\bibitem{Rajagopal:2000ff}
  K.~Rajagopal and F.~Wilczek,
  Phys.\ Rev.\ Lett.\  {\bf 86}, 3492 (2001).


\bibitem{Alford:2007xm}
  M.~G.~Alford, A.~Schmitt, K.~Rajagopal and T.~Schafer,
  Rev.\ Mod.\ Phys.\  {\bf 80}, 1455 (2008).

\bibitem{Ivanenko1969}
D. D. Ivanenko and D. F. Kurdgelaidze, Lett. Nuovo Cim. \textbf{II S1}, 13-16 (1969).


\bibitem{Ivanenko:1965dg}
  D.~D.~Ivanenko and D.~F.~Kurdgelaidze,
  Astrophysics {\bf 1}, 251 (1965)
  [Astrofiz.\  {\bf 1}, 479 (1965)];
  D.~Ivanenko and D.~F.~Kurdgelaidze,
  Lett.\ Nuovo Cim.\  {\bf 2}, 13 (1969).

\bibitem{Balian:1970fw}
  R.~Balian, C.~Bloch,
  Annals Phys.\  {\bf 60}, 401 (1970).


\bibitem{Heiselberg:1993dc}
  H.~Heiselberg,
  Phys.\ Rev.\  D {\bf 48}, 1418 (1993).

\bibitem{Endo:2005zt}
  T.~Endo, T.~Maruyama, S.~Chiba and T.~Tatsumi,
  Prog.\ Theor.\ Phys.\  {\bf 115}, 337 (2006).


\bibitem{Alford:2006bx}
  M.~G.~Alford, K.~Rajagopal, S.~Reddy and A.~W.~Steiner,
  Phys.\ Rev.\  D {\bf 73}, 114016 (2006).

\bibitem{CRMaite} Carlos Rodríguez Castellanos, María Teresa Pérez Maldonado, \textbf{Introducción a la Física
Estadística}, ed Félix Varela, La Habana (\textbf{2002}).


\bibitem{Kasuya1993}
  M.~Kasuya, et al,
  Phys.\ Rev.\ D.\  {\bf 47}, 2153 (1993).

\bibitem{Ichimura1993}
  M.~Ichimura, et al,
  Il Nuovo Cim.\ A.\  {\bf 106}, 843 (1993).


\bibitem{Saito1995}
  T.~Saito,
  Proc.\ 24 ICRC.\ Rome.\ {\bf 1}, 898 (1995).


\bibitem{Capdeville1996}
  J. N.~Capdeville,
  Il Nuovo Cim.\ C.\  {\bf 19}, 623 (1996).


\bibitem{Banerjee2006}
  S. Banerjee, S. K. Ghosh, S. Raha, D. Syam,
  arXiv:hep-ph  $0006286v1$

\bibitem{Lourenco:2002c}
  C.~Lourenco,
  Nucl. Phys. A\  {\bf 698}, 13c (2002).

\bibitem{Wiener2006}
  M.~Wiener,
  Int. J. Mod. Phys. E\  {\bf 15}, 37 (2006).

\bibitem{Angelis:2001c}
  A.L.S. Angelis, et al.,
  Nucl. Phys. B\ (Proc. Suppl.)  {\bf 97}, 227 (2001).

\bibitem{Armstrong:1997l}
  T.A. Armstrong, et al.,
  Phys. Rev. Lett. \textbf{79} 3612 (1997), Nucl. Phys. A \textbf{625} 494 (1997), Phys. Rev. C \textbf{63} 054903(2001) .

\bibitem{Klingenberg:1996a}
  R. Klingenberg et al.,
  Nucl. Phys. A {\bf 610}, 306 (1996).

\bibitem{Boligan} M. Boligán Expósito, Estrellas de Quarks Magnetizadas, Tesis de Maestría 2008.


\bibitem{Martinez:2003dz}
  A.~P.~Mart\'{\i}nez, H.~P.~Rojas and H.~J.~Mosquera Cuesta,
  Eur.\ Phys.\ J.\  C {\bf 29}, 111 (2003).


\bibitem{Alford:1999pb}
  M.~G.~Alford, J.~Berges and K.~Rajagopal,
  Nucl.\ Phys.\  B {\bf 571}, 269 (2000).

\bibitem{Gorbar:2000ms}
  E.~V.~Gorbar,
  Phys.\ Rev.\  D {\bf 62}, 014007 (2000).


\bibitem{Fukushima:2007fc}
  K.~Fukushima and H.~J.~Warringa,
  Phys.\ Rev.\ Lett.\  {\bf 100}, 032007 (2008).


\bibitem{Felipe:2010vr}
  R. Gonz\'{a}lez Felipe, D. Manreza Paret, A. P\'{e}rez Mart\'{\i}nez,
  Eur.\ Phys.\ J.\ A {\bf 47}, 1 (2011).


\bibitem{Ferrer:2005vd}
  E.~J.~Ferrer, V.~de la Incera and C.~Manuel,
  Phys.\ Rev.\ Lett.\  {\bf 95}, 152002 (2005);
  E.~J.~Ferrer, V.~de la Incera and C.~Manuel,
  Nucl.\ Phys.\  B {\bf 747}, 88 (2006);
  E.~J.~Ferrer and V.~de la Incera,
  Phys.\ Rev.\  D {\bf 76}, 045011 (2007).

\bibitem{Cristinam:2007}
  C. Manuel,
  Nuc.\ Phys.\ A {\bf 785}, (2007).


\bibitem{Noronha:2007wg}
  J.~L. Noronha and I.~A.~Shovkovy,
  Phys.\ Rev.\  D {\bf 76}, 105030 (2007).


\bibitem{Paulucci:2010uj}
  L.~Paulucci, E.~J.~Ferrer, V.~de la Incera and J.~E.~Horvath,
  Phys.\ Rev.\  D {\bf 83} (2011) 043009.

\bibitem{Schmitt:2002sc}
  A.~Schmitt, Q.~Wang, D.~H.~Rischke,
  Phys.\ Rev.\  D {\bf 66}, 114010 (2002).


\bibitem{Alford:2002kj}
  M.~Alford and K.~Rajagopal,
  JHEP {\bf 06}, 031 (2002).


\bibitem{Alford:2004pf}
  M.~Alford, M.~Braby, M.~W.~Paris and S.~Reddy,
  Astrophys.\ J.\  {\bf 629}, 969 (2005).








%
%
%
%
%
%





\end{thebibliography}
\end{document}